\documentclass[aps,prl,twocolumn,longbibliography,superscriptaddress,10pt,floatfix]{revtex4-2}
\PassOptionsToPackage{table,dvipsnames}{xcolor}
\usepackage{xcolor}
\usepackage{blindtext}
\usepackage{graphicx}
\usepackage{pgf}
\usepackage{url}
\usepackage{layouts}
\usepackage{gensymb}
\usepackage{amsmath}
\usepackage{textcomp}
\usepackage{textgreek}
\usepackage{tabularx}
\usepackage{array}
\usepackage[caption=false]{subfig}
\usepackage[shortlabels]{enumitem}
\usepackage{mathtools}
\usepackage{soul}
\usepackage{amssymb}
\usepackage{qip}
\usepackage{pifont}
\usepackage{multirow}
\usepackage[T1]{fontenc}
\usepackage{colortbl}
\usepackage{booktabs}
\usepackage{siunitx}

\usepackage{bbm}
\usepackage{braket}
\usepackage{ulem}
\usepackage{nicefrac}
\usepackage{makecell}
\usepackage{hyperref}

\usepackage{qip}

\usepackage{lipsum}

\newcommand*{\mline}[2]{%
\begingroup
    \renewcommand*{\arraystretch}{1.1}%
   \begin{tabular}[c]{@{}>{\centering\arraybackslash}p{#1cm}@{}}#2\end{tabular}%
  \endgroup
}

\newcolumntype{C}[1]{>{\centering\arraybackslash}m{#1}}

\definecolor{green1}{rgb}{0.33, 0.7, 0.69}
\hypersetup{colorlinks=true,linktoc=all,linkcolor=blue,breaklinks=true,citecolor=green1,urlcolor=blue}

\renewcommand{\thesection}{\Roman{section}}
\begin{document}

\title{Surpassing the loss-noise robustness trade-off in quantum key distribution}

\author{Hannah Seabrook}
\thanks{Corresponding author: hannah.seabrook@bristol.ac.uk}
\affiliation{Quantum Engineering Technology Labs, H. H. Wills Physics Laboratory and Department of Electrical \& Electronic Engineering, University of Bristol, BS8 1FD, United Kingdom}
\affiliation{Quantum Engineering Centre for Doctoral Training, Centre for Nanoscience and Quantum Information, University of Bristol, United Kingdom}

\author{Emilien Lavie}
\affiliation{Quantum Engineering Technology Labs, H. H. Wills Physics Laboratory and Department of Electrical \& Electronic Engineering, University of Bristol, BS8 1FD, United Kingdom}

\author{Teodor Str\"{o}mberg}
\affiliation{Institute of Science and Technology Austria, Klosterneuburg, Austria}

\author{Matthew P. Stafford}
\affiliation{Quantum Engineering Technology Labs, H. H. Wills Physics Laboratory and Department of Electrical \& Electronic Engineering, University of Bristol, BS8 1FD, United Kingdom}
\affiliation{Quantum Engineering Centre for Doctoral Training, Centre for Nanoscience and Quantum Information, University of Bristol, United Kingdom}

\author{Giulia Rubino}
\thanks{Corresponding author: giulia.rubino@bristol.ac.uk}
\affiliation{Quantum Engineering Technology Labs, H. H. Wills Physics Laboratory and Department of Electrical \& Electronic Engineering, University of Bristol, BS8 1FD, United Kingdom}
\affiliation{H. H. Wills Physics Laboratory, University of Bristol, Tyndall Avenue, Bristol, BS8 1TL, United Kingdom}

\date{\today}
\begin{abstract}
Quantum key distribution (QKD) offers a theoretically secure method to share secret keys, yet practical implementations face challenges due to noise and loss over long-distance channels. Traditional QKD protocols require extensive noise compensation, hindering their industrial scalability and lowering the achievable key rates. Alternative protocols encode logical qubits in noise-resilient states, but at the cost of using many physical qubits, increasing susceptibility to loss and limiting transmission distance. In this work, we introduce a logical qubit encoding that uses antisymmetric Bell-states in the continuous photonic degrees of freedom, frequency and time. By leveraging the continuous space, we overcome this noise-loss robustness trade-off by minimising the number of photons per logical qubit, whilst optimising the encoding resilience over noise fluctuations. We analyse the security of our encoding and demonstrate its robustness compared to existing state-of-the-art protocols. This approach provides a path towards scalable, efficient QKD implementations under realistic noise conditions.
\end{abstract}

\maketitle

\section*{Introduction}

Quantum information theory reached key milestones with the establishment of early protocols such as BB84~\cite{BB84_1} and advancements like Shor’s algorithm~\cite{Shor1994}, which showcased its transformative potential for real-world applications.
These developments solidified the field's focus on practical challenges, laying the foundation for quantum technologies today~\cite{Acín_2018}. 
Among these technologies, quantum key distribution (QKD) stands out as a leading candidate for near-term applications and commercialisation due to its potential for fundamentally secure communication~\cite{HoiKwong_1999, Mayers2001}. 
However, transitioning QKD from theoretical models to scalable, industrial implementations remains a significant challenge~\cite{Diamanti2016}. 
Achieving high key rates across a large-scale QKD network requires extensive stabilisation and compensation techniques to overcome the noise experienced by qubits in transmission \cite{Peranic2023, Liu2021, Zhou2023}.
For example, twin-field QKD (TF-QKD), while enabling state-of-the-art key rates, requires the stabilisation of two lasers' frequencies, polarisations, and field phases across long-distance noisy channels \cite{Lucamarini2018, Clivati2022, Li2023}. 
Impressive systems have been demonstrated experimentally that actively stabilise these elements \cite{Pittaluga2021,Chen2022,Wang2022,Zhou2023,Chen2024}. 
Nonetheless, this raises the question of whether an alternative approach to implementing QKD could offer a path forward without the substantial overheads required for compensated QKD.

Here, we investigate such an alternative solution: devising a QKD protocol that encodes information into inherently noise-resilient states. 
Such protocols have previously been proposed, e.g., for resilience against collective unitary noise channels \cite{XBWang2005, Boileau, efficient_collective_noise, Guo:20, 6state_DFS_improved}, and they typically encode logical qubits in the subspace of a higher-dimensional space formed from many photonic qubits.
However, these approaches inherently involve a trade-off between loss-robustness and noise-robustness (see Table~\ref{tab:tradeoff}).
On the one hand, increasing the number of qubits improves the noise-robustness of the logical subspace, and beyond a minimum qubit threshold, the logical states can be made perfectly noise-robust.
On the other hand, if even one photon is lost in transmission, the whole logical state is compromised.
Thus, increasing the number of photons in the logical state makes it exponentially more susceptible to photon loss, severely limiting the distance over which the secret key can be shared.

\setlength{\tabcolsep}{0pt}
\renewcommand{\arraystretch}{1.4}
\begin{table*}
    \centering
    \begin{tabular}{|C{5.2cm}|C{1.4cm}|C{2.8cm}|C{2.6cm}|C{3.46cm}|}
    \hline
    \rowcolor{gray!20} 
            \mline{5.2}{\textbf{Protocol}} & \mline{1.4}{\textbf{Loss scaling}} & \mline{2.8}{\textbf{No stabilisation needed?}} & \mline{2.6}{\textbf{Robust to collective unitary noise?}} & \mline{3.46}{\textbf{Robust to randomly varying unitary noise parameters?}}  \\ 
            \hline

        \rowcolor{gray!10} TF-QKD \cite{Lucamarini2018} & \cellcolor{gray!10} \textcolor[HTML]{1B9E77}{$\eta^{1/2}$}
        & \cellcolor{gray!10} \textcolor[HTML]{D73027}{\ding{55}}
        & \cellcolor{gray!10} \textcolor[HTML]{D73027}{\ding{55}}
        & \cellcolor{gray!10} \textcolor[HTML]{D73027}{\ding{55}} \\
    \hline
    
    BB84 \cite{BB84_1} & \textcolor[HTML]{1B9E77}{$\eta$}
        & \textcolor[HTML]{D73027}{\ding{55}}
        & \textcolor[HTML]{D73027}{\ding{55}} 
        & \textcolor[HTML]{D73027}{\ding{55}}\\
    \hline

    \rowcolor{gray!10} Wang (2005) \cite{XBWang2005} & \cellcolor{gray!10} \textcolor[HTML]{66C2A5}{$\eta^2$}
        & \cellcolor{gray!10} \textcolor[HTML]{D73027}{\ding{55}}
        & \cellcolor{gray!10} \textcolor[HTML]{1B9E77}{\ding{51}}
        & \cellcolor{gray!10} \textcolor[HTML]{D73027}{\ding{55}}\\
    \hline

    \textbf{This work} & \textcolor[HTML]{66C2A5}{$\eta^2$}
        & \textcolor[HTML]{1B9E77}{\ding{51}}
        & \textcolor[HTML]{1B9E77}{\ding{51}}
        & \textcolor[HTML]{1B9E77}{\ding{51}}\\
    \hline

    \rowcolor{gray!10} Boileau \textit{et al.} (2004) 3-photon \cite{Boileau} & \cellcolor{gray!10} \textcolor[HTML]{FC8D62}{$\eta^3$}
        & \cellcolor{gray!10} \textcolor[HTML]{1B9E77}{\ding{51}}
        & \cellcolor{gray!10} \textcolor[HTML]{1B9E77}{\ding{51}}
        & \cellcolor{gray!10} \textcolor[HTML]{1B9E77}{\ding{51}} \\
    \hline

    Boileau \textit{et al.} (2004) 4-photon \cite{Boileau} & \textcolor[HTML]{D73027}{$\eta^4$}
        & \textcolor[HTML]{1B9E77}{\ding{51}}
        & \textcolor[HTML]{1B9E77}{\ding{51}}
        & \textcolor[HTML]{1B9E77}{\ding{51}} \\
    \hline

    \rowcolor{gray!10} Li \textit{et al.} (2008) \cite{efficient_collective_noise} &  \cellcolor{gray!10} \textcolor[HTML]{D73027}{$\eta^4$}
        & \cellcolor{gray!10} \textcolor[HTML]{1B9E77}{\ding{51}}
        & \cellcolor{gray!10} \textcolor[HTML]{1B9E77}{\ding{51}}
        & \cellcolor{gray!10} \textcolor[HTML]{1B9E77}{\ding{51}} \\
    \hline
            
    \end{tabular}
    \caption{\textbf{Comparison of commonly employed protocols and existing noise-robust protocols in terms of loss- and noise-robustness.} The table illustrates the trade-off between loss and noise robustness in QKD protocols, which we further analyse in Appendix~\ref{sec:tradeoff}, and highlights how our protocol improves upon existing approaches. Protocols are listed with increasing susceptibility to photon loss down the table, where $\eta = 10^{-\text{loss(dB)}/10}$ is the channel transmissivity. Twin-Field QKD (TF-QKD) \cite{Lucamarini2018} and BB84 \cite{BB84_1} are not robust to noise fluctuations, requiring extensive stabilisation techniques for their effective operation. As we show in Section~\ref{sec:noisycomp}, Ref.~\cite{XBWang2005} shows better robustness to collective unitary noise than BB84. However, its key rate vanishes under time-varying noise parameters, thus necessitating noise compensation methods. As our protocol utilises both frequency-bin and time-bin qubits, in Section~\ref{sec:FBS} we outline the model used to assess the robustness of our protocol on conjugate qubits, in the form of a unitary on frequency-bin qubits.}
    \label{tab:tradeoff}
\end{table*}

In this work, we introduce an encoding scheme that reduces the number of photons in a logical qubit state, thereby minimising the impact of loss, whilst increasing the robustness of the logical state to noise. 
Inspired loosely by bosonic codes --- a field of quantum error correction that uses the redundancy of infinite-dimensional Hilbert spaces to encode logical qubits and mitigate physical errors~\cite{Berent2024} --- we exploit similar redundancies in a continuous photonic degree of freedom (DoF) to achieve noise resilience.
Specifically, our encoding represents logical qubit states in superpositions across two-photon antisymmetric Bell-states defined over frequency bins and time bins.
The antisymmetric structure of the states ensures intrinsic robustness to noise. 
At the same time, by utilizing superpositions of excitations across modes of a continuous photonic DoF, our encoding reduces the number of photons required for a logical qubit, thus offering a low-loss alternative to noise-robust QKD protocols.

We illustrate the robustness of our protocol under two continuous channel noise models: one experimentally motivated (chromatic dispersion) and the other motivated to enable a comparison with existing discrete variable (DV) encoded protocols (a unitary applied to frequency-bin qubits).
Additionally, we demonstrate that the parameters of our encoding can be optimised over fluctuations in noise parameters.

Finally, we perform a rigorous and unified key rate analysis of our protocol under the noise models and loss channel, and compare the key rates to those of existing noise-robust protocols \cite{XBWang2005, Boileau, efficient_collective_noise}.
We show that our protocol surpasses the noise-resilience of two-photon protocols \cite{XBWang2005} and the loss-robustness of perfectly noise-cancelling protocols \cite{Boileau, efficient_collective_noise}.
These results highlight how encoding in continuous DoFs can enhance the robustness of secure communication protocols.

In the following section, we present our encoding.
In Section \ref{sec:noiserob}, we demonstrate the noise-robust properties of our encoding. 
In Section \ref{sec:security}, we analyse and discuss the security of our protocol and existing protocols in lossy and collective unitary noise channels. 
We conclude by discussing our findings and their implications and suggesting possible directions for future work.

\section{Protocol} \label{sec:protocol}

\begin{figure}[h]
    \centering
    \includegraphics[width=\columnwidth]{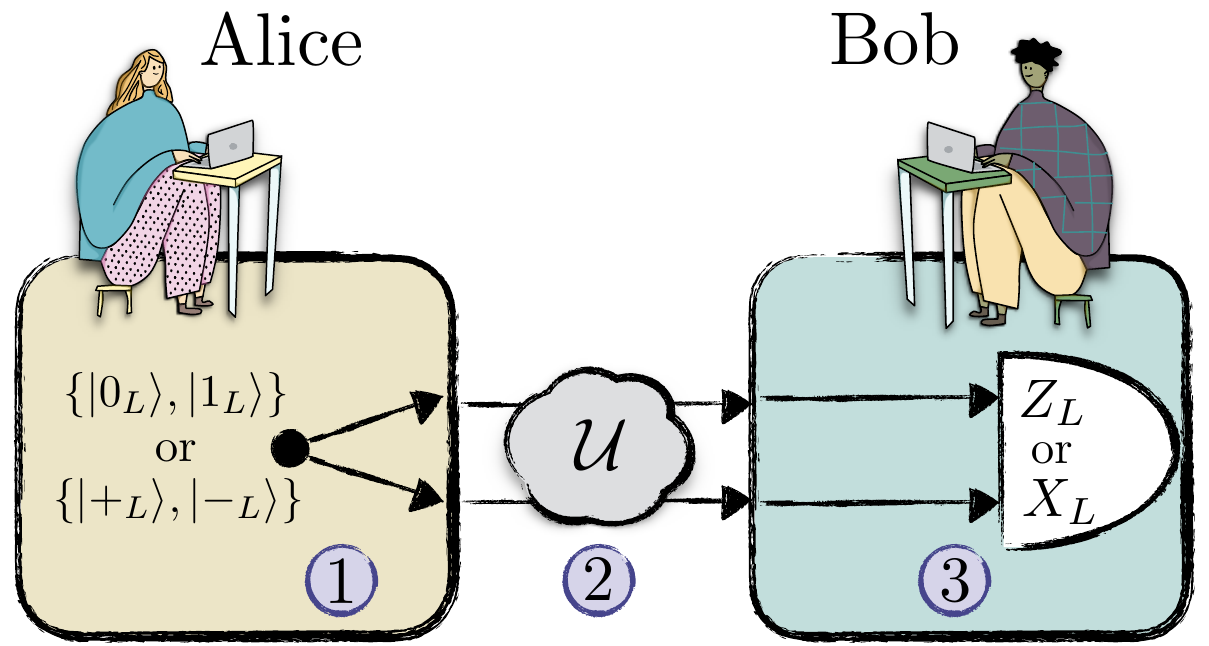}
    \caption{\textbf{Depiction of the BB84 protocol using logical noise-robust qubit states.} We propose a BB84 protocol using logical noise-robust states to improve key rates in noisy channels. Alice prepares a logical qubit in either the $Z_L$ or $X_L$ basis, randomly, encoding the bit value 0 in $\ket{0_L}$ or $\ket{+_L}$ and the bit value 1 in $\ket{1_L}$ or $\ket{-_L}$. She sends the state to Bob, who measures randomly in either the $Z_L$ or $X_L$ basis. After transmission, Alice and Bob discard cases with mismatched bases and measurement outcomes outside the logical subspace, and check for errors due to an eavesdropper on a subset of the remaining bits. If the error rate is below a certain threshold, they proceed with error correction and privacy amplification, otherwise they abort the protocol. We assume the channel induces `collective noise', applying the same unknown transformation $\mathcal{U}$ to each photon in the logical state.}
    \label{fig:BB84}
\end{figure}

Our protocol follows the BB84 scheme \cite{BB84_1}, with single photon states replaced by two-photon logical qubit states encoded in a continuous DoF for noise-resilience, as illustrated in Fig.~\ref{fig:BB84}.
In the protocol, a sender (Alice) prepares states randomly in either the logical $Z$ (which we label $Z_L$) basis $\{\ket{0_L}, \ket{1_L}\}$, or the logical $X$ ($X_L$) basis $\{\ket{\pm_L} = \bigl(\ket{0_L} \pm \ket{1_L}\bigr)/\sqrt{2}\}$. 
Similarly, a receiver (Bob) measures randomly in either $Z_L$ or $X_L$. Bob's projectors are given by $\{\ketbra{0_L}{0_L}, \ketbra{1_L}{1_L}, \ketbra{+_L}{+_L}, \ketbra{-_L}{-_L},  \ketbra{\perp_L}{\perp_L}\}$, where $\ket{\perp_L}$ corresponds to an inconclusive measurement due to loss or the measurement of a state outside of the logical subspace.

To describe our encoding, we simply need to establish the encoding of the $Z_L$ basis states $\{\ket{0_L}, \ket{1_L}\}$, as the $X_L$ basis states and the measurement operators follow trivially.
We use antisymmetric Bell-states to encode $\ket{0_L}$ and $\ket{1_L}$ in such a way that they are naturally robust to noise, as in the existing DV protocols \cite{Boileau,efficient_collective_noise, XBWang2005} (reviewed in Supplemental Material (SM), Sec.~\ref{sec:litrev}). These states are invariant under local collective unitary operations, i.e., $(U \otimes U)\ket{\Psi^-} = e^{i\Delta}\ket{\Psi^-}$ where $\ket{\Psi^-} = \frac{1}{\sqrt{2}}(\ket{0}\ket{1} - \ket{1}\ket{0})$ and $\Delta$ is some $U$-dependent global phase factor \cite{nielsen_chuang_2010} (see Appendix~\ref{sec:tradeoff}). 
To minimise the number of photons per logical state, we encode the $Z_L$ basis states as two-photon antisymmetric Bell-states in conjugate variables, frequency and time. 

Our encoding is built in two layers. 
The first layer is formed of single-photon physical qubits. 
In this layer, we define single-photon physical qubit states in conjugate variables: the single-photon frequency-bin qubit states that we label by $\{\ket{0_f}, \ket{1_f}\}$, and the single-photon time-bin qubit states that we label as $\{\ket{0_t}, \ket{1_t}\}$. 
The second layer is the logical encoding layer (labelled by subscript-$L$). 
In the logical layer, we construct two two-photon antisymmetric Bell-states, one from the frequency-bin qubits of the first layer, and the other from the time-bin qubits.
These two antisymmetric Bell-states act as the building blocks of our logical qubit space $\{\ket{0_L}, \ket{1_L}\}$.

We start by defining the first layer of the encoding using bosonic creation operators on the vacuum state $\ket{\varnothing}$.
We define the creation operator for a photon occupying a frequency-bin mode as
\begin{align}
    \hat{a}^{\dagger}_{\Omega_i,M} = \int d\omega f_{\Omega_i, \sigma_{\omega}}(\omega)\hat{a}^{\dagger}_{M}(\omega),
\end{align}
where $\hat{a}^{\dagger}_{M}(\omega)$ is the creation operator for a photon in some mode $M$ (for example, a polarisation mode or a spatial mode) with frequency $\omega$, and $f_{\Omega_i, \sigma_{\omega}}(\omega)$ is the frequency-bin mode amplitude function, which is a sharply peaked function centred around $\Omega_i$ with width $\sigma_{\omega}$. 

Similarly, we define the creation operator for a photon occupying a time-bin mode as
\begin{align}
 \hat{a}^{\dagger}_{\tau_i,M} = \int dt f_{\tau_i, \sigma_{t}}(t)\hat{a}^{\dagger}_{M}(t) = \int d\omega F_{\tau_i, \sigma_{t}}(\omega)\hat{a}^{\dagger}_{M}(\omega),
\end{align}
where $\hat{a}^{\dagger}_{M}(t)$ is the creation operator for a photon in some mode $M$ at time $t$ and $f_{\tau_i, \sigma_{t}}(t)$ is the time-bin mode amplitude function, sharply peaked in time, centred around $\tau_i$ with width $\sigma_{t}$.
The second equality shows that this can be expressed as a creation operator on frequency modes by applying the Fourier transform between frequency and time. 
In this context, $F_{\tau_i, \sigma_{t}}(\omega)$ is the Fourier transform of $f_{\tau_i, \sigma_{t}}(t)$, and $\omega$ is now defined relative to the central frequency of the states.

Defining two frequency bins centred at frequencies $\Omega_0$ and $\Omega_1$ and two time bins centred at times $\tau_0$ and $\tau_1$, we can now define single-photon frequency-bin qubit and time-bin qubit states as
\begin{equation} \label{eq:i_x}
    \ket{i_{x}}_M = \hat{a}^{\dagger}_{\mu_i,M}\ket{\varnothing},
\end{equation}
where $i\in \{0,1\}$ and $x \in \{f, t\}$, where $f$ stands for the frequency-bin states and $t$ is for the time-bin states, and $\mu_i$ is $\Omega_i$ or $\tau_i$, respectively. This assumes that $\Omega_0$ and $\Omega_1$, and $\tau_0$ and $\tau_1$ are sufficiently well separated such that $\braket{0_x|1_x}_M \approx 0$.
For brevity, we label the parameters of the frequency-bin and time-bin qubits by $\vec{\boldsymbol{\Omega}} = (\Omega_0, \Omega_1, \sigma_{\omega})$ and $\vec{\boldsymbol{\tau}} = (\tau_0, \tau_1, \sigma_t)$, respectively.

\begin{figure}[t]
    \centering
    \includegraphics[width=\columnwidth]{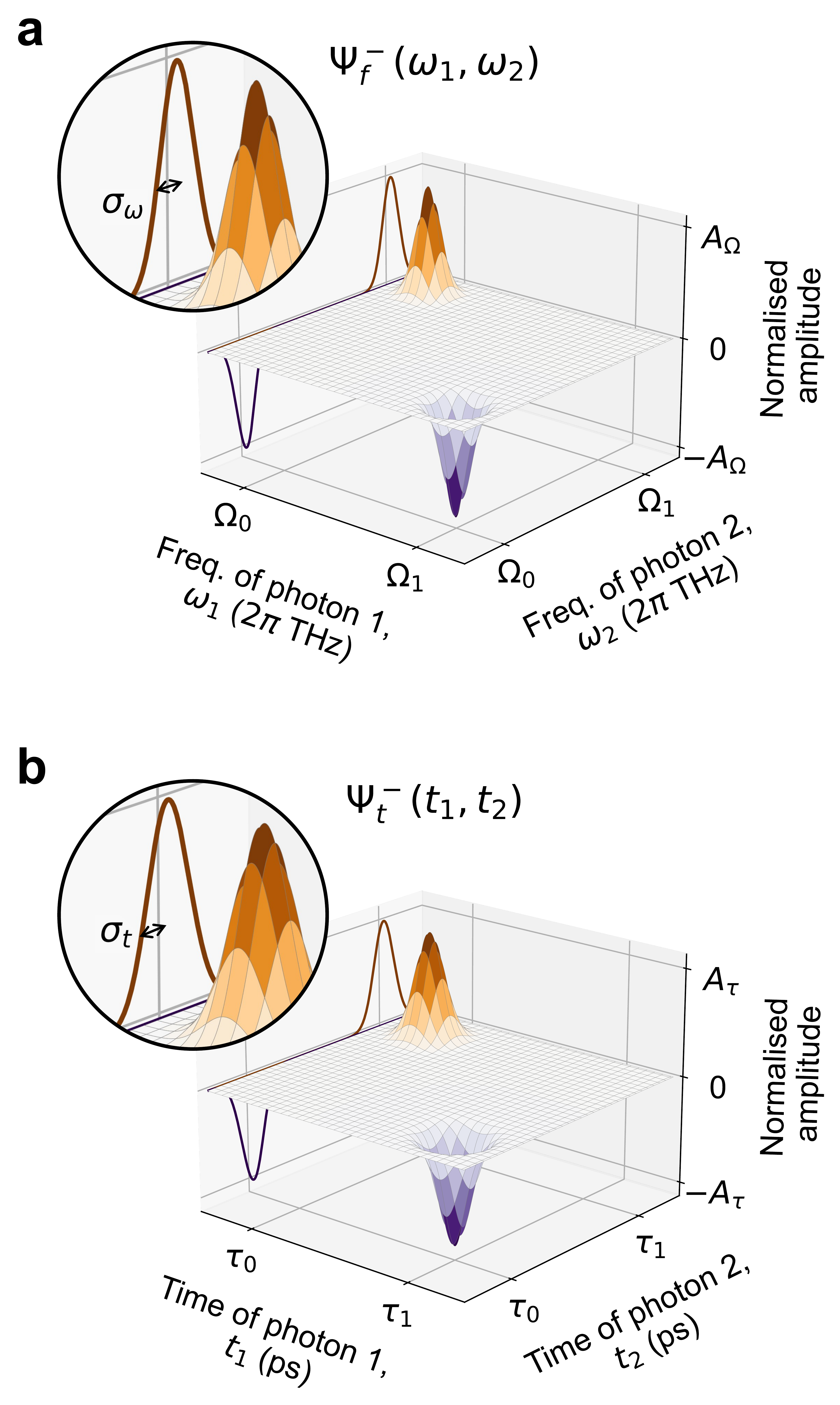}
    \caption{\textbf{2D joint frequency and temporal spectra of the frequency-bin and time-bin encoded antisymmetric Bell-states used to form the logical states $\ket{0_L}$ and $\ket{1_L}$.} (a) The normalised bi-photon wavefunction amplitude of the antisymmetric frequency-bin encoded Bell-state $\ket{\Psi^-_f}$ visualised in the frequency domain and (b) the normalised bi-photon wavefunction amplitude of the antisymmetric time-bin encoded Bell-state $\ket{\Psi^-_t}$ visualised in the time domain, for amplitude functions modelled as state-normalised Gaussians. $A_{\Omega}$ and $A_{\tau}$ are the peak amplitudes of the wavefunctions (set by the normalisation conditions and the encoding parameters $\vec{\boldsymbol{\Omega}}$ and $\vec{\boldsymbol{\tau}}$, respectively -- see SM, Sec.~\ref{sec:overlap}). The insets highlight the widths, $\sigma_{\omega}$ and $\sigma_t$, of the 1D frequency bins and time bins.}
    \label{fig:encoding}
\end{figure}

We now establish the logical encoding layer, formed from bi-photon antisymmetric Bell-states in frequency bins and time bins.
We define these as
\begin{equation} \label{eq:bellstate}
    \ket{\Psi^-_x} = \frac{\ket{0_x}_M\ket{1_x}_N - \ket{1_x}_M\ket{0_x}_N}{\sqrt{2}},
\end{equation}
where the labels $M$ and $N$ signify that the two photons are distinguished by some extra mode (e.g., spatial mode or polarisation). 
The states $\ket{\Psi^-_f}$ in the frequency domain and $\ket{\Psi^-_t}$ in the time domain are depicted in Fig.~\ref{fig:encoding}, with Gaussian amplitude functions used as an illustrative example.
$\ket{\Psi^-_f}$ and $\ket{\Psi^-_t}$ define a 2-dimensional logical Hilbert space.
The logical states in the $Z_L$ basis are given by
\begin{equation}
\ket{0_L} = \ket{\Psi^-_f} \qquad \ket{1_L} = \ket{\Psi^-_{\perp f}},
\end{equation}
where $\ket{\Psi^-_{\perp f}}$ is a state orthogonal to $\ket{\Psi^-_f}$, defined according to 
\begin{equation}\label{eq:psiperp}
    \ket{\Psi^-_{\perp f}} = \frac{(\ket{\Psi^-_t} - a\ket{\Psi^-_f})}{\sqrt{1-|a|^2}},
\end{equation} where $a$ is the overlap of the frequency-bin and time-bin Bell-states, $\braket{\Psi^-_f|\Psi^-_t}$. 
Typical experimental parameters give a negligible value for $a$. In fact, we show that $\ket{\Psi^-_f}$ and $\ket{\Psi^-_t}$ can be made perfectly distinguishable (i.e., $a=0$) for both Gaussian and Lorentzian amplitude functions by tuning $\vec{\boldsymbol{\Omega}}$ and $\vec{\boldsymbol{\tau}}$ (see SM, Sec.~\ref{sec:overlap}).
In Appendix~\ref{app:gen&meas}, we propose one possible method for preparing these states and measuring in the two bases, using domain-engineered non-linear crystals.

In the next section, we demonstrate the robustness of this encoding to different noise channels and their parameter fluctuations.

\section{Noise-robustness of encoding} \label{sec:noiserob}

We now showcase the resilience of our encoding against two example channel noise models, for state amplitudes modelled as Gaussian functions. (As justified in SM, Sec.~\ref{sec:CVnoise}, utilising more general peaked amplitude functions should yield results analogous to those of Gaussian states.)
First, we briefly demonstrate the robustness of our encoding under a common experimental noise model (dispersion), illustrating the applicability of our encoding to typical experimental scenarios. This initial analysis is heuristic, with a detailed quantitative study and figures provided in the SM, Sec.~\ref{sec:dispanalysis}.
The remainder of this section is dedicated to in-depth analysis of our encoding's resilience to a theoretically-motivated noise model (the ``frequency beamsplitter'' (FBS)). This model enables us to compare the efficacy of our encoding’s noise-robustness to that of the existing protocols in Refs~\cite{Boileau,XBWang2005, efficient_collective_noise}, which we subsequently present in Section~\ref{sec:security}.

\subsection{Dispersion noise} \label{sec:dispmain}

Chromatic dispersion is an optical property of transparent media in which the phase velocity and group velocity of light traversing a medium are frequency-dependent \cite{Paschottachromatic_dispersion} (typical in optical fibres, for example \cite{Willner2005}).
In this subsection, we present the analytical noise model of chromatic dispersion and summarise the effect of this noise on our encoding and its key rates. 

By expanding the frequency-dependent wavevector of light in a dispersive media about some frequency $\omega_0$, the effect of $n$-th order dispersion on the photon's spectral amplitude can be written as
\begin{equation}\label{eq:dispersion}
    g(\omega) \mapsto e^{i\alpha_n(\omega-\omega_0)^n} g(\omega),
\end{equation}
where $\alpha_n$ is the corresponding dispersion coefficient (as derived in the SM, Sec.~\ref{sec:dispanalysis}).
Typical materials only need consideration of the group delay dispersion (GDD), $n=1$ term, and the group velocity dispersion (GVD), $n=2$ term. 
When higher-order terms have to be considered, it is often more convenient to use direct numerical modelling \cite{Paschottachromatic_dispersion}.

GDD ($n=1$) physically corresponds to a simple time-delay of the pulses sent by Alice, as understood by the time-shift property of the Fourier transform. 
Both our protocol and current QKD experiments can cancel this using common synchronisation techniques based on the arrival time of a reference signal from Alice at Bob \cite{Bienfang:04, Vagniluca2020, TimeBinQKD, Wang:21}.
Synchronisation ensures Alice and Bob assign the same sequence number to each event \cite{Miller2023}, and that the time-bin basis state labels of Alice's encoding align with those of Bob's measurement for time-bin encoded QKD. 
However, our encoding relaxes the synchronisation requirements needed for typical time-bin encoded QKD.
Only synchronising the events' sequence number is necessary, as an offset between Alice and Bob's time-bin labels results in an incurred loss rather than a quantum bit error rate (QBER) for our protocol (see SM, Sec.~\ref{sec:n=1disp}). 

GVD ($n=2$) instead changes the shape of the temporal probability amplitude by broadening it. 
For single-qubit time-bin encoded QKD, this broadening effect compromises the security of the protocol over intermediate distances ($\approx$ 50 km of fibre), unless appropriate compensation is employed \cite{TimeBinQKD}.
This occurs due to the QBER rapidly increasing as the pulses spread into the neighbouring time bins.
In contrast, for our encoding, GVD increases the effective loss, due to an increasing rate of inconclusive outcomes as the photons' probability density ``leaks'' outside of Bob's logical measurement subspace.
However, this does not raise the QBER, allowing our protocol to maintain security under this noise at long distances.

Overall, the effect of chromatic dispersion on our encoding results in decaying oscillations of key rate as a function of $|\alpha_n|$.
The decay rate is due to the increased inconclusive outcomes and depends on the encoding parameters $\vec{\boldsymbol{\Omega}}$ and $\vec{\boldsymbol{\tau}}$. 
The oscillations arise from an $\alpha_n$-dependent decoherence phase imparted on the single-qubit frequency-bin states, due to the difference in dispersive phase between the two central frequencies. 
This corresponds to a decoherence phase at the logical layer of our encoding, resulting in an oscillating phase error rate, with a period depending again on $\vec{\boldsymbol{\Omega}}$ and $\vec{\boldsymbol{\tau}}$.
The encoding parameters can be optimised to maintain significant key rates across the channel, by ensuring the key rate oscillation period is large compared to the variation of the dispersion parameters, $\alpha_n$, (e.g., due to varying temperature \cite{Andre2005}). 
This allows the key rate to remain close to a maximum within the fluctuation range of $\alpha_n$.
Figures, further mathematical analysis, and the consideration of realistic noise and non-optimal experimental parameters are presented in the SM, Sec.~\ref{sec:dispanalysis}.

\subsection{Frequency beamsplitter (FBS) noise} \label{sec:FBS}
To compare the noise resilience of our protocol against existing protocols that leverage antisymmetric Bell-states (Refs.~\cite{Boileau,XBWang2005,efficient_collective_noise}), we now consider an alternative noise model based on frequency-bin qubits.
While the physical encoding of Refs.~\cite{Boileau,XBWang2005,efficient_collective_noise} can be expressed in a discrete Hilbert space, our logical qubit encoding utilises two distinct physical qubit types defined in conjugate variables, requiring representation in a continuous space.
To address this, we define the FBS noise model, which applies a unitary operation on frequency-bin qubits, enabling a direct comparison of the noise-robustness of these protocols and ours (assuming the former are encoded in frequency-bin qubits).
Although rooted in theoretical considerations, this noise model has potential experimental relevance, for example, in processes such as Bragg scattering four-wave mixing \cite{Lee2024}.

\subsubsection{Frequency beamsplitter model}

We first define a single-frequency beamsplitter (BS) operation. 
A single-frequency BS acts on pairs of single-frequency creation operators $\hat{a}^{\dagger}(\omega)$ to apply a BS unitary operation, $U_{BS}$, on the frequency modes in the form $\hat{a}^{\dagger}(\omega_1) \mapsto \cos \theta \hat{a}^{\dagger}(\omega_1) + e^{i\phi} \sin \theta \hat{a}^{\dagger}(\omega_2)$ and $\hat{a}^{\dagger}(\omega_2) \mapsto - e^{-i\phi}\sin \theta \hat{a}^{\dagger}(\omega_1) + \cos \theta \hat{a}^{\dagger}(\omega_2)$. 
This applies a unitary operation between a pair of single-frequency modes, with rotation noise parameter $\theta$, and dephasing noise parameter $\phi$.

\begin{figure}[h]
    \centering
    \includegraphics[width=0.9\columnwidth]{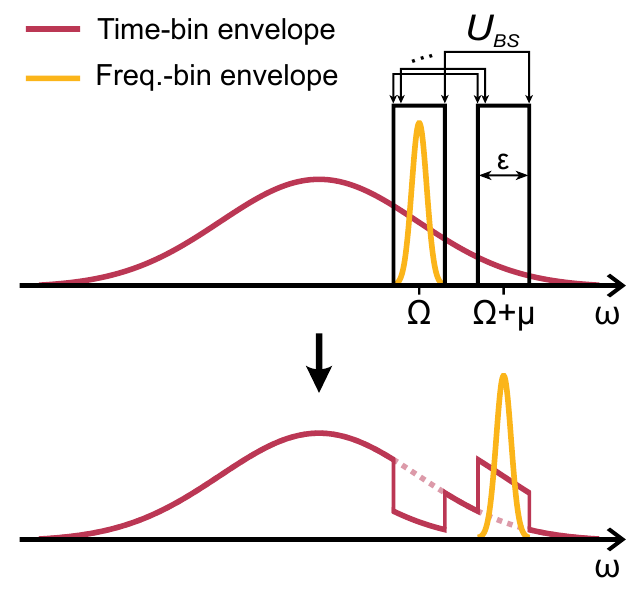}
    \caption{\textbf{Schematic of the frequency beamsplitter (FBS) noise model.} The FBS model applies a beamsplitter unitary ($U_{\mathrm{BS}}$) across pairs of frequency modes separated by $\mu$ over continuous frequency intervals of width $\epsilon$, as shown in the top diagram. For the example of a bit-flip ($X$) operation on single-photon time-bin and frequency-bin states, this operation transforms the pink and yellow states in the top diagram to the corresponding states in the bottom diagram. The pink curve represents the broad spectral amplitude envelope of time-bin states, whilst the orange curve depicts a frequency-bin state.}
    \label{fig:CVnoisemodel}
\end{figure}

We now generalise the single-frequency BS to define the general FBS noise model, illustrated in Fig.~\ref{fig:CVnoisemodel}. 
In the general FBS model, the same (but unknown) single-frequency BS operations act between pairs of frequencies, separated by $\mu$, over continuous frequency intervals of width $\epsilon$. 
The first interval is centred on frequency $\Omega$ and, correspondingly, the second interval on the frequency $\Omega + \mu$. 
For brevity, we label the FBS noise parameters by $\vec{\boldsymbol{\epsilon}} = (\Omega, \mu, \epsilon)$. 
The action of this model in the frequency domain is made mathematically explicit in the SM, Sec.~\ref{sec:CVnoise} and further extended in SM, Sec.~\ref{sec:multiU}.

Before analysing the impact of the FBS noise model on our encoding, it is helpful to summarise the key parameters introduced so far. 
The frequency-bin qubit states are parameterised by frequencies $\Omega_0$ and $\Omega_1$, and Gaussian width $\sigma_{\omega}$. 
The time-bin qubit states are parameterised by times $\tau_0$ and $\tau_1$, and Gaussian width $\sigma_t$. 
We then parameterise the FBS noise model as a unitary operation across frequency modes with rotation angle $\theta$ and dephasing angle $\phi$. 
This FBS noise acts on frequency bins centred around $\Omega$ and $\Omega + \mu$, with a top-hat filter shape of width $\epsilon$.

\subsubsection{Effect of frequency beamsplitter noise} \label{sec:FBSeffect}
To analyse the effect of the FBS on our encoding, we must first make assumptions about the relationship of parameters $\vec{\boldsymbol{\epsilon}}$ to $\vec{\boldsymbol{\Omega}}$.
We assume that the FBS acts exactly across the frequency-bin states of Alice's encoding (i.e. $\Omega = \Omega_0$ and $\Omega+\mu = \Omega_1$), and the noise bands are sufficiently wide to capture the frequency-bin amplitude function, as in Fig.~\ref{fig:CVnoisemodel} (requiring $\epsilon \gtrsim 6\sigma_{\omega}$ to contain all but a negligible fraction of a Gaussian function).
Under these assumptions, the FBS applies a unitary operation on the frequency-bin qubit states $\{\ket{0_f}, \ket{1_f}\}$. 
Thus, $\ket{0_L} = \ket{\Psi^-_f}$ is unaffected by this noise.

In contrast, the $\ket{1_L}$ state will be affected under the FBS. 
Due to the broad nature of the time bins in the frequency domain, both $\ket{0_t}$ and $\ket{1_t}$ span both of the FBS bins (pink curve in Fig.~\ref{fig:CVnoisemodel}).
Therefore, the FBS does not apply a logical unitary to the time-bin qubit states, but does change the shape of their distribution (with a dependence on $\vec{\boldsymbol{\epsilon}}, \theta, \phi$), as depicted in Fig.~\ref{fig:CVnoisemodel} and SM, Fig.~\ref{fig:timetrans}.
As with GVD, the changing time-bin state distribution increases the rate of inconclusive measurement outcomes at Bob due to the probability amplitude ``leaking'' outside of the logical subspace.

To quantitatively analyse the effect this has on the key rate, we consider the states measured by Bob ($\ket{\Psi^-_t}$ and $\ket{\Psi^-_f}$) and the time-bin state arriving at Bob after traversing the FBS noise, which we call $\ket{\Psi'^-_t}$.
As $a$ is typically small, we set $a=0$ here to provide a clearer, more intuitive picture of the scheme’s robustness. 
However, the security analysis in the following section fully accounts for the true, non-zero (albeit small) value of $a$ determined by the chosen encoding parameters.
We calculate the fidelities $\mathcal{F}(\Psi'^-_t, \Psi^-_t) = |\braket{\Psi'^-_t|\Psi^-_t}|^2$ and $\mathcal{F}(\Psi'^-_t, \Psi^-_f) = |\braket{\Psi'^-_t|\Psi^-_f}|^2$, and any phase accumulated by $\ket{\Psi'^-_t}$, as a function of $\vec{\boldsymbol{\epsilon}}, \theta, \phi$ and the encoding parameters $\vec{\boldsymbol{\tau}}$ and $\vec{\boldsymbol{\Omega}}$. 
In the aforementioned simplified case ($a = 0$), the sum of these fidelities equals the probability of Bob measuring a conclusive outcome given $\ket{\Psi'^-_t}$ arrived.
$\mathcal{F}(\Psi'^-_t, \Psi^-_f)$ gives the bit error rate in the logical encoding, and the phase accumulated by $\ket{\Psi'^-_t}$ relative to $\ket{\Psi^-_t}$ and $\ket{\Psi^-_f}$ gives rise to a phase error rate.

\begin{figure}[h]
    \centering
    \includegraphics[width=\columnwidth]{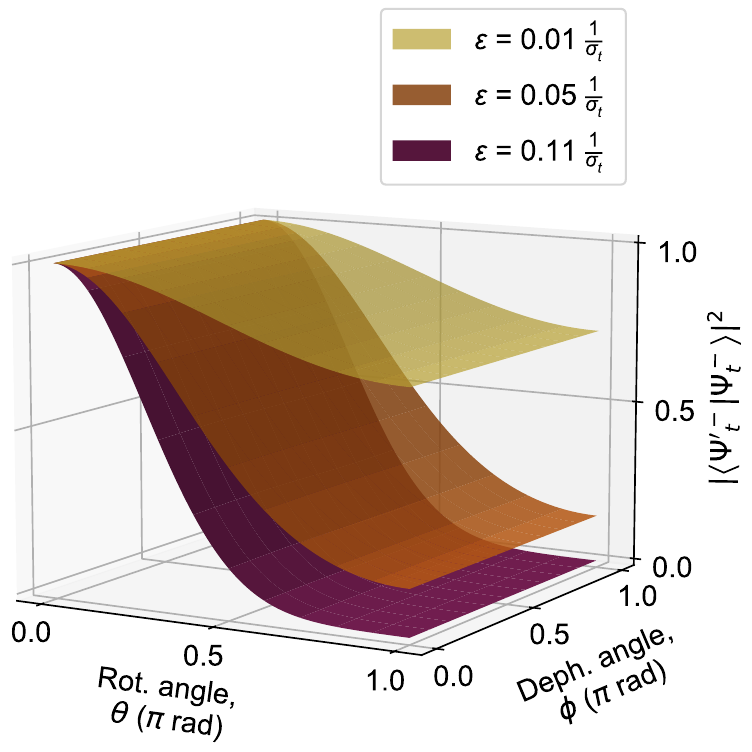}
    \caption{ \textbf{Robustness of time-bin Bell-states under the FBS noise model, for varying frequency noise bin widths.} Fidelity between the arriving state $\ket{\Psi'^-_t}$ sent by Alice through the FBS channel and the state $\ket{\Psi^-_t}$ measured by Bob. The fidelity is plotted as a function of the FBS channel parameters $\theta$ and $\phi$. For the simulation, we use a Gaussian time-bin state with parameters $\tau_0=0$ ps, $\tau_1=220$ ps, and $\sigma_t = 17$ ps (values experimentally reported in Ref~\cite{finco2024timebinentangledbellstate}) and FBS noise parameters $\Omega = 195.860$ THz, $\Omega +\mu = 195.879$ THz, with varying width $\epsilon =$ 0.6 GHz, 3.0 GHz, 6.6 GHz (close to experimentally reported values for a frequency-bin Bell-state encoding in Ref~ \cite{Clementi2023}, with $\epsilon = 6\sigma_{\omega}$). In the legend, we express $\epsilon$ as a fraction of the time-bin state's width in the frequency domain ($\propto \frac{1}{\sigma_t}$). We take the the central frequency of the time-bin state to equal $\Omega$.}
    \label{fig:staterobustness}
\end{figure}

The bit error is negligible here and thus has negligible effect on the key rate. 
For example, for the experimentally reported parameters used in Fig.~\ref{fig:staterobustness}, $\mathcal{F}(\Psi'^-_t, \Psi^-_f)$ is numerically found to be $~10^{-4}$ for varying $\theta, \phi$ (giving less than 0.3\% reduction in key rate).
Further, for typical parameters (such as those used in Fig.~\ref{fig:staterobustness}), the accumulated relative phase is small and has minimal effect on the key rate.
Therefore, the effect of the FBS on the key rate is dominated by $\mathcal{F}(\Psi'^-_t, \Psi^-_t)$.
Further details on the accumulated phase and bit error can be found in SM, Sec.~\ref{sec:phaseacc} and \ref{sec:biterror}, respectively.

Fig.~\ref{fig:staterobustness} shows the fidelity $\mathcal{F}(\Psi'^-_t, \Psi^-_t)$ of a time-bin Bell-state under an FBS, using experimentally reported values, for various noise bin widths $\epsilon$, as a function of $\theta$ and $\phi$. 
We observe that the state fidelity is approximately independent of the FBS dephasing angle $\phi$, and reduces with FBS rotation angle $\theta$. 
As $\epsilon$ decreases, the fidelity is higher for a given $\theta$ and $\phi$. 
If the time-bin states are sufficiently wide in the frequency domain ($\propto 1/\sigma_t$) compared to $\epsilon$, the effect of the noise is small on the time-bin states. 
Thus, our encoding will show a high rate of conclusive outcomes at Bob, giving high key rates, for any values of the noise parameters $\theta,\phi$.
The detailed analytical expression for $\mathcal{F}(\Psi'^-_t, \Psi^-_t)$ using Gaussian amplitude functions and the study of further parameter dependence is presented in the SM, Sec.~\ref{sec:fidelityanalytic}.

In this analysis, we demonstrated the robustness of our encoding against variations in noise parameters using fixed settings. Alternatively, our protocol can be adapted by optimising the encoding parameters to enhance robustness when noise fluctuations can be characterized, as detailed in Appendix~\ref{app:optimise}. The fixed-parameter approach is ideal when noise characterization is impractical, while the second strategy enables fine-tuned performance in scenarios where detailed noise characterization is possible.

Overall, the effect of the FBS noise channel can be effectively understood as an amplitude damping and decoherence channel on our logical states, where the damping probability and decoherence phase are dependent on the channel noise parameters $\vec{\boldsymbol{\epsilon}}, \theta, \phi$. 

In Appendix~\ref{app:compencoding}, we highlight the advantages of our proposed encoding compared to alternative continuous DoF encodings with only two photons. 
Specifically, we demonstrate how our use of both time-bin and frequency-bin qubits allows us to make fewer assumptions on the channel noise model.

\section{Security analysis} \label{sec:security}

In this section, we present a comparative security analysis of our protocol against existing ones that are robust against collective unitary noise \cite{Boileau, XBWang2005} and the standard BB84 protocol \cite{BB84_1} under the FBS model introduced above.
We extend this model to analyse the security under experimentally realistic noise fluctuations, by using a probabilistic value for the noise parameter $\theta$.
Subsequently, we compare these schemes under a lossy channel, additionally including TF-QKD in the analysis. 
Despite its alternative set-up --- featuring two senders and a central measuring device, unlike the prepare-and-measure BB84-type scheme used by the others --- the inclusion of TF-QKD allows for a rigorous comparison of our protocol with state-of-the-art approaches. 
We conclude by discussing how the stabilisation techniques required for TF-QKD and BB84 affect their key rates and implementation overheads.
We leave the analysis of Ref.~\cite{efficient_collective_noise} to SM, Sec.~\ref{sec:4photoncomp}, as it does not contribute further insights to the main discussion.

It is worth noting that the protocols in Refs.~\cite{Boileau, efficient_collective_noise, XBWang2005} were developed during the early years of QKD, when robustness against specific individual attacks was the primary focus. 
Here, we harness more recent developments in numerical techniques for QKD security analysis to provide asymptotic secure key rates valid against any collective attack. 
More precisely, we use the framework developed in Ref.~\cite{Winick2018reliablenumerical} to compute an asymptotic lower bound on the secure key rates for each of the protocols listed above, as well as our proposed implementation.
We provide an explanation and full details of the analysis methods used in the SM, Sec.~\ref{sec:securityframework}. 

\subsection{Noise-robust protocols under the FBS channel}\label{sec:noisycomp}

In this subsection, we consider Refs.~\cite{BB84_1, XBWang2005, Boileau, efficient_collective_noise} implemented using frequency-bin qubits (with a width less than the noise bin widths, i.e., $\epsilon \gtrsim 6\sigma_{\omega}$). With this encoding, their security can be analysed by examining the effects of a collective unitary operation (Appendix~\ref{sec:tradeoff}) on the physical qubit states, as described in the previous section.

By contrast, the FBS channel affects our logical encoding differently, causing amplitude damping and decoherence at the logical level (see Section~\ref{sec:FBSeffect}). 
To evaluate the key rate while accounting for the two-photon nature of our states, we reformulate the continuous DoF single-photon frequency-bin and time-bin qubit states of Eq.~\eqref{eq:i_x} into a discrete orthonormal ququart basis (see SM, Sec.~\ref{sec:ququartencoding}). 
By inverting this mapping, we can express our logical states in a two-ququart basis. 
Then, using the overlaps between the single-photon states comprising Bob's measurement projectors and the noise-transformed single-photon states arriving at him, we construct a channel acting on this ququart basis (as detailed this in SM, Sec.~\ref{sec:ququartchannel}). 

The resulting key rates are close to those of BB84 subject to an amplitude damping and decoherence channel, where the fidelities of our encoding under the FBS (Fig.~\ref{fig:staterobustness}) determine the amplitude damping, while the accumulated phases correspond to the decoherence phase for each set of noise parameters $(\vec{\boldsymbol{\epsilon}}, \theta, \phi)$ (see SM, Sec.~\ref{sec:ampdampmodellogical}). 
This confirms that considering the channel’s action on the states at the logical level in the previous section (i.e., without explicitly introducing the full ququart basis) provides a meaningful intuition, as it suffices to capture the impact of the noise on the key rate. 
Indeed, although a more detailed treatment is necessary for the completeness of the security analysis, this yields results that are effectively equivalent.

\begin{figure}[h]
    \centering
    \includegraphics[width=\columnwidth]{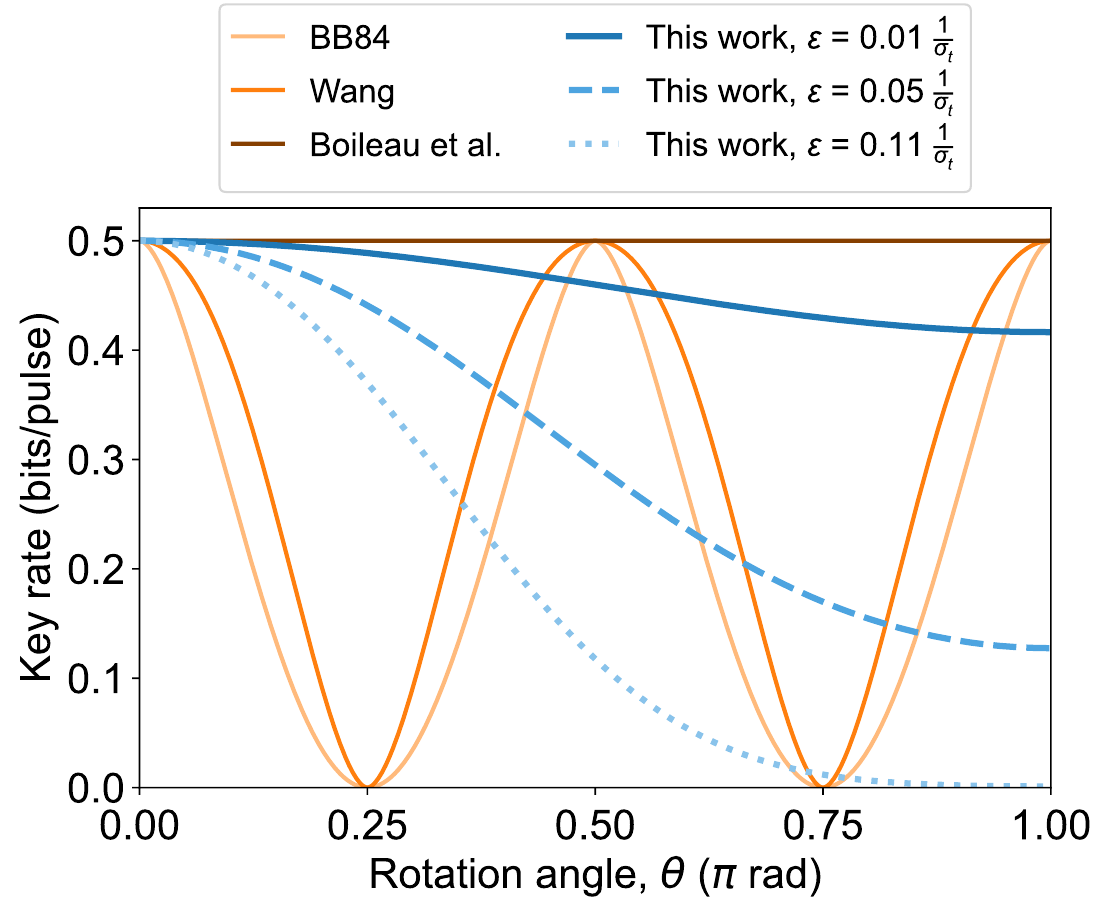}
    \caption{\textbf{Comparison of key rates as a function of FBS noise parameter $\theta$ (with $\phi=0$) for collective unitary noise robust protocols.} Key rate in an FBS channel with unitary parameter $\theta$ of BB84, Wang (2005) \cite{XBWang2005}, Boileau \textit{et al.} (2004) \cite{Boileau} and our protocol for the states and noise outlined in Fig.~\ref{fig:staterobustness} (Gaussian states with $\tau_0 = 0$ ps, $\tau_1 = 220$ ps, $\sigma_t = 17$ ps, and noise parameters $\Omega = 195.860$ THz, $\Omega + \mu = \Omega_1 = 195.879$ THz, $\epsilon = 6\sigma_{\omega} =$ 0.6, 3, 6.6 GHz). The action of the FBS on the DV protocols is analysed as a collective unitary operation on frequency-bin qubits. BB84 and Ref.~\cite{XBWang2005} show less robustness to the FBS noise than our approach. Unlike our scheme, the key rates of BB84 and Ref.~\cite{XBWang2005} vanish when $\theta$ varies with time due to their key rate oscillating with $\theta$ (see model in Eq.~\eqref{eq:mixedchan}). Ref.~\cite{Boileau} exhibits the highest key rate and robustness under FBS noise. However, it has the worst scaling with loss of the examined protocols (see Sec.~\ref{sec:lossycomp}, Fig.~\ref{fig:losscomp}).}
    \label{fig:noisycomp}
\end{figure}

Fig.~\ref{fig:noisycomp} shows the secure key rates of the different protocols in a lossless FBS channel as a function of the unitary rotation angle $\theta$, with other parameters fixed. 
The FBS noise and the encoding parameters match those of the surfaces in Fig.~\ref{fig:staterobustness}.
The key rate is symmetric in $\theta$ about $\theta=\pi$ for all protocols, due to the symmetry of the FBS operation under $\theta$.
In the SM, Sec.~\ref{sec:KR_U_prots} and \ref{sec:KRphi}, we provide an additional analysis showing the key rate's $\phi$-dependence (which is small for all protocols presented).

The three- and four-photon protocols in Ref.~\cite{Boileau} perfectly cancel collective unitary noise, therefore their key rates are maximal for both unitary parameters ($\theta, \phi$) in a lossless channel. 
We illustrate this with the flat-line reference plot in Fig.~\ref{fig:noisycomp}, and need not consider the noise robustness further here.

We now analyse the one- and two-photon protocols. 
Our protocol shows a substantial improvement in key rate compared to Refs.~\cite{BB84_1, XBWang2005} for $0<\theta\lesssim0.4\pi$. 
As the noise width reduces relative to the time-bin width in the frequency domain, the key rate robustness to $\theta$ increases (blue curves in Fig.~\ref{fig:noisycomp}).
With sufficiently narrow FBS noise bins or optimised $\sigma_t$ of our states, the key rate of our protocol remains non-zero for all unitary parameters ($\theta,\phi$), unlike Refs.~\cite{BB84_1, XBWang2005}. 
This, combined with preventing the accumulation of a bit error rate, is advantageous for our encoding when experimental time variations of noise parameters are accounted for.

We demonstrate this by updating the FBS channel model to consider random mixtures of parameters instead. 
We now model the channel as a FBS with fixed parameter values $\theta_i$, with probabilities $p_i$, so that the channel map becomes 
\begin{equation}\label{eq:mixedchan}
    \rho_N \rightarrow \sum_i p_i \hat{F}_{\theta_i}^{\otimes N}\rho_N \hat{F}_{\theta_i}^{\dagger \otimes N}, 
\end{equation}
where we define $\hat{F}_{\theta}$ as the action of the FBS with parameter $\theta$ and all the other unspecified parameters $\vec{\boldsymbol{\epsilon}}, \phi$ kept fixed. 
In our simulation, we consider points of $\theta_i$ uniformly sampled between $0$ and $2\pi$. 

Under Eq.~\eqref{eq:mixedchan}, we find the key rate for both Ref.~\cite{BB84_1} and~\cite{XBWang2005} vanishes. 
Indeed, while a positive key can be obtained both in the regime where $\theta<\frac{\pi}{4}$ and $\theta>\frac{\pi}{4}$, these two regimes correspond to Alice and Bob's bit value being correlated and anticorrelated, respectively. 
As a result, under a random mixture of $\theta$, no correlation can be established. 
In contrast, our protocol can maintain a positive key rate since the bit values remain correlated throughout the parameter range.
The dotted, dashed and solid blue lines in Fig.~\ref{fig:noisycomp} achieve rates of 0.124, 0.278, 0.457 bits/pulse, respectively. 
This also shows that a significant key rate can be achieved even with non-optimal encoding parameters (e.g., due to only partial channel characterisation) as long as an upper bound estimate on $\epsilon$ is available.

\subsection{Noise-robust protocols in lossy channels} \label{sec:lossycomp}

\begin{figure}[h]
    \centering
    \includegraphics[width=\columnwidth]{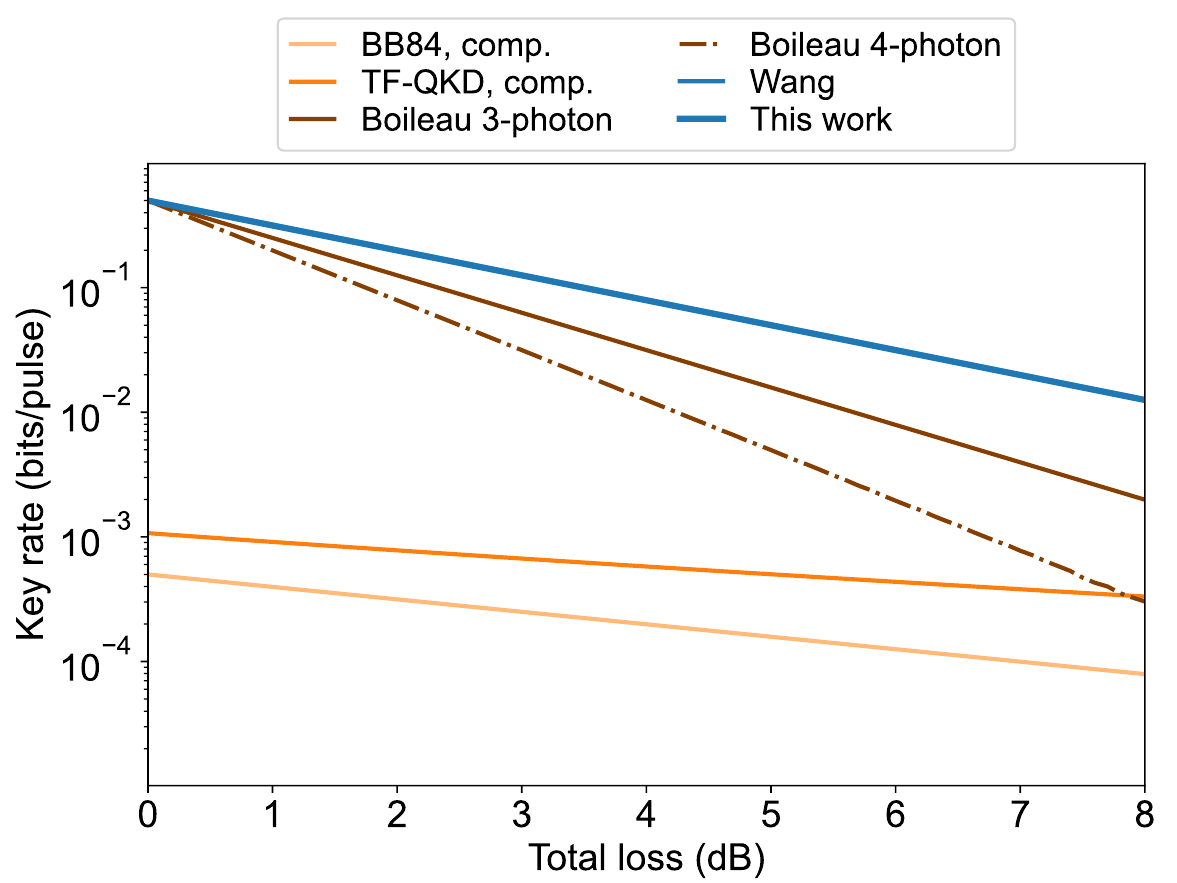}
    \caption{\textbf{Key rate as a function of channel loss in dB for collective unitary noise-robust protocols and commonly implemented QKD protocols.} Comparison of the key rate scaling with loss for various protocols in a lossy channel. Our scheme and the existing collective unitary noise-robust approaches \cite{Boileau, XBWang2005} show a BB84 scaling ($\sim\eta$) with loss. For these, the key rate for a given loss scales exponentially with the number of photons in the logical qubit state. Our protocol and Ref.~\cite{XBWang2005} use the same number of photons per logical state, and thus show the same loss dependence (thick blue line). However, Ref.~\cite{XBWang2005} shows worse performance than our scheme under channel noise (Sec.~\ref{sec:noisycomp}, cf. Fig.~\ref{fig:noisycomp}). For comparison, we also show the key rate for BB84 \cite{BB84_1} (simulated) and TF-QKD \cite{Lucamarini2018} (analytically calculated in Ref.~\cite{Lutkenhaus2018_TFQKDsecurity}, with $\mu = \mu_{opt} = 0.1146$), accounting for noise compensation techniques that are necessary for the operation these protocols, unlike the noise-robust schemes.}
    \label{fig:losscomp}
\end{figure}

We now present the analysis of different DV-QKD protocols across a loss-only channel.
Fig.~\ref{fig:losscomp} shows the key rate dependence of our protocol, noise-robust protocols with a varying number of photons per logical state~\cite{XBWang2005, Boileau}, BB84 \cite{BB84_1} and TF-QKD \cite{Lucamarini2018}.
These results highlight quantitatively the loss-noise robustness trade-off discussed earlier and presented in Table~\ref{tab:tradeoff} (and Appendix~\ref{sec:tradeoff}).
From Fig.~\ref{fig:losscomp} we confirm that, for a given channel loss, the key rate scales exponentially with the number of photons in a protocol.
However, as seen in Fig.~\ref{fig:noisycomp} above, reducing the number of photons in a logical state decreases the protocol's robustness to noise, underscoring the trade-off.

Our scheme requires only two photons per logical state, and thus scales better with loss than other noise-robust protocols. 
Meanwhile, by exploiting non-trivial encodings in continuous DoFs, it leads to an increased parameter robustness compared to the existing methods with two or less qubits (Fig.~\ref{fig:noisycomp}). 
Therefore, by optimising the encoding parameters to expected noise fluctuations (Fig.~\ref{fig:staterobustness} and Appendix~\ref{app:optimise}), our approach can surpass the key rates of previously proposed noise-robust schemes. 

In the SM, Sec.~\ref{sec:KR_DV_prots}, we present further discussion of the observed key rates for Refs.~\cite{Boileau, XBWang2005, efficient_collective_noise}, and compare the rates to the security predicted in their works.
In SM, Sec.~\ref{sec:multiphotonsec}, we also address security considerations relevant for realistic implementations that utilise photon-pair sources with imperfect photon number purity. We justify that decoy-state methods can be applied within our scheme, ensuring security against photon-number splitting (PNS) attacks.

\subsection{Compensation techniques for BB84 and TF-QKD} \label{sec:securitycomp}

By construction, the BB84 \cite{BB84_1} and TF-QKD \cite{Lucamarini2018} schemes scale better with loss than the noise-robust protocols, achieving higher key rates over longer distances.
However, implementing these protocols in practice relies on extensive compensation and stabilisation techniques to mitigate the effects of channel fluctuations (discussed further in Appendix~\ref{sec:tradeoff}) \cite{Zhou2023, Clivati2022, Wang2022, Chen2022, Pittaluga2021, Agnesi:20, Ding:17, Du2023}. 
These methods require allocating a portion of the pulses to estimating the noise, reducing the fraction available for establishing a shared secret key \cite{Agnesi:20, Ding:17, Du2023}.

To account for compensation, in Fig.~\ref{fig:losscomp}, we present key rates for BB84 and TF-QKD as experimentally reported using these techniques.
For BB84 with polarisation compensation, Ref.~\cite{Agnesi:20} presents a secret key rate of 80 bits/second with 40 dB loss and a transmission rate of 50 MHz. 
Converting this to a key rate in bits/pulse, and accounting for detection losses, this corresponds to an order $10^3$ reduction in key rate.
For TF-QKD, Ref.~\cite{Chen2022} reports a secret key rate of $9.22\times10^{-10}$ bits/pulse at 106 dB loss.
Accounting for detection and component losses, this corresponds to an order $10^2$ reduction in key rate.
We note that experiments employing different stabilisation methods will result in slight variations of the key rate reduction factor.
An exhaustive analysis of the experimentally achieved key rates would require a more open discussion on experimental downtimes, duty cycles and other factors affecting the total average key rate. 
However, detailed data in this area is currently limited.
In Appendix~\ref{app:KRlosstheoryexpt}, we include the theoretical key rates of BB84 and TF-QKD for a complete comparison.
Additionally, we present the key rate in bits/second (another important figure of merit for assessing real-world practicality), based on the estimated rate at which our logical states can be generated experimentally.

\section*{Discussion}
In this work, we introduced a novel approach to encoding logical qubits for QKD, where they are represented within subspaces of continuous photonic DoFs. 
Specifically, we define a non-trivial orthonormal basis in the frequency domain from superpositions of antisymmetric Bell-states defined in frequency-bin and time-bin qubits. 
Antisymmetry is employed for its noise-resilient properties, while the use of a continuous DoF enhances loss-robustness by reducing the number of photons per logical state.
Our analysis demonstrates that this scheme offers superior resilience to loss and noise compared to existing methods under realistic noise fluctuations.
Furthermore, it eliminates the need for extensive stabilisation typically required in state-of-the-art protocols, which often entails significant operational complexity and imposes limited duty cycles.
This passive robustness may be especially advantageous in environments where active compensation is impractical or unstable, such as channels with rapidly varying noise.
We also outlined a viable experimental implementation of our protocol, showcasing the near-term applicability of these findings for scalable QKD networks.
More broadly, our work highlights the importance of considering industrial scalability --- not just loss scaling or one-off implementation overhead --- when evaluating QKD protocols. By avoiding active compensation and relying on compact, replicable components, our approach is well-positioned for deployment at scale.

Recent works have explored architectures for quantum repeaters based on concatenated bosonic error-correcting codes, leveraging Gottesman-Kitaev-Preskill (GKP) qubits for their inherent loss-resilient properties~\cite{Rozpedek2021, Schmidt2024, Fukui2021, Rozpedek2023}. 
While our protocol does not rely on error correction, it shares a conceptual connection with these approaches by exploiting redundancies in continuous photonic DoFs to achieve resilience against noise. 
Practical realisation of GKP states, however, remains a significant experimental challenge, particularly in optical implementations, with only recent progress reported \cite{OpticalGKPStates}. 
Alternative strategies, such as that proposed by Ref.~\cite{Descamps2024} for optical quantum computing, address these challenges by encoding GKP qubits in a fixed number of excitations superposed across frequency and time modes, rather than the typical bosonic codes which encode qubits in superpositions of the number of excitations within a single mode. 
This approach provides a more experimentally feasible route to the implementation of optical bosonic error-correcting codes.

Similarly to Ref.~\cite{Descamps2024}, our approach employs a fixed number of excitations in a superposition across modes of a continuous photonic DoF to improve the practical feasibility. 
However, the encoding we propose is specifically tailored for QKD applications, enabling the use of simpler states since only noise-robustness, rather than full error correction, is required. 
Our approach opens new directions for exploring alternative encodings in continuous photonic DoFs for quantum information tasks, extending beyond the scope of QKD.

A natural extension of this research would be to adapt our encoding strategy to develop a general methodology for designing implementation-specific encodings. Such a framework could be optimised for robustness against experimentally characterised noise channels and their expected noise fluctuations, while minimising the impact of loss. For example, by modelling and parameterising the noise channel, and characterising the expected fluctuations of its parameters, it could be possible to develop tools that output optimal encoding spectral functions and their parameters for a given implementation. Additionally, while our encoding is not compatible with some widely used protocols like TF-QKD, it could hold promise for other QKD protocols, such as device-independent QKD \cite{Zapatero2023}.

Beyond these immediate extensions, our work reveals a counter-intuitive result: antisymmetric Bell-states in frequency-bin and time-bin modes, despite their conjugate nature, can be made perfectly orthogonal. This property may have deeper connections to foundational topics in quantum mechanics, such as the nature of EPR steering \cite{EPRColloqium}. While our results clearly establish the utility of this encoding for QKD, they also suggest broader implications for leveraging foundational principles in practical quantum technologies. We envision that these ideas could inspire new approaches across the field of quantum information, bridging foundational insights with real-world applications.


\section*{Acknowledgements}
We thank B.~Baragiola, A.~Boubriak, M.~Clark, M.~Jones, and P.~Skrzypczyk for useful discussions. 
\textbf{Funding:} H.S.~acknowledges financial support from EPSRC Quantum Engineering Centre for Doctoral Training grant EP/SO23607/1. E.L.~acknowledges support from the Engineering and Physical Sciences Research Council (EPSRC) Hub in Quantum Computing and Simulation (EP/T001062/1). T.S.~acknowledges that they received no funding in support for this research. M.P.S.~acknowledges support from EPSRC Quantum Engineering Centre for Doctoral Training EP/SO23607/1 and the European Commission through Starting Grant ERC-2018-STG803665 (PEQEM). G.R.~acknowledges support from the Royal Commission for the Exhibition of 1851 through a Research Fellowship, from the European Commission through Starting Grant ERC-2018-STG803665 (PEQEM) and Advanced Grant ERC-2020-ADG101021085 (FLQuant), from EPSRC through Standard Proposal Grant EP/X016218/1 (Mono-Squeeze).
\textbf{Competing interests:} The authors declare that they have no competing interests. 
\textbf{Data Availability:} All codes used to generate the data presented in this article are available at \href{https://github.com/hmis4/SurpassingLNTradeOffQKD}{https://github.com/hmis4/SurpassingLNTradeOffQKD}. 

\section*{Author contributions} The conceptualisation and methodology for this project were developed by H.S. and G.R.. All authors contributed to the validation and interpretation of this work. H.S. and E.L. carried out the formal analysis. Visualisation was conducted by H.S. and G.R.. G.R. supervised the work. All the authors contributed to scientific discussion, reviewed the draft and approved the final manuscript. 

\begin{appendix}

\section{Fundamental trade-off between loss-robust and noise-robust QKD} \label{sec:tradeoff}

In this appendix, we outline the collective unitary noise channel and the loss channel models considered for the analysis of Refs.~\cite{XBWang2005, Boileau, efficient_collective_noise}. We then discuss the overheads of employing compensation techniques to mitigate noise fluctuations and the limitations these present for the industrial scalability of QKD implementations. Finally, we explain the loss-noise robustness trade-off in more detail. We highlight this with the specific example of protocols that exploit antisymmetry for robustness to collective unitary noise, such as those in Refs.~\cite{XBWang2005, Boileau, efficient_collective_noise}.

 \subsection{Collective unitary noise channel}
A collective unitary noise channel applies the same unknown unitary $U$ to all qubits in a given $N$-qubit state, $\rho_N$, transforming the state as
\begin{equation}
    \rho_N \rightarrow U^{\otimes N} \rho_N U^{\dagger \otimes N}. 
\end{equation}

\subsection{Loss channel}
A lossy channel absorbs or loses a photon in transmission with probability $p$. 
All $N$ photons in a logical qubit state must arrive at the receiver for the state to be successfully transmitted. 
Under a lossy channel, the state transforms as
\begin{equation}\label{eq:loss}
    \rho_N \rightarrow p^N \rho_N + (1-p^N)\rho_{\perp}.
\end{equation}
We define $\rho_{\perp}$ as a state in the subspace in which one or more photons are lost to the channel, and will result in an inconclusive measurement outcome due to loss.

\subsection{Compensation techniques}
Compensation techniques are required by protocols that do not have inherent noise-resilience, such as BB84 (including decoyed-BB84) and TF-QKD. 
Without stabilisation against channel fluctuations, noise leads to an overwhelming error rate, rapidly rendering key sharing insecure.
Compensating noise admits a large resource cost (such as long servo channels and active feedback loops \cite{Pittaluga2021}) which limits the practical scalability of real-world implementations of QKD.
For example, TF-QKD experiments require stabilising the phase, frequency, and polarisation of two sources arriving at a central node from separate locations.
These stabilisation tasks have been addressed in various ways, although each introduces its own level of complexity.
Frequency locking can be achieved using common laser sources \cite{Wang2022,Chen2022}, while polarisation stabilisation was demonstrated using active feedback from the central detectors \cite{Zhou2023, Clivati2022, Wang2022, Chen2022, Pittaluga2021, Li2023}. 
Phase stabilisation typically relies on techniques such as time- and wavelength-multiplexing to track phase fluctuations via reference signals \cite{Pittaluga2021, Zhou2023}, or post-processing approaches that estimate and correct phase drift using reference frame time blocks \cite{Li2023}. 
However, all these methods come at the cost of increased system complexity, requiring multiple active control loops and tightly coordinated reference frames.
While feasible in well-controlled, point-to-point links, as noted in Ref.~\cite{Peranic2023}, such schemes pose serious challenges for scalability, particularly in dynamic or multi-user network settings, where the overhead of independent stabilising controllers quickly becomes prohibitive.

In addition to the physical resources required, stabilisation techniques require continuous monitoring of channel fluctuations.
Typical methods involve interleaving data and reference pulses in time \cite{Pittaluga2021, Wang2022, Li2023}, or monitoring the QBER using a significant fraction of the sifted shared key \cite{Agnesi:20, Ding:17, Du2023}.
These methods inevitably reduce the protocol's effective duty cycle, as a fraction of pulses must be reserved for stabilisation rather than key generation. 
Specifically, once a channel's stability degrades, reference pulses must be sent (in place of signal pulses) to estimate the current noise parameters of the channel and inform adjustments to the stabilising components (e.g., motorised waveplates for polarisation compensation). 
Consequently, only a fraction of operational time contributes to secret key generation, with the remainder consumed by channel monitoring and correction. 
As discussed in Sec.~\ref{sec:securitycomp} of the main text, this leads to orders-of-magnitude reductions in the effective key rate in state-of-the-art decoy state BB84 and TF-QKD implementations.

While a complete survey of the compensation burdens across all QKD architectures is beyond the scope of this manuscript, it is well supported in the literature that such requirements --- in terms of both hardware complexity and active control overhead --- are a persistent challenge for large-scale and cost-effective deployment.

\subsection{Loss-noise robustness tradeoff}

To circumvent the extensive overheads associated with compensating noise fluctuations, encoding inherently noise-resilient states can be considered.
Designing a secure QKD scheme requires at least two non-orthogonal states and measurements in their respective bases \cite{B92}. 
Collective unitary noise-robust protocols proposed to date \cite{Boileau, XBWang2005, efficient_collective_noise, Guo:20} use the invariant property of the antisymmetric Bell-state ($\ket{\Psi^-} = \frac{1}{\sqrt{2}}(\ket{0}\ket{1} - \ket{1}\ket{0})$) under local unitary transformations, i.e., $(U \otimes U)\ket{\Psi^-} = e^{i\Delta}\ket{\Psi^-}$, where $\Delta$ is some unobservable $U$-dependent global phase factor \cite{nielsen_chuang_2010}.
However, $\ket{\Psi^-}$ is the only two-qubit state satisfying invariance under the action of a collective unitary \cite{6stateDFS}.
Thus, two perfectly noise-cancelling non-orthogonal bases using only two qubits cannot be found. 
By introducing more qubits into a logical state, noise-robust subspaces of at least dimension 2 can be found.
A QKD protocol perfectly noise-cancelling against collective unitary noise can be devised when the logical state contains three or four qubits, as in Ref~\cite{Boileau}.

Qubits in QKD are typically encoded in photonic degrees of freedom, since photons are natural information carriers \cite{AlJuboori2023, Slussarenko2019}.
QKD protocols devised with higher noise-robustness thus use more photons per logical state.
However, key rate scales exponentially with the channel loss probability of a logical qubit and loss probability scales exponentially with the number of photons (i.e. physical qubits) in the logical state [Eq.~\eqref{eq:loss}]. 
Therefore, we arrive at a trade-off between loss- and noise-robustness: increasing the number of physical qubits (i.e. photons) per logical qubit state for the purpose of noise-cancellation leads to an exponential decrease in the key rate for a given transmission distance. We qualitatively illustrate this trade-off in Table~\ref{tab:tradeoff}.
To overcome this trade-off, we encode a fixed number of photons in a continuous DoF, rather than the tensor product space of many qubit Hilbert spaces, in a conceptually similar manner to bosonic codes \cite{Rozpedek2021, Descamps2024}.

\section{State generation and measurement}\label{app:gen&meas}

\begin{figure*}[htbp]
    \centering
    \includegraphics[width=\textwidth]{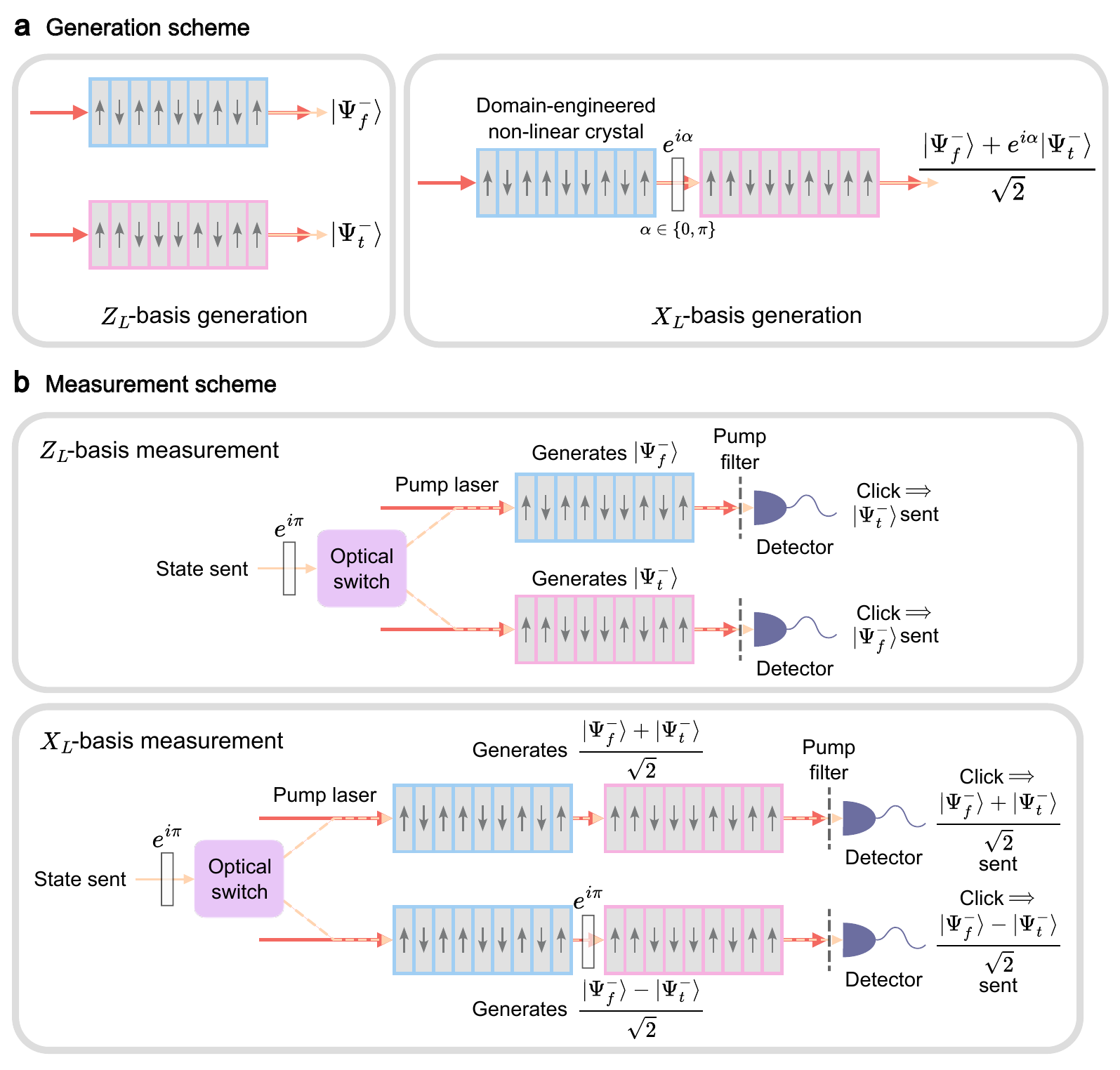}
    \caption{\textbf{Illustration of the proposed measurement and generation schemes for the logical Z and X basis states.} (a) Antisymmetric frequency-bin and time-bin Bell-states can be generated through domain engineering non-linear crystals and pump engineering. Superpositions of the frequency-bin and time-bin antisymmetric Bell-states can be generated by precisely tuning the phase difference $\alpha$ of the pump between the two domain-engineered crystals for the time-bin and frequency-bin state. (b) A logical $Z$ basis measurement and logical $X$ basis measurement can be made with a 50\% incurred loss, by destructively interfering the incoming states with one of the basis states. An optical switch is used to randomly measure for either the $\ket{0_L}$ or the $\ket{1_L}$ states in the $Z_L$ basis or randomly measure for either the $\ket{+_L}$ or the $\ket{-_L}$ states in the $X_L$ basis. If the incoming state is in the same state as the photon pair being generated, there will be no click at the detector (this cannot be distinguished from channel loss). Conversely, if there is a click at the detector, Bob infers that the state sent by Alice is opposite to the state generated by the detector's associated set-up. (In this figure, we have assumed $\braket{\Psi^-_f|\Psi^-_t}\approx 0$.)}
    \label{fig:genmeas}
\end{figure*}

Implementing our protocol requires generating antisymmetric Bell-states in frequency bins and time bins, and superpositions of these. Antisymmetric Bell-states in frequency bins can be produced through spontaneous parametric down-conversion \cite{Seshadri2022} or spontaneous four-wave mixing \cite{Clementi2023}, i.e. processes that involve pumping non-linear crystals. Entangled time-bin qubit states can be created using optical switches \cite{Lo_2023}. However, similarly to frequency-bin states, time-bin Bell-states can also be generated by pumping a domain-engineered non-linear crystal \cite{Dosseva2016, Graffitti_2017_domainengineering} (or by performing quantum optical synthesis in two-dimensional time-frequency space \cite{QOS_2021}). By producing the antisymmetric time-bin Bell-state in this way, superpositions of the frequency-bin and time-bin states can be made by coherently pumping crystals with different domain engineering, and implementing a set phase relation between these crystals \cite{coherentpumping_theory, Hochrainer2022} (see Fig. \ref{fig:genmeas}(a)).
It should be noted that this state generation method uses a probabilistic process that can produce multi-pair emissions, introducing a vulnerability to PNS attacks, where an eavesdropper could exploit extra photon pairs to gain information. However, in SM, Sec.~\ref{sec:multiphotonsec}, we clarify how a decoy-state extension of our scheme can mitigate this risk, ensuring security in such implementations.

To estimate the state generation rates for our scheme, we extrapolate from existing experimental work on photon-pair generation using SPDC. Specifically, we model a type-0 SPDC process in ppLN, and normalize our photon rates to the values reported for such a source in \cite{Neumann2022experimental}. Using a \SI{5}{\centi\meter} long crystal with a poling period of \SI{19.1}{\micro\meter}, we find that antisymmetric frequency states with \SI{1}{\nano\meter} bandwidth could be generated at a rate of $10^9$ pairs/s~\cite{github}. Our simulation assumes lossless spectral filtering of both photons using \SI{1}{\nano\meter} broad filters with a Gaussian shape, which in a real source would be replaced by domain engineering in the crystals, a technique which has been shown in \cite{chen2017efficient}. This would simultaneously allow for the generation of the time-bin states.

Performing measurements in the logical $Z_L$ and $X_L$ bases is a harder challenge to tackle. Here, we outline one possible method for performing the measurements, based on reverse-engineering the state generation process. For example, consider making a measurement to distinguish $\ket{-_L} = 1/\sqrt{2}(\ket{\Psi^-_f} - \ket{\Psi^-_{\perp f}})$ from $\ket{+_L} = 1/\sqrt{2}(\ket{\Psi^-_f} + \ket{\Psi^-_{\perp f}})$. To do this, Bob generates the state $\ket{+_L}$ by coherently pumping domain-engineered crystals with the correct relative phase (as proposed above for state preparation). This set-up is aligned such that it has the opposite phase to the incoming signal and thus destructively interferes with the incoming state. If the state $\ket{+_L}$ has been sent by Alice, the two states will destructively interfere \cite{coherentpumping_theory} and a loss will be measured. If instead the state $\ket{-_L}$ is sent on the set-up generating state $\ket{+_L}$, then part of the amplitude will constructively interfere and photons will be measured at the detector. Therefore, if photons are detected out of the set-up generating the state $\ket{+_L}$, Bob concludes that the state sent was $\ket{-_L}$, whereas if a loss is measured, the result is inconclusive. To make this a balanced measurement, Bob should generate $\ket{+_L}$ only 50\% of the time. The other 50\% of the time, Bob generates $\ket{-_L}$ to infer if the state $\ket{+_L}$ was sent or not. Identically, this process could be used to perform the $Z_L$ basis measurement. This measurement process works with an incurred 50\% loss, thus the key rate is reduced by a factor of a half, but the scaling of key rate with channel loss is identical ($\sim \eta^2$). This measurement scheme is illustrated in Fig. \ref{fig:genmeas}(b).

The proposed scheme requires a simple experimental set-up, though is not perfectly faithful to our theoretical model. Specifically, it allows $Z_L$ and $X_L$ basis measurements to be made, but does not distinguish inconclusive outcomes (i.e., the projector $\ketbra{\perp_L}{\perp_L}$) which instead manifest as bit errors for this implementation. 

This set-up underscores the experimental feasibility of our protocol. Its simplicity highlights the practical potential for near-term applications. Optimizing this measurement scheme to account for inconclusive outcomes due to noise is beyond the present scope, and would merit a dedicated study. This proposal provides a valuable foundation for further progress towards this goal.

\section{Optimisation of encoding parameters of FBS noise fluctuations}\label{app:optimise}

\begin{figure}[h]
    \centering
    \includegraphics[width=0.95\columnwidth]{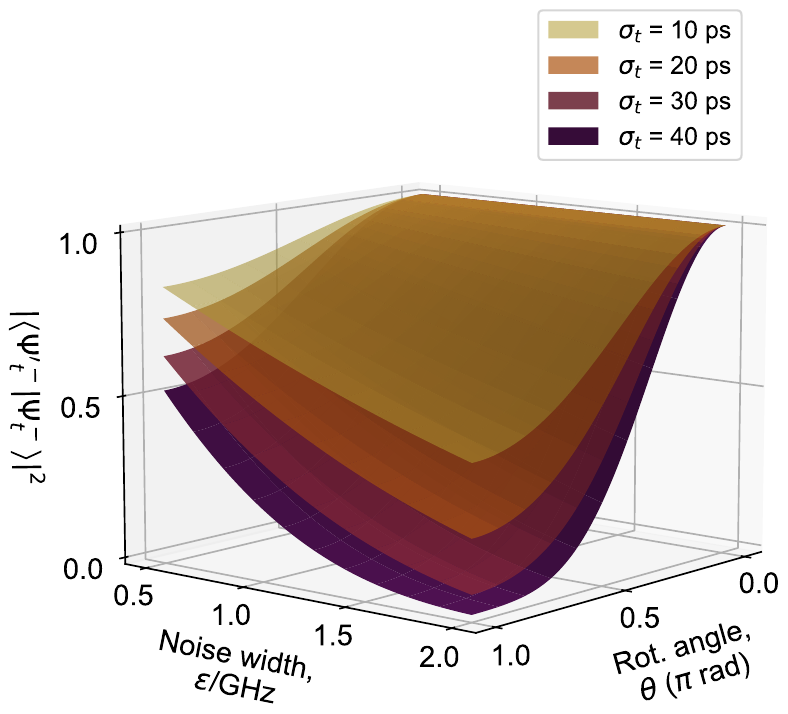}
    \caption{\textbf{Robustness of time-bin Bell-states with varying bin width under the FBS noise model, as a function of noise bin width and rotation angle.} Fidelity between the arriving state $\ket{\Psi'^-_t}$ sent by Alice through the FBS channel and the state $\ket{\Psi^-_t}$ measured by Bob. The fidelity is plotted as a function of the FBS channel parameters $\epsilon$ and $\theta$. For the simulation, we use Gaussian time-bins with parameters $\tau_0=0$ ps, $\tau_1=220$ ps, and (values experimentally reported in Ref~\cite{finco2024timebinentangledbellstate}), with a central frequency assumed at 195.860 THz and varying $\sigma_t = 10, 20, 30, 40$ ps. The fixed FBS noise parameters are $\Omega = 195.860$ THz, $\Omega +\mu = 195.879$ THz, and $\phi = 0$ and varying width $\epsilon =$ 0.6 GHz, 3.0 GHz, 6.6 GHz (close to experimentally reported values for a frequency-bin Bell-state encoding in Ref~ \cite{Clementi2023}, with $\epsilon = 6\sigma_{\omega}$). The key rate can be optimised by reducing $\sigma_t$.}
    \label{fig:stateoptimise}
\end{figure}

In Fig.~\ref{fig:stateoptimise}, we present the fidelity $\mathcal{F}(\Psi'^-_t, \Psi^-_t)$ of a time-bin Bell-state with varying widths, $\sigma_t$, under the FBS for a range of noise bin widths $\epsilon$, and rotation angles $\theta$.
By finding the fidelity over a range of FBS variables and varying the Bell-state parameters, this highlights how the encoding can be optimised for robustness to characterised fluctuations in the noise parameters.
Here, we observe increasing fidelity over the range of FBS variables given as $\sigma_t$ decreases.
Therefore, for noise fluctuates over the ranges given for $\epsilon$ and $\theta$ in Fig.~\ref{fig:stateoptimise}, $\sigma_t$ is optimised when it is sufficiently small to ensure the fidelity never falls to zero, giving non-zero key rates without compensation.

\section{Comparison of our encoding to alternative continuous DoF encoding schemes}\label{app:compencoding}

Given that the time-bin states are not perfectly robust against the FBS noise model, it is worth comparing our encoding scheme to alternative but similar encodings in the continuous space. The advantage of our protocol is that fewer assumption on the noise model are needed compared to other continuous DoF encodings, whilst still achieving a robust key rate. We highlight this with a couple of example cases.

Firstly, let us consider instead encoding purely in Bell-states of two pairs of frequency bins, such that $\ket{0_L} = \ket{\Psi^-_f}_{01,MN}$ and $\ket{1_L} = \ket{\Psi^-_f}_{23,MN}$, where $\ket{\Psi^-_f}_{ab,MN} = \frac{\ket{a_f}_M \ket{b_f}_N - \ket{b_f}_M \ket{a_f}_N}{\sqrt{2}}$. $\ket{0_L}$ is thus the antisymmetric frequency-bin Bell-state on modes $M$ and $N$ over frequency-bin states $\ket{0_f}=\ket{\Omega_0}$ and $\ket{1_f} = \ket{\Omega_1}$ and $\ket{1_L}$ is the antisymmetric frequency-bin Bell-state on modes $M$ and $N$ over new frequency-bin states $\ket{2_f}=\ket{\Omega_2}$ and $\ket{3_f} = \ket{\Omega_3}$. To have no logical bit error rate on this basis, this encoding makes the assumption that the FBS acts between states $\ket{0_f}$ and $\ket{1_f}$ and between $\ket{2_f}$ and $\ket{3_f}$ only (or outside of this subspace). If we instead allow the noise to act across these the two frequency-bin Bell-states (e.g. between states $\ket{0_f}$ and $\ket{2_f}$ and between states $\ket{1_f}$ and $\ket{3_f}$), the protocol accrues a bit error rate and an incurred loss at the measurement stage. In contrast, if we allow this model of unitary noise on our proposed encoding (i.e. the FBS noise acting between the encoded frequency bins and frequency bins outside of that space) negligible bit error rate and only an extra loss at measurement is incurred. We quantify this in SM, Sec.~\ref{sec:diffbinencoding}.

Our chosen encoding also allows us to alleviate another assumption on the noise channel. We assumed in Section~\ref{sec:FBS} that the noise acts between frequency bins. By encoding one basis state in frequency bins and one basis state in time bins, we allow the possibility that the noise acts between time bins instead, whilst maintaining the robustness of the encoding. In contrast, the fidelity of the above encoding would degrade more than our proposed encoding. Thus, our encoding is more robust under fewer noise constraints.

If the channel takes the states $\ket{0_f}$ to $\ket{0_t}$ and $\ket{1_f}$ to $\ket{1_t}$, our encoding is no longer robust. In this case however, a simple change in measurement choice at Bob allows for a perfectly noise-robust single-qubit encoding to be appropriate. Designing an encoding robust against such a model is beyond the scope considered in this paper.

\section{Key rate under lossy channels -- theoretical and experimental comparisons} \label{app:KRlosstheoryexpt}

\begin{figure*}[htbp]
    \centering
    \includegraphics[width=\textwidth]{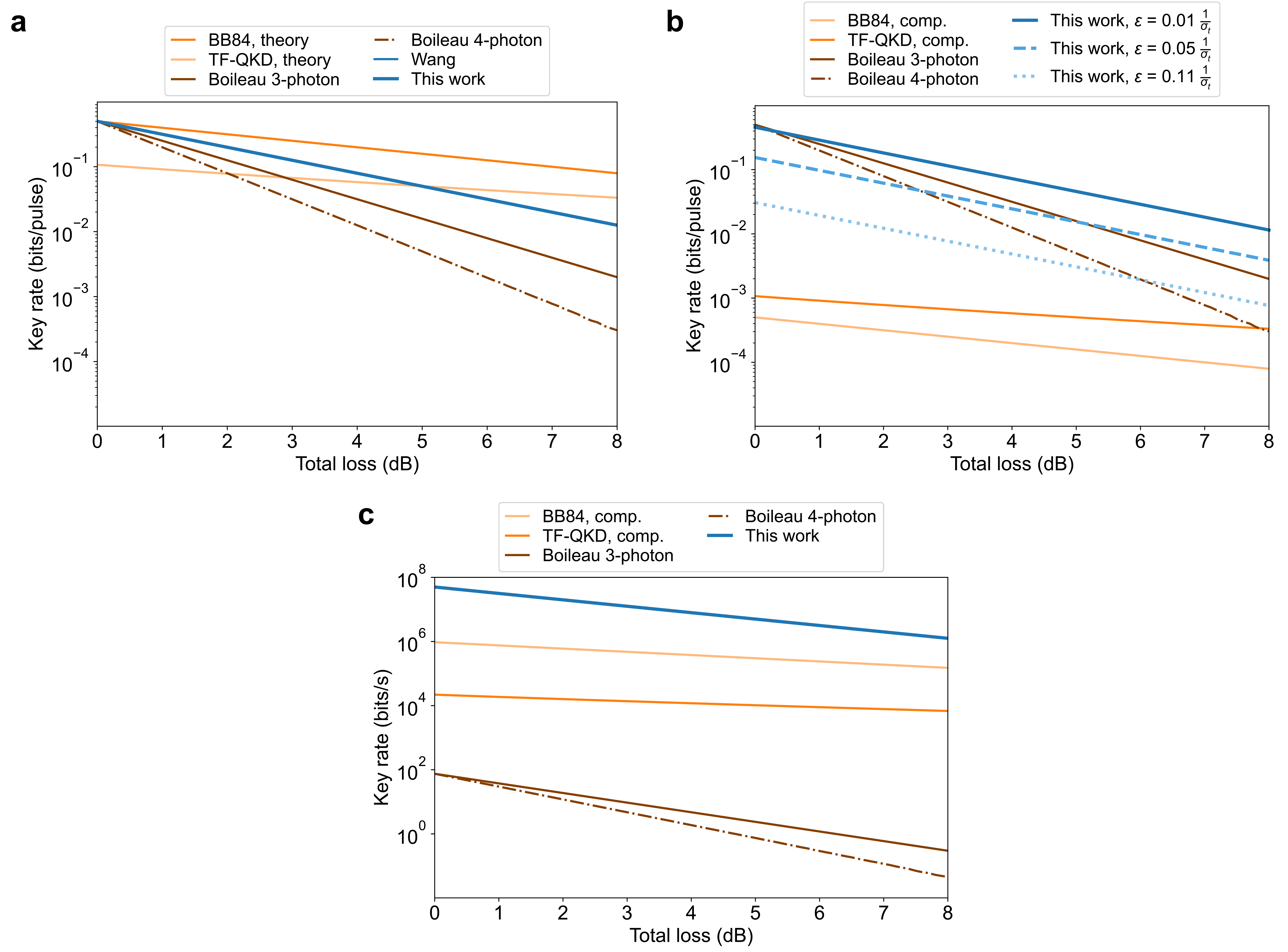}
    \caption{\textbf{Key rate as a function of channel loss for collective unitary noise robust and commonly implemented QKD protocols under theoretical and experimental conditions.} (a) Comparison of the key rate in a lossy but otherwise ideal channel. TF-QKD and BB84 perform efficiently due to using low photon numbers per shared bit. However, under experimental noise conditions, BB84, TF-QKD and Ref.~\cite{XBWang2005} require extensive stabilisation techniques to mitigate the effects of noise. These lower the shared key rates and come with substantial overheads, as shown in (b). In contrast, our protocol and those in Ref.~\cite{Boileau} do not require noise compensation. (b) Comparison of the key rate in a lossy channel when the effects of noise are accounted for. Noise scales the key rate of our noise-optimised protocol by only an order of 1 (see Sec.~\ref{sec:noisycomp}). Conversely, the compensation techniques required for BB84, TF-QKD and Ref.~\cite{XBWang2005} lower the key rates by orders of magnitude. (c) Comparison of the key rates in bits/s in a lossy but otherwise ideal channel. Rates for BB84 and TF-QKD are extracted from experimental literature, while those for our protocol and Ref.~\cite{Boileau} are estimated based on simulations informed by experimental data. Our protocol achieves significantly higher key rates than existing approaches under realistic assumptions, maintaining competitive performance even when considering experimentally feasible generation rates of our time- and frequency-entangled two-photon states.}
    \label{fig:losstheorycomp}
\end{figure*}

In this Appendix, we present the key rate of all protocols in Fig.~\ref{fig:losscomp} under loss in both an ideal theoretical case and under experimental considerations. We demonstrate these in Figs.~\ref{fig:losstheorycomp}(a)-(c).

Under ideal conditions (loss-only channel), Fig.~\ref{fig:losstheorycomp}(a) shows that TF-QKD and BB84 perform well, since they use fewer photons per bit shared by Alice and Bob. However, under experimental noise conditions, extensive noise compensation techniques are required. Not only do these methods require complex feedback systems that reduce the industrial scalability of the protocols, but they also require a significant portion of the pulses to be used for estimating and stabilising against the channel noise fluctuations. This greatly lowers the overall rate of establishing a key between Alice and Bob (by orders of magnitude), as shown in Fig.~\ref{fig:losstheorycomp}(b). In contrast, our scheme does not require stabilisation mechanisms due to its inherent noise resilience. By optimising the encoding parameters for the expected noise fluctuations, our protocol will see only an order of one reduction in key rate under channel noise without employing any compensation methods (e.g., as seen for the channel model in Eq.~\eqref{eq:mixedchan}).

In addition to Figs.~\ref{fig:losstheorycomp}(a) and (b), which present key rates in units of bits/pulse, it is also important to evaluate performance in terms of bits/second, which is a critical figure of merit for practical implementations. 
Accordingly, in Fig.~\ref{fig:losstheorycomp}(c), we compare the key rates of the discussed protocols in bits/second as a function of channel loss.

To facilitate a comparison, we first consider all protocols at 0 dB loss and then apply the appropriate loss-dependent scaling. 
For BB84 and TF-QKD, we extract key rates from Refs.~\cite{Agnesi:20} and~\cite{Chen2022}, respectively. 
Ref.~\cite{Agnesi:20} reports 80 bits/s at 40 dB loss for BB84, which corresponds to approximately $9.5 \times 10^5$ bits/s at 0 dB after accounting for detection and transmission losses. 
Similarly, Ref.~\cite{Chen2022} reports a rate of 0.092bits/s at 106 dB loss for TF-QKD, which scales to roughly $2.2 \times 10^4$bits/s at 0dB. 

By contrast, to our knowledge, the existing noise-cancelling schemes in Ref.~\cite{Boileau} are yet to be realised experimentally. 
Nonetheless, we can estimate performance by considering the generation of the four-photon entangled states needed by the protocol. 
These states remain experimentally challenging to produce, with recent demonstrations reporting generation rates of approximately 150 Hz for four-photon GHZ states~\cite{Pont2024}, corresponding to a key rate of around 75 bits/s at 0 dB.

To estimate the key rate in bits per second for our protocol, we use the methods described in Appendix~\ref{app:gen&meas}, drawing on experimental parameters from Ref.~\cite{Neumann2022experimental}. 
Under optimised conditions, our protocol can achieve generation rates on the order of $10^9$ logical states per second at 0 dB loss, enabled by domain-engineered nonlinear crystals.
To account for further experimental constraints (such as limitations due to multiphoton emissions), in Fig.~\ref{fig:losstheorycomp}(c) we consider a more conservative generation rate of $10^8$ pairs per second. 
Multiplying this rate by the bits/pulse obtained from the previous figures also yields key rates on the order of $10^8$.

This comparison highlights that our protocol is competitive with state-of-the-art approaches both in terms of bits/pulse and bits/second. 

\end{appendix}

\bibliographystyle{ieeetr}
\bibliography{references}

\clearpage
\newpage

\onecolumngrid
\begin{center}
\textbf{\large Supplemental Material}
\end{center}


\onecolumngrid


\setcounter{secnumdepth}{4}
\setcounter{section}{0}
\setcounter{equation}{0}
\setcounter{figure}{0}
\setcounter{table}{0}
\makeatletter
\renewcommand{\theequation}{S\arabic{equation}}
\renewcommand{\thefigure}{S\arabic{figure}}
\renewcommand{\thetable}{S\arabic{table}}
\renewcommand{\thesection}{S.\Roman{section}}
\@removefromreset{equation}{section}
\makeatother

\section{Review of collective unitary noise-robust protocols} \label{sec:litrev}
We provide a brief description of key protocols that are robust to collective unitary noise. In the main text, we studied the schemes in Refs.~\cite{Boileau, efficient_collective_noise, XBWang2005}, which we will describe here. Further protocols were presented by Sun \textit{et al.} (2009) \cite{6state_DFS_improved} who made improvements on Cabello (2007) \cite{6stateDFS}. These methods, however, require more qubits than the ones we analysed, making them far more susceptible to channel loss. Therefore, we present only Refs.~\cite{Boileau, efficient_collective_noise, XBWang2005}, as the others do not contribute further insights to the discussion. Additionally, we will outline another scheme presented by Guo \textit{et al.} (2020) \cite{Guo:20} that requires only two photons for the logical encoding. However, as we will explain, this requires strict assumptions on the noise model, limiting its applicability to real scenarios, and uses a specific photonic encoding that cannot be compared under the FBS model. Hence, we do not analyse it further.

\subsection{Boileau \textit{et al.} (2004)}\label{sec:boileau}
Two protocols were proposed in Ref.~\cite{Boileau} --- the first requiring four physical qubits (i.e. photons), and the second requiring only three physical qubits. For the first, Ref.~\cite{Boileau} proposed encoding logical states in four-qubit states generated by pairs of antisymmetric Bell-states. By varying the pairing of the four physical qubits (distinguished by a temporal or spatial label, for example), three distinct noise robust states can be generated:
\begin{subequations}
\begin{align}
    \ket{\psi_1} &= \ket{\Psi^-}_{12} \otimes \ket{\Psi^-}_{34},\\
    \ket{\psi_2} &= \ket{\Psi^-}_{13} \otimes \ket{\Psi^-}_{24},\\
    \ket{\psi_3} &= \ket{\Psi^-}_{14} \otimes \ket{\Psi^-}_{23}.
\end{align}
\end{subequations}
These states perfectly cancel collective unitary noise. Furthermore, these states are mutually-unbiased, i.e., $|\braket{\psi_i|\psi_j}|=1/2$ for $i\neq j$ and thus can be used to generate two non-orthogonal, mutually unbiased bases: a key basis $\{ \ket{\psi_1}, \ket{\psi_2} \}$ and a test basis $\{ \ket{\psi_2}, \ket{\psi_3} \}$. By measuring the four physical qubits either all in the $Z$ or all in the $X$ basis, randomly, the state sent by Alice can be determined by Bob with a 50\% conclusive probability once Alice has announced her basis. 
The second protocol presented in Ref.~\cite{Boileau} uses the same encoding as above, however, Alice discards the fourth qubit (creating mixed states) before sending the state on to Bob. 

\subsection{Li \textit{et al.} (2008)}\label{sec:li}
Ref.~\cite{efficient_collective_noise} proposed a modified version of the protocol in Ref.~\cite{Boileau}, that increases the measurement efficiency by eliminating inconclusive outcomes at Bob. This comes at the cost of offering either rotation-only noise cancellation (independence of $\theta$ in the unitary in Eq.~\eqref{eq:U} in SM, Sec.~\ref{sec:securityframework}) or dephasing-only noise cancellation (independence of $\phi$ in Eq.~\eqref{eq:U} in SM, Sec.~\ref{sec:securityframework}) but not the cancellation of a general unitary, as in Ref.~\cite{Boileau}. Ref.~\cite{efficient_collective_noise} achieves this by encoding four physical qubits in pairs of rotation-only ($\ket{\Psi^-}, \ket{\Phi^+}$) or dephasing-only ($\ket{\Psi^-}, \ket{\Psi^+}$) noise-cancelling two-qubit states. In the rotation-only noise-robust scheme, the key basis states ($\ket{\Psi^{\text{rot}}_0}, \ket{\Psi^{\text{rot}}_1}$) and the test basis states ($\ket{\Phi^{\text{rot}}_0}, \ket{\Phi^{\text{rot}}_1}$) are given by:
\begin{subequations}
\begin{align}
    \ket{\Psi^{\text{rot}}_0} &= \ket{\Phi^+}_{12} \otimes \ket{\Phi^+}_{34},\\
    \ket{\Psi^{\text{rot}}_1} &= \ket{\Psi^-}_{12} \otimes \ket{\Psi^-}_{34},\\
    \ket{\Phi^{\text{rot}}_0} &= \ket{\Phi^+}_{13} \otimes \ket{\Phi^+}_{24},\\
    \ket{\Phi^{\text{rot}}_1} &= \ket{\Psi^-}_{13} \otimes \ket{\Psi^-}_{24}.
\end{align}
\end{subequations}
The dephasing-only noise-robust protocol, replaces $\ket{\Phi^+}$ with $\ket{\Psi^+}$. 

As in Ref.~\cite{Boileau}, Bob measures the four physical qubits either all in the $Z$ or all in the $X$ basis randomly. Bob can determine the state sent by Alice perfectly once Alice has announced her basis. The advantage of using this protocol is its efficient encoding scheme which doubles the key rate (see Fig.~\ref{fig:li_losscomp} in SM, Sec.~\ref{sec:KR_U_prots}) compared to those which discard measurement outcomes due to a mismatch in sent and received bases (BB84) or due to inconclusive outcomes (Ref.~\cite{Boileau}). The drawbacks are the more restrictive noise assumptions, and highly loss-dependent key rate.

\subsection{Wang (2005)}\label{sec:wang}

In contrast to the use of perfectly noise-cancelling states in Refs.~\cite{Boileau, efficient_collective_noise}, Ref.~\cite{XBWang2005} presented an error rejection scheme that uses fault-tolerant states. Ref.~\cite{XBWang2005} proposed a logical state BB84 protocol, requiring two physical qubits (i.e. photons) per logical state. Here, the logical qubit states are encoded in the two-qubit subspace $\{\ket{01}, \ket{10}\}$. The key basis is given by $\{\ket{01}, \ket{10}\}$, with a computational basis measurement at Bob, and the test basis is given by the Bell-states $\{\ket{\Psi^{\pm}} = \frac{1}{\sqrt{2}}(\ket{01} \pm \ket{10}) \}$, with a Bell-state measurement at Bob. Unitary rotation noise can push these states outside of the $\{\ket{01}, \ket{10}\}$ two-qubit subspace. At the measurement stage, Alice and Bob discard outcomes corresponding to lying outside of this subspace. 

The original version of the protocol is perfectly independent of dephasing-only noise, but not of rotations. Ref.~\cite{XBWang2005} additionally presented a rotation-only noise-robust version of their scheme (similarly to Ref.~\cite{efficient_collective_noise}), at the expense of dephasing robustness. In this, they replace the logical qubit basis with the Bell-states $\{\ket{\Phi^+},\ket{\Psi^-}\}$. For the logical key basis measurements, both qubits are measured in the $Z$ basis. For the logical test basis, the first qubit is measured in the $Z$ basis, and the second is measured in the $X$ basis, allowing the states to be distinguished. We focus on the dephasing-only version of Ref.~\cite{XBWang2005}, since it allows a more clear comparison to the BB84 and our method, which also both show minimal dephasing dependence.

Ref.~\cite{XBWang2005} uses only two photons to encode logical qubits, reducing the susceptibility of the protocol to channel loss compared to Refs.~\cite{Boileau, efficient_collective_noise}. However, unlike Ref.~\cite{Boileau}, this does not perfectly cancel the rotation noise. Nonetheless, this scheme is more robust to rotation noise than BB84 (see Fig.~\ref{fig:noisycomp} in the main text).

\subsection{Guo \textit{et al} (2020)}\label{sec:guo}

Ref~\cite{Guo:20} proposed an efficient protocol robust against collective rotation noise. This is identical to 4-photon scheme in Ref~\cite{efficient_collective_noise}, however, it requires only two photons by encoding the physical qubits in hyperentangled orbital angular momentum (OAM) and polarisation states. This requires the assumption that both the OAM and polarisation degrees of freedom experience the same rotation. Whilst this is true for a rotation in free space or for a misalignment of reference frames, this assumption does not hold in general. For example, in optical fibres, birefringence acts only on the polarisation degree of freedom, and thus the OAM modes and polarisation modes will experience different noise \cite{Yao:11}. Similarly, astigmatism affects OAM modes, but has no effect on the polarization \cite{Yao:11}. Thus, the implementation of this protocol is limited to free space channels, meaning it can be applied to fewer scenarios than Refs~\cite{Boileau,efficient_collective_noise,XBWang2005} and our protocol.

\section{Overlap of antisymmetric frequency-bin and time-bin Bell-states} \label{sec:overlap}
Here, we demonstrate that the frequency-bin and time-bin antisymmetric Bell-states can be made perfectly orthogonal. We show this first for Gaussian amplitude functions (which we utilise for the results in the main text). Subsequently, we demonstrate result also holds for Lorentzian amplitude functions and comment on extension to more arbitrary peaked functions.

\subsection{Gaussian amplitude functions}\label{sec:gaussstates}
First, for Gaussian states, we start by defining the Gaussian amplitude functions that we use for the frequency-bin and time-bin states. The state-normalised Gaussian function of the frequency-bin states in the frequency domain is given by
\begin{equation}\label{eq:gauss1d}
    f_{\Omega_i, \sigma_{\omega}}(\omega) = \pi^{-1/4}\sigma_{\omega}^{-1/2} \exp{\left(- \frac{(\omega-\Omega_i)^2}{2\sigma_{\omega}^2} \right)}.
\end{equation}
and the Fourier transform of the time-bin state-normalised Gaussian functions (used such that the time states can also be represented in the frequency domain) is given by
\begin{equation}\label{eq:ftgauss1d}
\begin{split}
     F_{\tau_i, \sigma_{t}}(\omega) &= \frac{1}{\sqrt{2\pi}} \int_{-\infty}^{\infty} dt f_{\tau_i, \sigma_{t}}(t) e^{-i\omega t} \\
    &=  \pi^{-1/4}\sigma_{t}^{1/2}\exp{(-i \omega \tau_i)}\exp{\left(- \frac{\sigma_t^2 \omega^2}{2} \right)},
\end{split}
\end{equation}
where $\omega$ is defined relative to the central frequency of the time-bin states.

Using the bosonic creation operators defined in Section~\ref{sec:protocol} of the main text, the time-bin and frequency-bin bi-photon antisymmetric Bell-states can be generally written as
\begin{align} \label{eq:psix}
\begin{split}
    \ket{\Psi^-_x} = A(\mu_0, \mu_1, \sigma)(a^{\dagger}_{\mu_0,M}a^{\dagger}_{\mu_1,N} - a^{\dagger}_{\mu_1,M}a^{\dagger}_{\mu_0,N})\ket{\varnothing}
\end{split}
\end{align}
where $x$ is f or t and $\mu_i$ represents $\Omega_i$ for $x=f$ and $\tau_i$ for $x=t$. $\sigma$ is the bin width and $A(\mu_0, \mu_1, \sigma)$ is the normalisation factors for the continuous DoF state, depending on the frequencies and times of the frequency and time bins defining the qubit and their widths. For the model of Gaussian pulses that we consider above, the normalisation factors are
\begin{align}
    A(\mu_0, \mu_1, \sigma) &= \frac{1}{\sqrt{2\left[1 - \exp\left[ - \frac{(\mu_0 -\mu_1)^2}{2\sigma^2}\right] \right]}}
\end{align}
which tend to $\frac{1}{\sqrt{2}}$ when the frequency bins and time bins are sufficiently well separated (i.e. $\mu_1 - \mu_0 >> \sigma$ such that $\braket{0_x|1_x}_M \approx 0$), in agreement with Eq.~\eqref{eq:bellstate} the main text.

Expanding out the expressions for the mode operators and applying the single frequency mode operators to the vacuum state, we can write these states as
\begin{align}
\begin{split}
    \ket{\Psi^-_f} &= \int_{-\infty}^{\infty} d\omega_1 \int_{-\infty}^{\infty} d\omega_2 \Psi^-_f(\omega_1, \omega_2) \ket{\omega_1}_M\ket{\omega_2}_N\\
    &=A(\Omega_0, \Omega_1, \sigma_{\omega}) \int_{-\infty}^{\infty} d\omega_1 \int_{-\infty}^{\infty} d\omega_2 \left[f_{\Omega_0, \sigma_{\omega}}(\omega_1)f_{\Omega_1,\sigma_{\omega}}(\omega_2)- f_{\Omega_1, \sigma_{\omega}}(\omega_1) f_{\Omega_0,\sigma_{\omega}}(\omega_2)\right] \ket{\omega_1}_M \ket{\omega_2}_N,
\end{split}
\end{align}
\begin{align}
\begin{split}
    \ket{\Psi^{-}_t} &= \int_{-\infty}^{\infty} d\omega_1 \int_{-\infty}^{\infty} d\omega_2 \Psi^-_t(\omega_1, \omega_2) \ket{\omega_1}_M\ket{\omega_2}_N\\
    &=A(\tau_0, \tau_1, \sigma_t)\int_{-\infty}^{\infty} d\omega_1 \int_{-\infty}^{\infty} d\omega_2 \left[F_{\tau_0, \sigma_t}(\omega_1)F_{\tau_1,\sigma_t}(\omega_2)- F_{\tau_1, \sigma_t}(\omega_1)F_{\tau_0,\sigma_t}(\omega_2)\right] \ket{\omega_1}_M\ket{\omega_2}_N.\\
\end{split}
\end{align}
For this, $\Omega_0$ and $\Omega_1$ are defined relative to the central frequency of the time-bin state, such that the frequency variable $\omega$ is zero at the central frequency of the time-bin state. Here, we have implicitly defined the bi-photon wavefunctions $\Psi^-_f(\omega_1, \omega_2)$ and $\Psi^-_t(\omega_1, \omega_2)$ of the frequency-bin and time-bin antisymmetric Bell-states, respectively, in the frequency domain.

The overlap between the 2D states can be calculated using the single photon overlap functions
\begin{equation} \label{eq:1doverlapft}
\begin{split}
    \braket{i_f|j_t} &= \int_{-\infty}^{\infty} d\omega f_{\Omega_i, \sigma_{\omega}}(\omega)^* F_{\tau_j, \sigma_t}(\omega)\\
    &= \sqrt{\frac{2\sigma_{\omega}\sigma_t}{1+\sigma_{\omega}^2\sigma_t^2}} \exp \left(- \frac{\Omega_i^2\sigma_t^2 + \tau_j^2\sigma_{\omega}^2}{2(1+\sigma_{\omega}^2\sigma_t^2)} \right)\ \exp \left(-\frac{i\Omega_i\tau_j}{1+\sigma_{\omega}^2\sigma_t^2} \right),
\end{split}
\end{equation}
where $i,j \in \{0,1\}$.
The 2D overlap function is then given by
\begin{equation}
    \begin{split}
        \braket{\Psi^-_f|\Psi^-_t} &= A(\Omega_0, \Omega_1, \sigma_{\omega})^* A(\tau_0, \tau_1, \sigma_t) [\braket{0_f|0_t}\braket{1_f|1_t} - \braket{0_f|1_t}\braket{1_f|0_t}]  \\
        &= A(\Omega_0, \Omega_1, \sigma_{\omega}) A(\tau_0, \tau_1, \sigma_t) \frac{4\pi \sigma_{\omega}^2 \sigma_t^2 }{1+\sigma_{\omega}^2\sigma_t^2} \exp{\left[ -\frac{(\Omega_0^2+\Omega_1^2)\sigma_t^2 + (\tau_0^2 +\tau_1^2)\sigma_{\omega}^2}{2(1+\sigma_{\omega}^2\sigma_t^2)} \right]} \times \\
        & \hspace{5cm}\left( \exp{\left[ -i\frac{\Omega_0\tau_0 +\Omega_1 \tau_1}{(1+\sigma_{\omega}^2\sigma_t^2)} \right]} - \exp{\left[ -i\frac{\Omega_0\tau_1 +\Omega_1 \tau_0}{(1+\sigma_{\omega}^2\sigma_t^2)} \right]} \right).
    \end{split}
\end{equation}

\begin{figure*}[t]
    \centering
    \includegraphics[width=\textwidth]{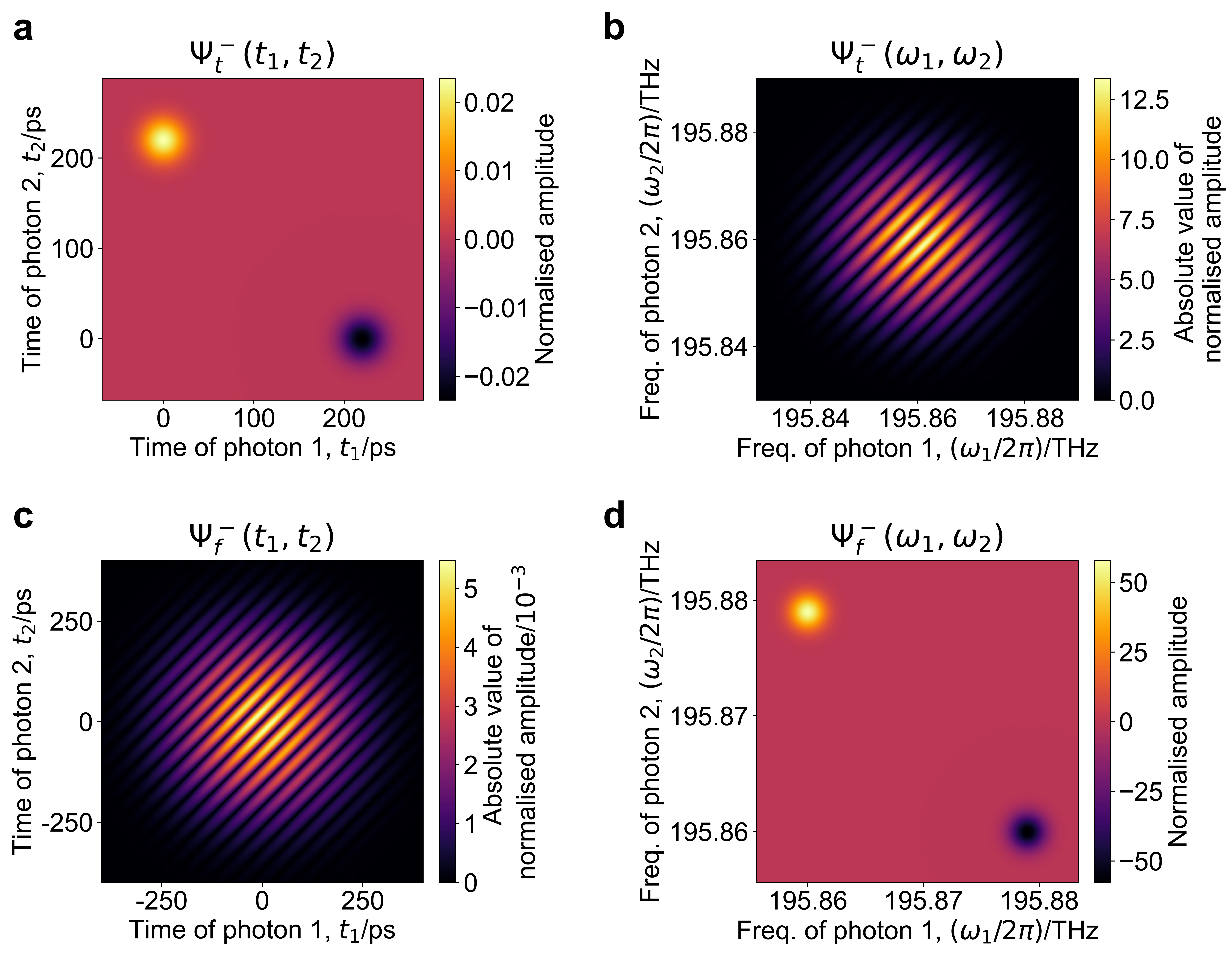}
    \caption{\textbf{2D joint temporal and frequency spectra of the time-bin and frequency-bin encoded antisymmetric Bell-states used to form the logical qubit states $\ket{0_L}$ and $\ket{1_L}$ in our encoding.} (a) Normalised bi-photon wavefunction amplitude of the antisymmetric time-bin encoded Bell-state visualised in the time domain, with $\tau_0=0$ ps, $\tau_1=220$ ps, and $\sigma_t = 17$ ps as in the main text and reported in \cite{finco2024timebinentangledbellstate}. (b) Absolute value of the normalised wavefunction of the antisymmetric time-bin encoded Bell-state in (a) visualised in the frequency domain, offset to the $\Omega_0$ of the states in (c) and (d). (c) Absolute value of the normalised bi-photon wavefunction amplitude of the antisymmetric frequency-bin encoded Bell-state visualised in the time domain, with $\Omega_0 = 195.860$ THz, $\Omega_1 = 195.879$ THz, $\sigma_{\omega} = 1.1$ GHz (as in the main text and close to the experimentally reported values in \cite{Clementi2023}). (d) Normalised bi-photon wavefunction amplitude of the antisymmetric frequency-bin encoded Bell-state from (c) visualised in the frequency domain.}
    \label{fig:allstates}
\end{figure*}

Taking the modulus squared and substituting in the expressions for $A(\Omega_0, \Omega_1, \sigma_{\omega})$ and $A(\tau_0, \tau_1, \sigma_t)$, the probability of measuring $\ket{\Psi^-_f}$ given $\ket{\Psi^-_t}$ (or vice versa) is
\begin{equation} \label{eq:overlapfunc}
\begin{split}
    |a|^2 = |\braket{\Psi^{-}_{f}|\Psi^{-}_{t}}|^2 &= \frac{1}{\left[1 - \exp{\left[- \frac{(\Omega_0-\Omega_1)^2}{2\sigma_{\omega}^2} \right]} \right]} \frac{1}{\left[1 - \exp{\left[- \frac{(\tau_0-\tau_1)^2}{2\sigma_t^2} \right]} \right]} \frac{8\sigma_{\omega}^2 \sigma_t^2 }{(1+\sigma_{\omega}^2\sigma_t^2)^2} \times \\
    & \hspace{1cm}\exp{\left[ -\frac{(\Omega_0^2+\Omega_1^2)\sigma_t^2 + (\tau_0^2 +\tau_1^2)\sigma_{\omega}^2}{(1+\sigma_{\omega}^2\sigma_t^2)} \right]} \left(1 - \cos{\left[\frac{(\Omega_1 - \Omega_0)(\tau_1 - \tau_0)}{(1+\sigma_{\omega}^2\sigma_t^2)} \right]} \right). \\
\end{split}
\end{equation}
Thus, if the frequency-bin and time-bin state parameters are selected such that $\frac{(\Omega_1 - \Omega_0)(\tau_1 - \tau_0)}{(1+\sigma_{\omega}^2\sigma_t^2)} = 2n\pi$ where $n \in \mathbbm{Z}$, the overlap of the time-bin and frequency-bin states is zero, and the states are perfectly orthogonal. If this condition is satisfied, $\ket{\Psi^-_f}$ and $\ket{\Psi^-_t}$ therefore form an orthonormal basis. 

We can also gain intuition about this result by observing the 2D temporal or frequency spectra of the two states (see Fig.~\ref{fig:allstates}). For example, let us observe both states in the frequency domain, as in Figs.~\ref{fig:allstates}(b) and (d). By varying the parameters $\Omega_0, \Omega_1, \sigma_{\omega}$, the peaks of the frequency-bin states can be positioned on the zero points of the time-bin fringes in the frequency domain, resulting in zero overlap and an orthonormal basis. Furthermore, the exponential term in Eq.~\ref{eq:overlapfunc} typically results in only a very small overlap between the two states, meaning $\ket{\Psi^-_f}$ and $\ket{\Psi^-_t}$ are approximately orthonormal in general.

\subsection{Lorentzian and arbitrary amplitude functions}
We now repeat the analysis above for Lorentzian states. The 1D state-normalised Lorentzian amplitude function is given by
\begin{equation}\label{eq:1DGauss}
    f'_{\Omega_i, \sigma_{\omega}}(\omega) = \sqrt{\frac{\sigma_{\omega}}{\pi}} \frac{\sigma_{\omega}/2}{(t-\Omega_i)^2 + (\sigma_{\omega}/2)^2}
\end{equation}
where,, as before, $\omega$ is defined relative to the central frequency of the time-bin states, $\Omega_i$ is the frequency at which the frequency-bin state is peaked, and $\sigma_{\omega}$ is the full width in frequency at half-maximum of the Lorentzian function.
The Fourier transform of the time-bin state-normalised Lorentzian function is
\begin{equation}
\begin{split}
    F'_{\tau_i, \sigma_t}(\omega) &= \frac{1}{\sqrt{2\pi}}\int_{-\infty}^{\infty} f'_{\tau_i, \sigma_t}(t) \exp[-i\omega t] dt = \sqrt{\frac{\sigma_t}{2}}\exp{\left[- i \omega \tau_i \right]} \exp{\left[- \frac{\sigma_t |\omega|}{2} \right]}.
\end{split}
\end{equation}
where $\tau_i$ is again the time at which the time-bin state is peaked, and $\sigma_{t}$ is the full width in time at half-maximum of the Lorentzian function.
The normalisation factors (from Eq.~\eqref{eq:psix}) for Lorentzian amplitude functions are
\begin{align}
    A'(\mu_0, \mu_1, \sigma) &= \frac{(\mu_0 - \mu_1)^2 + \sigma^2}{\sqrt{2\left[2\sigma^2(\mu_0-\mu_1)^2 + (\mu_0-\mu_1)^4\right]}},
\end{align}
which, as for the Gaussian case, tend to $\frac{1}{\sqrt{2}}$ when the frequency bins and time bins are sufficiently well separated (i.e. $\mu_1 - \mu_0 >> \sigma$ such that $\braket{0_x|1_x}_A \approx 0$), in agreement with the main text.

As before, we find the overlap between the antisymmetric Bell-states of frequency-bins and time-bins, now for Lorentzian amplitudes. This is given by
\begin{equation}
    \begin{split}
        \braket{\Psi^-_f|\Psi^-_t} &= A'(\Omega_0, \Omega_1, \sigma_{\omega}) A'(\tau_0, \tau_1, \sigma_t) \pi \sigma_{\omega} \sigma_t \exp{\left[ -\frac{\sigma_t\left(\sqrt{\Omega_0^2 + (\sigma_{\omega}/2)^2} + \sqrt{\Omega_1^2 + (\sigma_{\omega}/2)^2}\right)}{2} \right]} \times \\
        & \hspace{2cm} \exp{\left[ -\frac{\sigma_{\omega}(|\tau_0| + |\tau_1|)}{2} \right]} \left[ \exp{\left[ -i(\Omega_0\tau_0 +\Omega_1 \tau_1) \right]} - \exp{\left[ -i(\Omega_0\tau_1 +\Omega_1 \tau_0) \right]}\right],
    \end{split}
\end{equation}
corresponding to an overlap probability of 
\begin{equation} \label{eq:overlapfunc2}
\begin{split}
    |a|^2 = |\braket{\Psi^{-}_{f}|\Psi^{-}_{t}}|^2 &= \frac{((\Omega_0 - \Omega_1)^2 + \sigma_{\omega}^2)^2}{\left[2\sigma_{\omega}^2(\Omega_0-\Omega_1)^2 + (\Omega_0-\Omega_1)^4\right]} \frac{((\tau_0 - \tau_1)^2 + \sigma_t^2)^2}{\left[2\sigma_t^2(\tau_0-\tau_1)^2 + (\tau_0-\tau_1)^4\right]} \times \\
    & \hspace{1.5cm} \frac{\pi^2 \sigma_{\omega}^2 \sigma_t^2 }{2}
    \exp{\left[ -\sigma_t\left(\sqrt{\Omega_0^2 +(\sigma_{\omega}/2)^2} + \sqrt{\Omega_1^2 + (\sigma_{\omega}/2)^2}\right) \right]} \times \\
    & \hspace{3cm}\exp{\left[ -\sigma_{\omega}(|\tau_0| + |\tau_1|)\right]} \left(1 - \cos{\left[(\Omega_1 - \Omega_0)(\tau_1 - \tau_0) \right]} \right). \\
\end{split}
\end{equation}
Just as for Gaussian amplitude functions, an oscillatory term in the overlap probability emerges. Thus, the overlap probability goes to zero, now when $(\Omega_1 - \Omega_0)(\tau_1 - \tau_0) = 2n\pi$ (where $n \in \mathbbm{Z}$). Therefore, the antisymmetric Bell-states using Lorentzian functions can also be made perfectly orthogonal. As for Gaussian states, the exponential term and the normalisation prefactor typically result in a very small overlap between the two states, such that $\ket{\Psi^-_f}$ and $\ket{\Psi^-_t}$ approximately orthonormal in general.

The study of more general peaked amplitude functions is beyond the present scope. However, the overlap between the antisymmetric Bell-states in time bins and frequency bins for arbitrary peaked functions will typically be small such that they are approximately orthonormal. This is understood from the fact that states peaked sharply in time (the time-bin states) will be broad in the frequency domain and giving a small overlap with the sharply peaked states in frequency (the frequency-bin states).

Additionally, while we do not formally investigate here if the antisymmetric Bell-states can be made orthogonal for general classes of peaked amplitude functions, it is likely that this orthogonality condition could apply more broadly. 
The two-photon states can be made perfectly orthogonal when a sinusoidal factor in the overlap probability is present. As seen with the previous examples, this generally arises when the 1D overlap function between the 1D time-bin and frequency-bin amplitude functions (centred at time $\tau$ and frequency $\Omega$, respectively) has a factor of the form $\exp \left( i b \Omega \tau \right)$, where $b$ is some arbitrary factor that does not depend on $\Omega$ and $\tau$. A time-bin state centred at $\tau$ will take the form $f_{\sigma_t}(t-\tau)$ (where $f_{\sigma}(t)$ is the general peaked state function centred around zero). By the shift property, its Fourier transform is therefore given by $\exp(-i\omega\tau)F_{\sigma_t}(\omega)$ (where  $F_{\sigma}(\omega)$ is the Fourier transform of $f_{\sigma}(t)$). When finding the overlap of this with a frequency-bin state that is sharply peaked around a frequency $\Omega$, the exponential term may be approximately constant at $\exp(-i\Omega\tau)$ and thus factored out, resulting in the required sinusoidal probability.
Future investigations would be of interest to address this question more rigorously.

\section{CV model of FBS noise on the time-bin states} \label{sec:CVnoise}

We present the mathematical model of the FBS noise model, described in Section~\ref{sec:FBS} of the main text. We then use this model to present mathematically the impact of FBS noise on the time-bin states $\ket{\Psi^-_t}$ with Gaussian profiles, as defined in SM, Sec.~\ref{sec:overlap}. We relate this to the effect on our logical qubit encoding by analytically deriving the fidelity of the transformed state with the original time-bin state $\ket{\Psi^-_t}$ and frequency-bin state $\ket{\Psi^-_f}$ which form Bob's measurement basis (corresponding to the logical amplitude damping and bit error rate, respectively) and numerically calculating the global phase acquired by the transformed state (logical state decoherence). 

A detailed mathematical investigation of the FBS model for Lorentzian profiles and other general peaked functions is left for future work. However, as previously mentioned, we predict that utilising general amplitude functions, which may be experimentally easier to realise, will give similar results to the Gaussian case.
Sharply peaked general frequency-bin qubit states that lie within the FBS noise bins will leave the antisymmetric frequency-bin state unaffected by FBS noise. Meanwhile, general sharply peaked time-bin states will be broad in frequency, just as for the Gaussian states, thus suggesting comparable results will appear.

\subsection{Mathematical description of FBS noise model}\label{sec:FBSmathdesc}

As described in Section~\ref{sec:FBS} of the main text, we assume that the BS acts on bins in the frequency domain that have width $\epsilon$, and frequency separation $\mu(>\epsilon)$, with the central frequency of the first bin at $\Omega$. We assume that the regions outside of the relevant bins are unaffected. Further, we assume that the functions within the bins (in the ranges $\Omega-\epsilon/2 < \omega < \Omega + \epsilon/2$ and $\Omega-\epsilon/2 < \omega < \Omega + \epsilon/2$), become superposed according to the $U_{BS}$ mode operator described in the main text. Thus, the FBS acts a BS operation between single frequency modes at frequencies $\Omega-\epsilon/2$ and $\Omega+\mu-\epsilon/2$ and on each single frequency mode pair (separated by $\mu$) up to the pair $\Omega+\epsilon/2$ and $\Omega+\mu+\epsilon/2$ (Fig.~\ref{fig:CVnoisemodel} in the main text). 

Overall, the action of this noise mathematically takes a general wavefunction $y(\omega)$ to the function $z(\omega)$ according to
\begin{equation}
    z(\omega) = \begin{cases}
    y(\omega) & \omega \leq \Omega - \epsilon/2\\
    \cos \theta y(\omega) - e^{-i\phi} \sin \theta y(\omega+\mu) & \Omega - \epsilon/2 < \omega \leq \Omega + \epsilon/2\\
    y(\omega) & \Omega + \epsilon/2 < \omega  \leq \Omega +\mu - \epsilon/2\\
    \cos \theta y(\omega) + e^{i\phi} \sin \theta y(\omega-\mu) & \Omega +\mu - \epsilon/2 < \omega \leq \Omega + \mu + \epsilon/2\\
    y(\omega) & \omega > \Omega + \mu + \epsilon/2.
    \end{cases}
\end{equation} 

It would be interesting to extend this to a more general and more physically realised model, in which a smooth tail-off of the frequency noise bins, rather than a discontinuous jump, is considered. This analysis is outside our current focus and is left for subsequent research. However, we predict that a smooth tail-off would have a less degrading effect on the time-bin state, since only a fraction of the original state in the tails would be transformed by the noise.

\subsection{Overlap of transformed time-bin state with original time-bin state} \label{sec:fidelityanalytic}

We now derive the effect of this noise on the antisymmetric Bell-states in time-bins, with Gaussian profiles. We must calculate the fidelity between the transformed (sent by Alice) and original (measured by Bob) states $|\braket{\Psi'^-_t|\Psi^-_t}|^2$ (corresponding to amplitude damping on the logical encoding). Similarly, we must deduce the global phase accumulated by the state relative to the measured state under the transformation (corresponding to decoherence on the logical encoding, since $\ket{\Psi^-_f}$ is unaffected by the noise, as explained in the main text). Hence, we derive the overlap function $\braket{\Psi'^-_t|\Psi^-_t}$.

In the frequency domain, the antisymmetric time-bin Bell-state is given by
\begin{equation}
\begin{split}
    \Psi^-_t(\omega_1, \omega_2) = A(\tau_0, \tau_1, \sigma_t)\left[F_{\tau_0, \sigma_t}(\omega_1)F_{\tau_1,\sigma_t}(\omega_2) - F_{\tau_1, \sigma_t}(\omega_1)F_{\tau_0,\sigma_t}(\omega_2)\right] 
\end{split}
\end{equation}
where $F_{\tau_i, \sigma_t}(\omega)$ is defined in Eq.~\eqref{eq:ftgauss1d} of the previous section and describes the spectral probability amplitude of a single time-bin qubit state $\ket{i_t}$ (for $i=\{0,1\}$).

\begin{figure*}[t]
    \centering
    \includegraphics[width=0.7\textwidth]{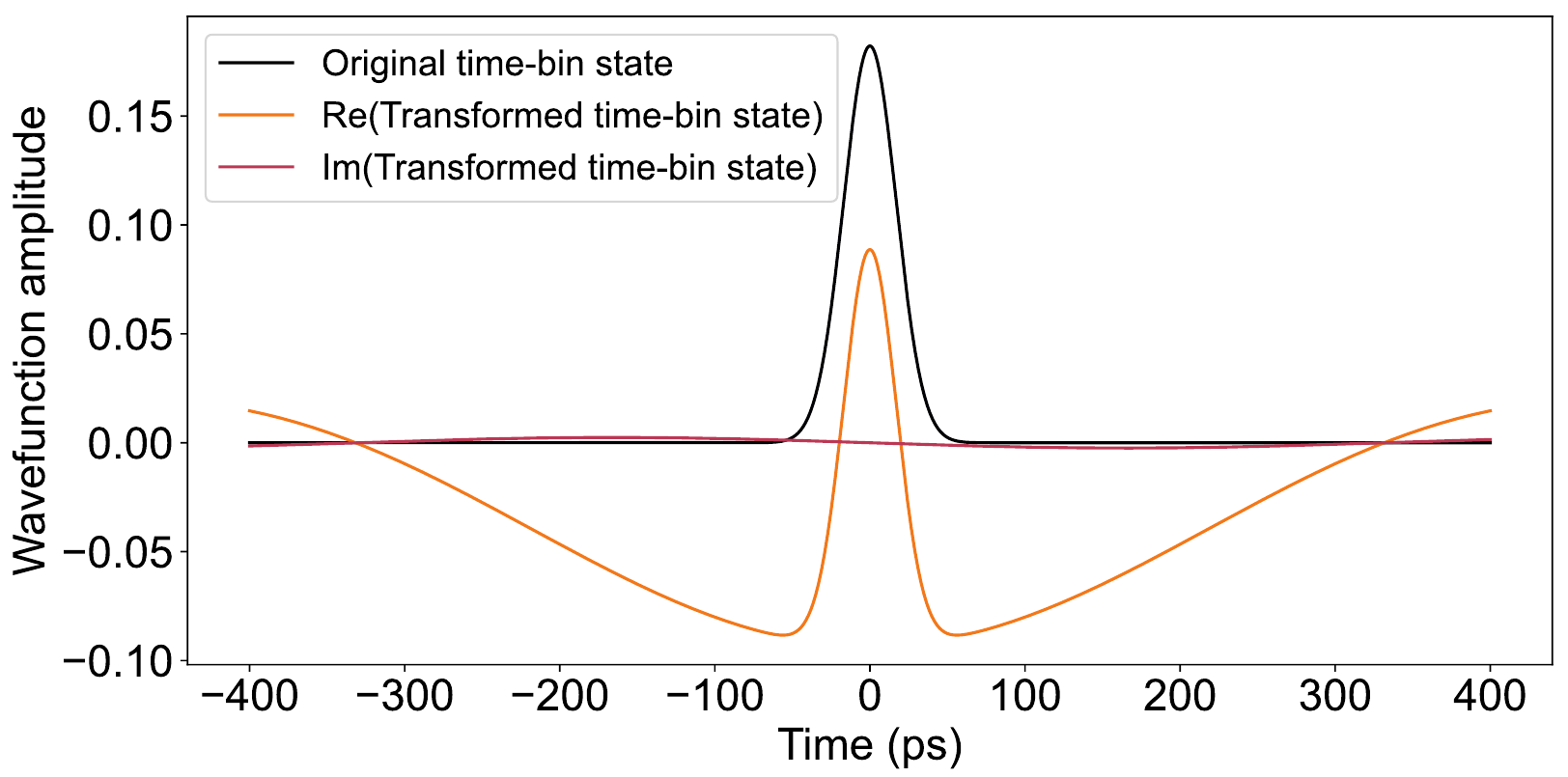}
    \caption{\textbf{Effect of the frequency beamsplitter operation on the time-bin state.} The black curve shows the wavefunction as a function of time for the $\ket{0_t}$ time-bin state with a Gaussian profile and parameters $\tau=0$ ps and $\sigma_t=17$ ps (with central frequency 194.86 THz). The orange and red curves show the real and imaginary parts respectively of the wavefunction in time for the black state transformed by a frequency beamsplitter operation with noise bins at $\Omega_0 = 195.86$ THz, $\Omega_1 = 195.879$ THz, $\sigma_{\omega} = 0.5$ GHz, with $\theta = \pi$, $\phi=\pi/2$. }
    \label{fig:timetrans}
\end{figure*}

The FBS channel transforms the single qubit states $F_{\tau, \sigma}(\omega)$ into states $G_{\tau, \sigma, \Omega, \mu, \epsilon, \theta, \phi}(\omega)$, as described in Section~\ref{sec:FBSmathdesc}, such that
\begin{equation}
    G_{\tau, \sigma_t, \Omega, \mu, \epsilon, \theta, \phi}(\omega) = \begin{cases}
    F_{\tau, \sigma_t}(\omega) & \omega \leq \Omega - \epsilon/2\\
    \cos \theta F_{\tau, \sigma_t}(\omega) - e^{-i\phi} \sin \theta F_{\tau, \sigma_t}(\omega+\mu) & \Omega - \epsilon/2 < \omega \leq \Omega + \epsilon/2\\
    F_{\tau, \sigma_t}(\omega) & \Omega + \epsilon/2 < \omega  \leq \Omega +\mu - \epsilon/2\\
    \cos \theta F_{\tau, \sigma_t}(\omega) + e^{i\phi} \sin \theta F_{\tau, \sigma_t}(\omega-\mu) & \Omega +\mu - \epsilon/2 < \omega \leq \Omega + \mu + \epsilon/2\\
    F_{\tau, \sigma_t}(\omega) & \omega > \Omega + \mu + \epsilon/2,
    \end{cases}
\end{equation} 
where again, $\Omega$ is defined relative to the central frequency of the time-bin state, such that $F_{\tau, \sigma_t}(\omega)$ is centred at $\omega = 0$ THz.
The effect of the FBS on a single photon time-bin qubit states in the time domain is visualised in Fig. \ref{fig:timetrans}.
When the time-bin Bell-state has travelled down the channel, it becomes:
\begin{equation}
\begin{split}
    \Psi'^-_t(\omega_1, \omega_2) = A(\tau_0, \tau_1, \sigma_t)[G_{\tau_0, \sigma_t, \Omega, \mu, \epsilon, \theta, \phi}(\omega_1)& G_{\tau_1, \sigma_t, \Omega, \mu, \epsilon, \theta, \phi}(\omega_2)\\
    &- G_{\tau_1, \sigma_t, \Omega, \mu, \epsilon, \theta, \phi}(\omega_1)G_{\tau_0, \sigma_t, \Omega, \mu, \epsilon, \theta, \phi}(\omega_2)]
\end{split}
\end{equation}
It can be verified that the normalisation factor $A(\tau_0, \tau_1, \sigma_t)$ does not change.

The overlap function $\braket{\Psi'^-_t|\Psi^-_t}$ therefore depends on the overlap between the transformed and original single qubit states $G_{\tau, \sigma, \Omega, \mu, \epsilon, \theta, \phi}(\omega)$ and $F_{\tau', \sigma}(\omega)$. We calculate explicitly the 1D overlap between these single photon amplitudes to be,
\begin{equation} \label{eq:1doverlap}
\begin{split}
    H(&\tau, \tau', \sigma, \Omega, \mu, \epsilon, \theta, \phi) \\
    =& \int_{-\infty}^{\infty} F_{\tau', \sigma}(\omega)^* G_{\tau, \sigma, \Omega, \mu, \epsilon, \theta, \phi}(\omega) d\omega \\
    =& \frac{1}{2}\exp \left(-\frac{(\tau-\tau')^2}{4\sigma^2}\right) \\
    & \hspace{0.5cm} \times\left[ \left[2+ \text{erf}\left(\sigma\left(\Omega -\epsilon/2 + \frac{i}{2\sigma^2}(\tau - \tau') \right)\right) - \text{erf}\left(\sigma\left(\Omega +\mu +\epsilon/2 + \frac{i}{2\sigma^2}(\tau - \tau') \right) \right) \right. \right. \\
    & \hspace{1.5cm} + \left.\text{erf}\left(\sigma\left(\Omega +\mu -\epsilon/2 + \frac{i}{2\sigma^2}(\tau - \tau') \right)\right) - \text{erf}\left(\sigma\left(\Omega +\epsilon/2 + \frac{i}{2\sigma^2}(\tau - \tau') \right) \right) \right] \\
    & \hspace{1cm} + \cos \theta \left[\text{erf}\left(\sigma\left(\Omega +\epsilon/2 + \frac{i}{2\sigma^2}(\tau - \tau') \right)\right) - \text{erf}\left(\sigma\left(\Omega - \epsilon/2 + \frac{i}{2\sigma^2}(\tau - \tau') \right) \right) \right] \\
    & \hspace{1cm} + \cos \theta \left[\text{erf}\left(\sigma\left(\Omega + \mu + \epsilon/2 + \frac{i}{2\sigma^2}(\tau - \tau') \right)\right) - \text{erf}\left(\sigma\left(\Omega + \mu - \epsilon/2 + \frac{i}{2\sigma^2}(\tau - \tau') \right) \right) \right] \\
    & \hspace{1cm} - e^{-i\phi}\sin\theta \exp \left(-i\frac{\mu}{2}(\tau + \tau')\right)\exp \left(-\frac{\sigma^2\mu^2}{4} \right) \left[\text{erf}\left(\sigma\left(\Omega + \epsilon/2 + \frac{\mu}{2} + \frac{i}{2\sigma^2}(\tau - \tau') \right)\right) \right. \\
    &\hspace{8cm} \left. - \text{erf}\left(\sigma\left(\Omega - \epsilon/2 + \frac{\mu}{2} + \frac{i}{2\sigma^2}(\tau - \tau') \right) \right) \right] \\
    &\hspace{1cm}+ e^{i\phi}\sin\theta \exp \left(+i\frac{\mu}{2}(\tau + \tau')\right)\exp \left(-\frac{\sigma^2\mu^2}{4} \right) \left[\text{erf}\left(\sigma\left(\Omega + \epsilon/2+ \frac{\mu}{2} + \frac{i}{2\sigma^2}(\tau - \tau') \right)\right) \right. \\
    &\hspace{8cm} \left. \left.- \text{erf}\left(\sigma\left(\Omega -\epsilon/2 + \frac{\mu}{2} + \frac{i}{2\sigma^2}(\tau - \tau') \right) \right) \right] \right],
\end{split}
\end{equation}
where erf($z$) is the error function of the Gaussian integral, defined as,
\begin{equation}
    \text{erf}(z) = \frac{2}{\sqrt{\pi}} \int_{0}^{z} dx \exp (-x^2).
\end{equation}

The total 2D overlap between the transformed and original state is then
\begin{equation} \label{eq:2Doverlap}
\begin{split}
    \braket{\Psi'^-_t|\Psi^-_t} = \int_{-\infty}^{\infty} &d\omega_1 \int_{-\infty}^{\infty} d\omega_2 \Psi'^-_t(\omega_1, \omega_2)^* \Psi^-_t(\omega_1, \omega_2) 
    \\&= 2|A(\tau_0, \tau_1, \sigma_t)|^2[H(\tau_0, \tau_0, \sigma_t, \Omega, \mu, \epsilon, \theta, \phi)H(\tau_1, \tau_1, \sigma_t, \Omega, \mu, \epsilon, \theta, \phi) \\
    & \hspace{3.5cm}- H(\tau_0, \tau_1, \sigma_t, \Omega, \mu, \epsilon, \theta, \phi)H(\tau_1, \tau_0, \sigma_t, \Omega, \mu, \epsilon, \theta, \phi)].
\end{split}
\end{equation}
This 2D overlap function is complex and 8-variate involving the error function, making it challenging to analyse. Nonetheless, we can draw some general conclusions about the dependence of the measurement probability of the time-bin state on different parameters, which we do in the following section. 

For a fully general and non-approximate treatment of the FBS noise on our encoding, the same analysis above, and in the remainder of this supplementary section, can be extended to the frequency-bin states. 
We perform this calculation for our scheme, and the results presented in Fig.~\ref{fig:noisycomp} incorporate this effect. 
However, the impact of the FBS noise bins on the frequency-bin states can be well understood using the intuition outlined in Sec.~\ref{sec:FBSeffect}, where the FBS is seen to act approximately as a unitary transformation on the frequency bins, leaving the two-photon state effectively unchanged.

\begin{figure*}[h]
    \centering
    \includegraphics[width=0.85\textwidth]{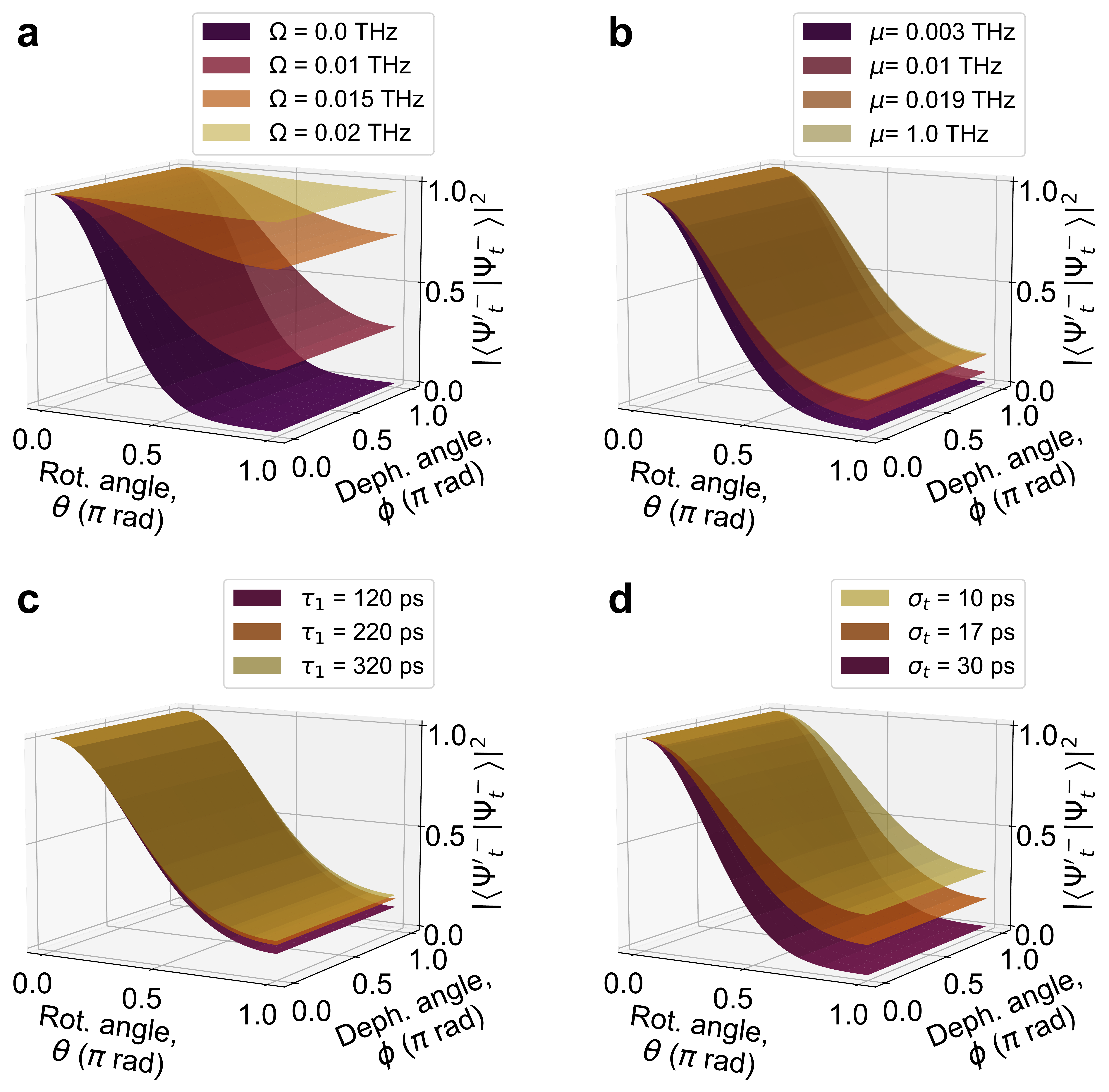}
    \caption{\textbf{Robustness of time-bin Bell-state to the frequency beamsplitter operation with different parameters.} (a) Robustness of the time-bin Bell-state with Gaussian parameters $\tau_0=0$ ps, $\tau_1 = 220$ ps, $\sigma_t = 17$ ps, to a frequency beamsplitter operation $U(\theta,\phi)$ acting on frequency bins with $\epsilon = 6.6$ GHz, a frequency separation of $\mu = 19$ GHz, and varying frequency of the first bin $\Omega_0$ (relative to the central frequency of the time-bin state in frequency), with $\Omega_0 = 0, 10, 15, 20$ GHz. (b) Robustness for $\tau_0=0$ ps, $\tau_1 = 220$ ps, $\sigma_t = 17$ ps, with $\epsilon = 3.0$ GHz, $\Omega_0 = 0$ GHz, and varying frequency separation $\mu = 3, 10, 19, 1000$ GHz. (c) Robustness for parameters $\tau_0=0$ ps, $\sigma_t = 17$ ps, and varying $\tau_1 = 120, 220, 320$ ps, to the noise parameters $\epsilon = 6.6$ GHz, $\Omega_0 = 0$ GHz, and $\mu = 19$ GHz. (d) Robustness of the time-bin Bell-state with $\tau_0=0$ ps, $\tau_1 = 220$ ps, and varying $\sigma_t = 10, 17, 30$ ps to an FBS on bins with $\epsilon = 6.6$ GHz, $\Omega_0 = 0$ GHz, and $\mu = 19$ GHz.}
    \label{fig:robustness}
\end{figure*}

\subsubsection{Fidelity between arriving and measured time-bin state under the FBS}

In Fig.~\ref{fig:robustness}, we present the robustness of the time-bin Bell-state to different FBS noise parameters. In Fig.~\ref{fig:staterobustness} in the main text, and Fig.~\ref{fig:robustness} here, we see that, generally, the time-bin Bell-state robustness to the FBS is approximately independent of the dephasing angle $\phi$ and dominated by the rotation angle $\theta$ of the unitary. 

Fig.~\ref{fig:robustness}(a) shows that there is a large increase in robustness against noise as the offset of the frequency noise bins from the central frequency of the time-bin increases (i.e. as $\Omega$ increases). We can understand this heuristically, as follows. As $\Omega$ increases, the noise bins move further into the tails of the time-bin state in the frequency domain. The time-bin wavefunction amplitude is lower in the tails, so the FBS has a smaller effect on the state amplitude. 

Fig.~\ref{fig:robustness}(b) shows that the robustness generally increases only marginally as the separation between the noise bins $\mu$ increases. We can understand this again by relating the gap $\mu$ to the ``second'' noise bin moving further into the tail of the time-bin state. However, since Fig.~\ref{fig:robustness}(b) is modelled with the first bin at the central frequency of the time-bin state, the effect of varying $\mu$ has a smaller effect on the fidelity than varying $\Omega$ (which shifts both of the noise bins). Further, we verify this by noticing that there is negligible effect on the fidelity with $\theta$ above $\mu \approx 18$ GHz, which we relate to the second noise bin being far into the time-bin tail where the amplitude is roughly constant, at zero.

We now look at the effect of changing the time-bin Bell-state parameters. Fig.~\ref{fig:robustness}(c) shows that the fidelity is barely affected by a changing the delay of the second time-bin relative to the first. This can be understood as $\tau_1$ does not alter the absolute value of the time-bin state in the frequency domain. Changing $\tau_1$ simply varies the phase modulation across the bin in the frequency-bin state. 

Finally, Fig.~\ref{fig:robustness}(d) shows that the fidelity with $\theta$ generally decreases as the time-bin width increases. Increasing the time-bin width makes the corresponding state in the frequency domain narrower. Thus, similarly to the variation of $\Omega$, the amplitude of the state within the noise bins is higher as $\sigma_t$ increases, resulting in a larger impact on the time-bin states.

\subsubsection{Phase between arriving and measured time-bin state under the FBS} \label{sec:phaseacc}

\begin{figure*}[h]
    \centering
    \includegraphics[width=\textwidth]{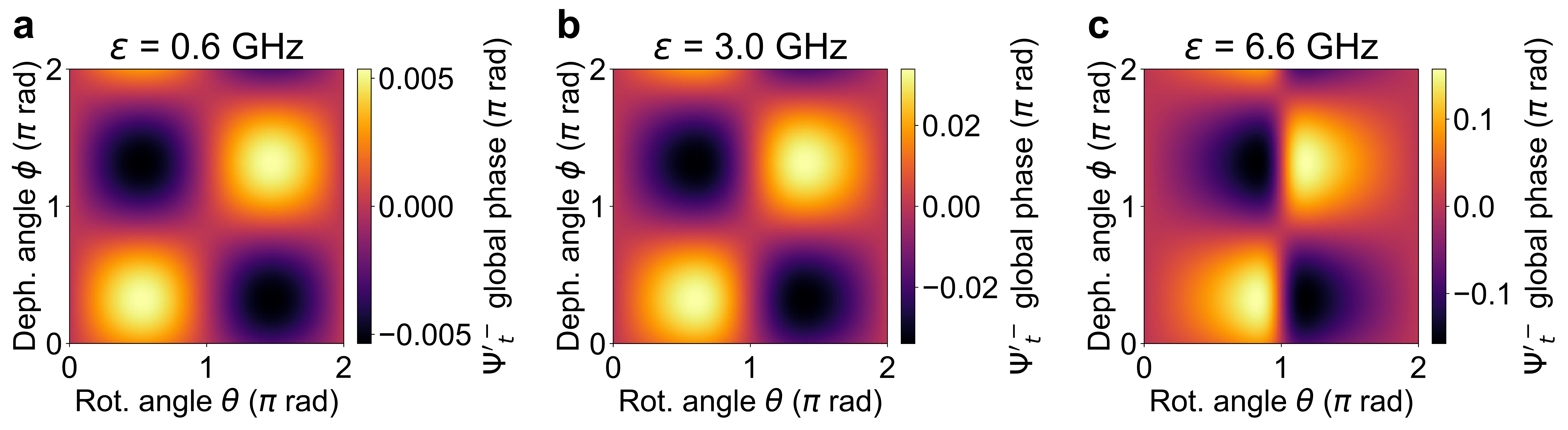}
    \caption{\textbf{Accumulated phase on the time-bin Bell-state under the frequency beamsplitter model as a function of $\theta$ and $\phi$ for varying noise bin widths.} The phase accumulated by the antisymmetric time-bin Bell-state $\Psi^-_t$, with Gaussian profile parameters $\tau_0= 0$ ps, $\tau_1 = 220$ ps and $\sigma_t= 17$ ps, as it traverses the FBS channel with $\Omega = 0$ THz relative to the central frequency of the time-bin state, $\mu = 0.019$ THz and varying noise bin widths (a) $\epsilon = 0.6$ GHz, (b) $\epsilon = 3.0$ GHz, (c) $\epsilon = 6.6$ GHz.}
    \label{fig:globalphase}
\end{figure*}

In Fig.~\ref{fig:globalphase}, we present the phase accumulated by the time-bin under the FBS noise channel relative to the untransformed basis states for the time-bin Bell-states and noise considered in Fig.~\ref{fig:staterobustness} of the main text. To find this, we numerically calculate the phase on the 2D overlap function $\braket{\Psi'^-_t|\Psi^-_t}$, given in Eq.~\eqref{eq:2Doverlap}, for varying parameters $\theta$ and $\phi$.

For the parameters used in Fig.~\ref{fig:globalphase}, we see that the phase shows an oscillating pattern. The maximum value the phase increases as the noise bin width increases for the parameters considered here. However, the maximum phase is only a fraction of $\pi/2$, i.e. the value of phase at which the key rate would fall to zero. Therefore, for small values of $\epsilon$, the phase imparted by the FBS has a negligible effect on the overall key rate for given $\theta$ and $\phi$ values.

\subsection{Overlap of transformed time-bin state with frequency-bin state} \label{sec:biterror}

In the previous subsection, we analysed the action of the FBS on the time-bin Bell-state to find the fidelity between the original and transformed state and the (logical decoherence) phase. We now perform a similar analysis to find the fidelity between the transformed time-bin Bell-state and the frequency-bin Bell-state, $|\braket{\Psi^-_f|\Psi'^-_t}|^2$. This gives us the bit error rate of our logical encoding. 

To calculate the fidelity, we first find an expression for the 1D overlap between the single photon frequency-bin state centred at frequency $\Delta$ (relative to the central frequency of the time-bin state), with the frequency-bin wavefunction given by $f_{\Delta, \sigma_{\omega}}(\omega)$, as defined in Eq.~\eqref{eq:gauss1d} in the previous section, and the transformed single photon time-bin state centred at $\tau$, given by $G_{\tau, \sigma, \Omega, \mu, \epsilon, \theta, \phi}(\omega)$ from the previous subsection. We find the 1D overlap function to be
\begin{equation}\label{eq:1dconjoverlap}
\begin{split}
    I(&\Delta, \tau, \sigma_{\omega}, \sigma_t, \Omega, \mu, \epsilon, \theta, \phi) = \int_{-\infty}^{\infty} f_{\Delta, \sigma_{\omega}}(\omega)^* G_{\tau, \sigma_t, \Omega, \mu, \epsilon, \theta, \phi}(\omega) d\omega \\
    &= \sqrt{\frac{\sigma_{\omega}\sigma_t}{2(1+\sigma_t^2 \sigma_{\omega}^2)}} \exp \left(-\frac{i\Delta\tau}{1+\sigma_{\omega}^2\sigma_t^2}\right)\exp \left(-\frac{\Delta^2\sigma_t^2}{2(1+\sigma_{\omega}^2\sigma_t^2)}\right) \exp \left(- \frac{\sigma_{\omega}^2\tau^2}{2(1+\sigma_{\omega}^2\sigma_t^2)} \right)   \\
    & \hspace{1cm} \times \left[ \left[2 + \text{erf}\left( \sqrt{\frac{1+\sigma_t^2\sigma_{\omega}^2}{2\sigma_{\omega}^2}}\left(\Omega -\epsilon/2 + \frac{(i\sigma_{\omega}^2\tau + \alpha\sigma_{\omega}^2\sigma_t^2 - \Delta)}{1+ \sigma_{\omega}^2\sigma_t^2} \right)\right) \right.\right. \\
    & \hspace{1.5cm}- \text{erf}\left( \sqrt{\frac{1+\sigma_t^2\sigma_{\omega}^2}{2\sigma_{\omega}^2}}\left(\Omega + \mu +\epsilon/2  + \frac{(i\sigma_{\omega}^2\tau + \alpha\sigma_{\omega}^2\sigma_t^2 - \Delta)}{1+ \sigma_{\omega}^2\sigma_t^2} \right)\right) \\
    & \hspace{1.5cm} + \text{erf}\left( \sqrt{\frac{1+\sigma_t^2\sigma_{\omega}^2}{2\sigma_{\omega}^2}}\left(\Omega +\mu -\epsilon/2 + \frac{(i\sigma_{\omega}^2\tau + \alpha\sigma_{\omega}^2\sigma_t^2 - \Delta)}{1+ \sigma_{\omega}^2\sigma_t^2} \right)\right) \\
    &\hspace{1.5cm} \left.- \text{erf}\left( \sqrt{\frac{1+\sigma_t^2\sigma_{\omega}^2}{2\sigma_{\omega}^2}}\left(\Omega +\epsilon/2  + \frac{(i\sigma_{\omega}^2\tau + \alpha\sigma_{\omega}^2\sigma_t^2 - \Delta)}{1+ \sigma_{\omega}^2\sigma_t^2} \right)\right) \right] \\
    & \hspace{1.5cm} + \cos\theta \left[\text{erf}\left( \sqrt{\frac{1+\sigma_t^2\sigma_{\omega}^2}{2\sigma_{\omega}^2}}\left(\Omega +\epsilon/2 + \frac{(i\sigma_{\omega}^2\tau + \alpha\sigma_{\omega}^2\sigma_t^2 - \Delta)}{1+ \sigma_{\omega}^2\sigma_t^2} \right)\right) \right. \\
    & \hspace{4cm} \left.- \text{erf}\left( \sqrt{\frac{1+\sigma_t^2\sigma_{\omega}^2}{2\sigma_{\omega}^2}}\left(\Omega - \epsilon/2  + \frac{(i\sigma_{\omega}^2\tau + \alpha\sigma_{\omega}^2\sigma_t^2 - \Delta)}{1+ \sigma_{\omega}^2\sigma_t^2} \right)\right)  \right] \\
    & \hspace{1.5cm} + \cos\theta \left[\text{erf}\left( \sqrt{\frac{1+\sigma_t^2\sigma_{\omega}^2}{2\sigma_{\omega}^2}}\left(\Omega +\mu+\epsilon/2 + \frac{(i\sigma_{\omega}^2\tau + \alpha\sigma_{\omega}^2\sigma_t^2 - \Delta)}{1+ \sigma_{\omega}^2\sigma_t^2} \right)\right) \right.\\
    &\hspace{4cm} \left. - \text{erf}\left( \sqrt{\frac{1+\sigma_t^2\sigma_{\omega}^2}{2\sigma_{\omega}^2}}\left(\Omega +\mu - \epsilon/2  + \frac{(i\sigma_{\omega}^2\tau + \alpha\sigma_{\omega}^2\sigma_t^2 - \Delta)}{1+ \sigma_{\omega}^2\sigma_t^2} \right)\right)  \right] \\
    & \hspace{1.5cm} - e^{-i\phi}\sin\theta \exp \left(- \frac{\Delta\mu\sigma_t^2}{(1+\sigma_{\omega}^2\sigma_t^2)} \right) \exp \left(-\frac{\sigma_t^2\mu^2(1-\sigma_{\omega}^2 + \sigma_t^2\sigma_{\omega}^2)}{2(1+\sigma_{\omega}^2\sigma_t^2)} \right) \exp \left(-\frac{i\mu\tau}{1+\sigma_{\omega}^2\sigma_t^2}\right) \\
    & \hspace{3cm} \times\left[ \text{erf}\left( \sqrt{\frac{1+\sigma_t^2\sigma_{\omega}^2}{2\sigma_{\omega}^2}}\left(\Omega + \epsilon/2 + \frac{(i\sigma_{\omega}^2\tau + \mu\sigma_{\omega}^2\sigma_t^2 - \Delta)}{1+ \sigma_{\omega}^2\sigma_t^2} \right)\right) \right. \\
    & \hspace{5cm} \left. - \text{erf}\left( \sqrt{\frac{1+\sigma_t^2\sigma_{\omega}^2}{2\sigma_{\omega}^2}}\left(\Omega - \epsilon/2 + \frac{(i\sigma_{\omega}^2\tau + \mu\sigma_{\omega}^2\sigma_t^2 - \Delta)}{1+ \sigma_{\omega}^2\sigma_t^2} \right)\right)\right] \\
    & \hspace{1.5cm} + e^{i\phi}\sin\theta \exp \left(+ \frac{\Delta\mu\sigma_t^2}{(1+\sigma_{\omega}^2\sigma_t^2)} \right) \exp \left(-\frac{\sigma_t^2\mu^2(1-\sigma_{\omega}^2 + \sigma_t^2\sigma_{\omega}^2)}{2(1+\sigma_{\omega}^2\sigma_t^2)} \right) \exp \left(+\frac{i\mu\tau}{1+\sigma_{\omega}^2\sigma_t^2}\right) \\
    & \hspace{3cm} \times\left[ \text{erf}\left( \sqrt{\frac{1+\sigma_t^2\sigma_{\omega}^2}{2\sigma_{\omega}^2}}\left(\Omega +\mu +\epsilon/2 + \frac{(i\sigma_{\omega}^2\tau - \mu\sigma_{\omega}^2\sigma_t^2 - \Delta)}{1+ \sigma_{\omega}^2\sigma_t^2} \right)\right) \right. \\
    & \hspace{5cm} \left.\left. - \text{erf}\left( \sqrt{\frac{1+\sigma_t^2\sigma_{\omega}^2}{2\sigma_{\omega}^2}}\left(\Omega +\mu - \epsilon/2 + \frac{(i\sigma_{\omega}^2\tau - \mu\sigma_{\omega}^2\sigma_t^2 - \Delta)}{1+ \sigma_{\omega}^2\sigma_t^2} \right)\right)\right] \right].
\end{split}
\end{equation}

From this, we calculate the total 2D overlap between the transformed time-bin state and the original frequency state measured by Bob. This is given by
\begin{equation}
\begin{split}
    \braket{\Psi^-_f|\Psi'^-_t} = \int_{-\infty}^{\infty} d\omega_1 \int_{-\infty}^{\infty} d\omega_2 &\Psi^-_f(\omega_1, \omega_2) \Psi'^-_t(\omega_1, \omega_2) 
    \\&= 2A^*(\Omega_0, \Omega_1, \sigma_{\omega})A(\tau_0, \tau_1, \sigma_t)[I(\Omega_0, \tau_0, \sigma_{\omega}, \sigma_t, \Omega, \mu, \epsilon, \theta, \phi)I(\Omega_1, \tau_1, \sigma_{\omega}, \sigma_t, \Omega, \mu, \epsilon, \theta, \phi) \\
    & \hspace{3.5cm}- I(\Omega_0, \tau_1, \sigma_{\omega}, \sigma_t, \Omega, \mu, \epsilon, \theta, \phi)I(\Omega_1, \tau_0, \sigma_{\omega}, \sigma_t, \Omega, \mu, \epsilon, \theta, \phi)].
\end{split}
\end{equation}

\begin{figure*}[h]
    \centering
    \includegraphics[width=\textwidth]{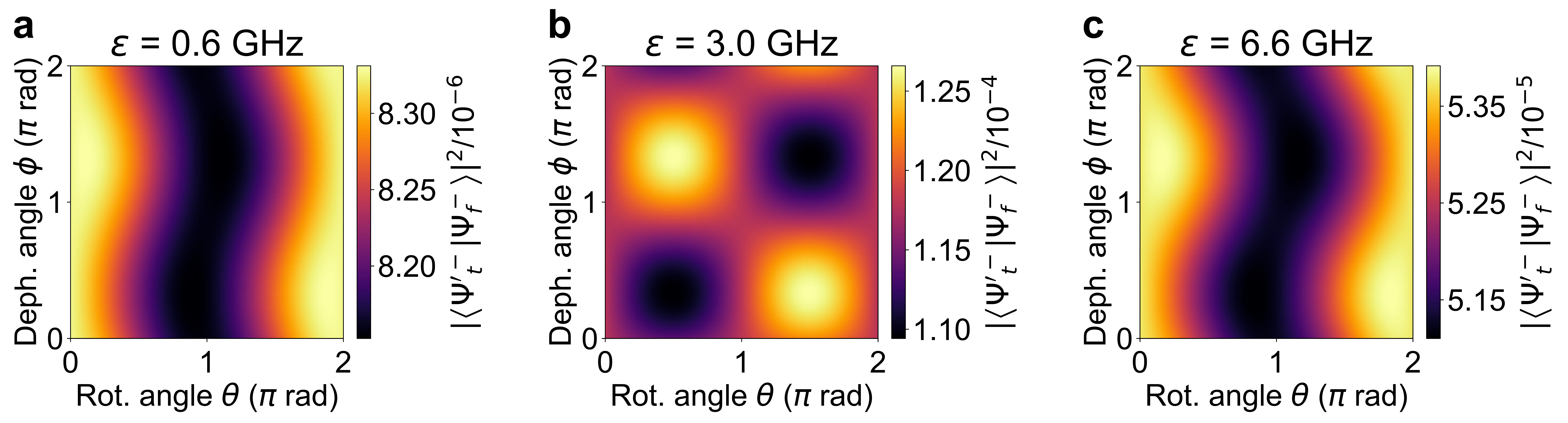}
    \caption{\textbf{Fidelity $\mathcal{F}(\Psi'^-_t, \Psi^-_f)$ of the encoding as a function of the beamsplitter channel parameters $\theta$ and $\phi$ for varying noise bin widths.}  The bit error rate of our encoding when sending the antisymmetric time-bin Bell-state $\Psi^-_t$, with parameters $\tau_0= 0$ ps, $\tau_1 = 220$ ps and $\sigma_t= 17$ ps, sent down a the frequency beamsplitter channel with $\Omega = 0$ THz relative to the central frequency of the time-bin state, $\mu = 0.019$ THz and varying noise bin widths (a) $\epsilon = 0.6$ GHz, (b) $\epsilon = 3.0$ GHz, (c) $\epsilon = 6.6$ GHz. The antisymmetric frequency-bin Bell-state measured here has parameters matching the frequency beamsplitter parameters, i.e. $\Omega_0=195.86$ THz (which we take to be the central frequency of the time-bin states), $\Omega_1=195.879$ THz, and $\sigma_{\omega}= \epsilon/6$ for each $\epsilon$.}
    \label{fig:biterror}
\end{figure*}

Calculating the modulus-squared value of this expression for different parameters, we find that the fidelity $\mathcal{F}(\Psi'^-_t, \Psi^-_f)$ giving rise to a bit error probability is typically negligible. For example, in Fig.~\ref{fig:biterror} we show that $\mathcal{F}(\Psi'^-_t, \Psi^-_f)$ is $\lesssim10^{-4}$ for the states and noise considered in Fig.~\ref{fig:staterobustness} of the main text. Therefore, the impact of bit error on the overall key rate is negligible and outweighed by the effect of amplitude damping.

In Fig.~\ref{fig:biterror}, we used $\epsilon = 6\sigma_{\omega}$ and $\Omega=\Omega_0$, $\Omega_1 -\Omega_0 = \mu$. In this way, we were assuming that the FBS would also be acting a unitary on the frequency-bin Bell-states $\ket{\Psi^-_f}$, if Alice had sent that instead of $\ket{\Psi^-_t}$.

\section{Dispersion noise model} \label{sec:dispanalysis}
\subsection{Outline of chromatic dispersion model derivation}
Single photons can be defined as positive frequency excitations of the electromagnetic field \cite{Milburn2015}.
As a result, the quantum state $\ket{\psi}$ of a photon as a function of space and time $(\textbf{x},t)$ can be written as
\begin{equation}\label{eq:planewave}
    \ket{\psi} = \int_0^{\infty} d\omega g(\omega) e^{i(\omega t-\textbf{k.x})} \hat{a}^{\dagger}(\omega) \ket{\emptyset},
\end{equation}
where $g(\omega)$ is a wavefunction as a function of frequency $\omega$, and $\textbf{k}$ is the wavevector, given by $k = |\textbf{k}| = n\omega/c$. Here, $n$ is the refractive index of the medium being traversed \cite{Smith_2007}.

Chromatic dispersion is an effect that arises when the refractive index (and correspondingly the wavevector) are frequency-dependent (i.e. $n(\omega), k(\omega)$ respectively). 
This occurs when a medium has electronic resonances \cite{Hecht2017}.
Expressions for $k(\omega)$ for different media can be found, for example, through Sellmeier equations \cite{saleh_teich_2019}.
By performing a Taylor expansion of $k(\omega)$ about a central frequency $\omega_0$, such that we write 
\begin{equation}
    \left. k(\omega) = \sum_{n=0}^{\infty} \frac{1}{n!}\frac{\partial^n k}{\partial \omega^n} \right\vert_{\omega_0} (\omega-\omega_0)^n,
\end{equation}
we see that the wavefunction of a photon state [Eq.~\eqref{eq:planewave}] will accumulate a phase that depends on the frequency to different polynomial orders \cite{Paschottachromatic_dispersion}. This is the effect of dispersion. Therefore, if only the $n$-th term is non-zero, we arrive at Eq.~\eqref{eq:dispersion} in the main text, to describe the effect of dispersion on the spectral wavefunction, with $\left.\alpha_n = \frac{1}{n!}\frac{\partial^n k}{\partial \omega^n} \right\vert_{\omega_0}$.

Using the equation for the channel model given in Eq.~\eqref{eq:dispersion} in the main text, we can write the effect of the channel noise model on the frequency-bin and time-bin Bell-state wavefunctions analytically. The transformed Bell-states are given by
\begin{align}
\begin{split}
    \Psi'^-_f &= e^{i\alpha_n((\omega_1-\omega_0)^n + (\omega_2-\omega_0)^n)} \Psi^-_f \\
    &= e^{i\alpha_n((\omega_1-\omega_0)^n + (\omega_2-\omega_0)^n)} A(\Omega_0, \Omega_1, \sigma_{\omega})\left[f_{\Omega_0, \sigma_{\omega}}(\omega_1)f_{\Omega_1,\sigma_{\omega}}(\omega_2) - f_{\Omega_1, \sigma_{\omega}}(\omega_1)f_{\Omega_0,\sigma_{\omega}}(\omega_2)\right],
\end{split}
\\[2ex]
\begin{split}
    \Psi'^-_t &= e^{i\alpha_n((\omega_1-\omega_0)^n + (\omega_2-\omega_0)^n)} \Psi^-_t \\
    &= e^{i\alpha_n((\omega_1-\omega_0)^n + (\omega_2-\omega_0)^n)} A(\tau_0, \tau_1, \sigma_t)\left[F_{\tau_0, \sigma_t}(\omega_1)F_{\tau_1,\sigma_t}(\omega_2) - F_{\tau_1, \sigma_t}(\omega_1)F_{\tau_0,\sigma_t}(\omega_2)\right],
\end{split}
\end{align}
where $f_{\Omega, \sigma_{\omega}}(\omega)$ and $F_{\tau, \sigma_{t}}(\omega)$ are the single-photon frequency-bin and time-bin states given by Eqs.~\eqref{eq:gauss1d} and \eqref{eq:ftgauss1d}, respectively, and $\Omega_0, \Omega_1, \omega_0$ are defined relative to the central frequency of the time-bin states, such that $\omega=0$ THz at the centre of the time-bin state as defined in the frequency domain.
Analysing the effect of this noise on our encoding requires calculating the fidelities and global phases (as for the FBS model), such that we can find key rates.
Calculating the fidelities of the transformed states with the states measured by Bob ($|\braket{\Psi'^-_x|\Psi^-_x}|^2$ where $x$ is $f$ or $t$) and the bit error under the operation of the channel ($|\braket{\Psi'^-_t|\Psi^-_f}|^2$) therefore requires evaluating the 1D overlap terms
\begin{align*}
    &\int_{-\infty}^{\infty} d\omega \exp (i\alpha_n (\omega-\omega_0)^n) f_{\Omega, \sigma_{\omega}}(\omega) f_{\Omega', \sigma_{\omega}}(\omega)^*,\\
    &\int_{-\infty}^{\infty} d\omega \exp (i\alpha_n (\omega-\omega_0)^n) F_{\tau, \sigma_{t}}(\omega) F_{\tau', \sigma_{t}}(\omega)^*, \\
    &\int_{-\infty}^{\infty} d\omega  \exp (i\alpha_n (\omega-\omega_0)^n) f_{\Omega, \sigma_{\omega}}(\omega) F_{\tau, \sigma_{t}}(\omega)^*.
\end{align*}

In the next subsections, we calculate these terms for the first order dispersion ($n=1$) with Gaussian amplitude functions, as considered in the main text and defined in SM, Sec.~\ref{sec:overlap}, and use these to calculate the fidelities and phases. We analyse these and convert them into key rates. Subsequently, we perform similar calculations for the second order dispersion term ($n=2$). We give a heuristic understanding of the effect of chromatic dispersion on the states in the small $\alpha_n$ regime. We anticipate similar results will hold for non-Gaussian functions, using arguments related to those in SM, Sec.~\ref{sec:CVnoise}, and leave the explicit calculations of such functions for the focus of future studies.

\subsection{Linear dispersion (\texorpdfstring{$n=1$}{n=1})} \label{sec:n=1disp}

For the $n=1$ case, we calculate the 1D overlap functions to be
\begin{equation} \label{eq:overlapff1}
\begin{split}
    \braket{i_f|j_f'} = \int_{-\infty}^{\infty} d\omega &\exp (i\alpha_1 (\omega-\omega_0)) f_{\Omega_j, \sigma_{\omega}}(\omega) f_{\Omega_i, \sigma_{\omega}}(\omega)^* \\
    &= \exp{(-i\alpha_1\omega_0)}\exp{\left( - \frac{(\Omega_j-\Omega_i)^2}{4\sigma_{\omega}^2}\right)} \exp{\left(-\frac{\alpha_1^2\sigma_{\omega}^2}{4} \right)} \exp{\left( \frac{i\alpha_1 (\Omega_j + \Omega_i)}{2}\right)},
\end{split}
\end{equation}
\begin{equation} \label{eq:prob1tt}
\begin{split} 
    \braket{i_t|j_t'} = \int_{-\infty}^{\infty} d\omega &\exp (i\alpha_1 (\omega-\omega_0)) F_{\tau_j, \sigma_{t}}(\omega) F_{\tau_i, \sigma_{t}}(\omega)^* = \exp{(-i\alpha_1\omega_0)} \exp{\left( - \frac{(\alpha_1 + \tau_i - \tau_j)^2}{4\sigma_t^2}\right)}.
\end{split}
\end{equation}
\begin{equation}\label{eq:overlapft1}
\begin{split}
    \braket{i_t|j_f'} = \int_{-\infty}^{\infty}& d\omega \exp (i\alpha_1 (\omega-\omega_0)) f_{\Omega_j, \sigma_{\omega}}(\omega) F_{\tau_i, \sigma_{t}}(\omega)^* \\
    &= \sqrt{\frac{2\sigma_{\omega}\sigma_t}{1+\sigma_{\omega}^2\sigma_t^2}} \exp{(-i\alpha_1\omega_0)} \exp{\left( - \frac{\sigma_t^2\Omega_j^2}{2(1+\sigma_{\omega}^2\sigma_t^2)}\right)} \exp{\left( - \frac{\sigma_{\omega}^2(\alpha_1 + \tau_i)^2}{2(1+\sigma_{\omega}^2\sigma_t^2)}\right)} \exp{\left( \frac{i\Omega(\alpha_1+\tau_i)}{(1+\sigma_{\omega}^2\sigma_t^2)}\right)},
\end{split}
\end{equation}

\begin{figure*}[h]
    \centering
    \includegraphics[width=\textwidth]{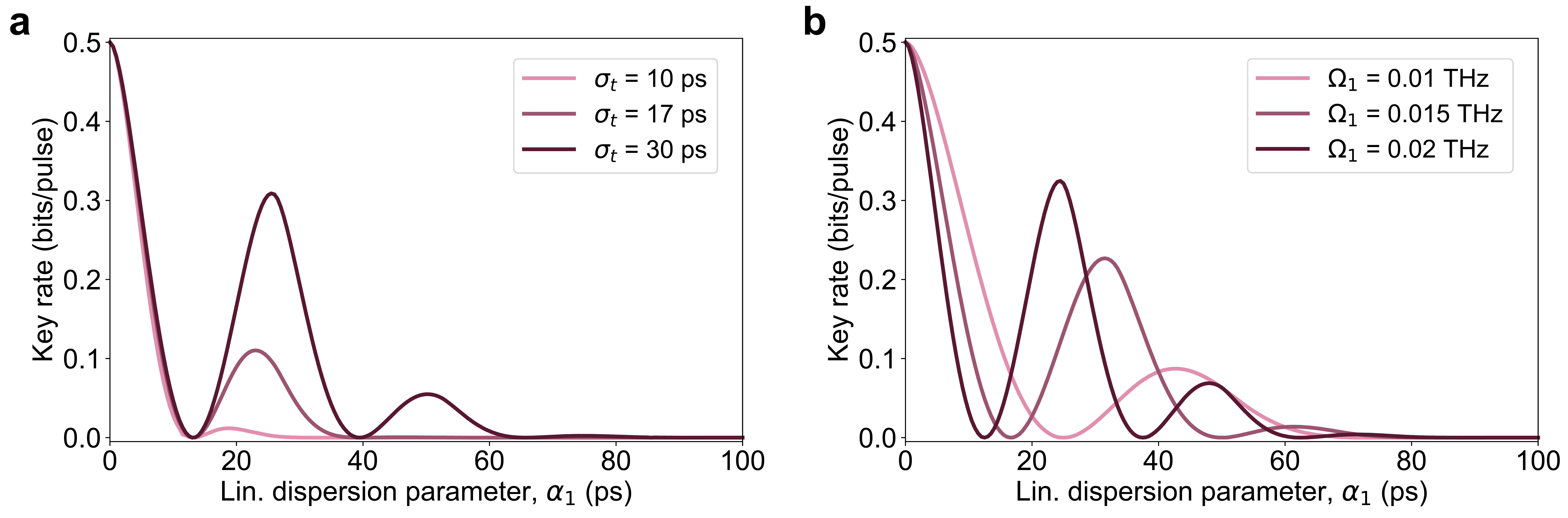}
    \caption{\textbf{Key rate as a function of the linear dispersion parameter.} The key rate as a function of the linear dispersion parameter $\alpha_1$ with varying Gaussian encoding variables. (a) Key rate for varying $\sigma_t = 10, 17, 30$ ps. The other encoding parameters are fixed, with $\Omega_0 = 19.86$ THz $\Omega_1 = 195.879$ THz, $\sigma_{\omega} = 1.1$ GHz, $\tau_0 = 0$ ps and $\tau_1 = 220$ ps, and $\omega_0$ and $\Omega_0$ are set equal to the central time-bin state frequency. (b) Key rate for varying $\Omega_1 = 0.01, 0.015, 0.02$ THz, with respect to the central time-bin frequency, keeping the other parameters from (a) the same, and setting $\sigma_t = 30$ ps.}
    \label{fig:linphaseKR}
\end{figure*}

These expressions correspond directly to the overlap parameters defined in Eq.~\eqref{eq:transformed1doverlaps} in SM, Sec.~\ref{sec:ququartchannel}, which are used to describe the action of the channel on an orthonormal ququart basis constructed from the frequency-bin and time-bin states defined in the main text (see SM, Sec.~\ref{sec:ququartencoding}). 
Using this framework, the expressions above can thus be directly applied to calculate the effect of the channel on the encoded states and, subsequently, the key rate, by following the procedure outlined in SM, Sec.~\ref{sec:ququartchannel}.
We present the key rates in Fig.~\ref{fig:linphaseKR} for varying time-bin widths $\sigma_t$ and varying frequency-bin parameter $\Omega_1$.

Fig.~\ref{fig:linphaseKR}(a) shows the key rate decays faster as $\sigma_t$ decreases. Further, the key rate shows oscillations, taking zeros values. In Fig.~\ref{fig:linphaseKR}(b), we show that, by varying $\Omega_1$, the key rate can be optimised over a fluctuation range of $\alpha_1$, such that the average $\alpha_1$ value is centred on a key rate peak, and avoid zeros in key rate over a selected range of $\alpha_1$. Therefore, if $\alpha_1$ is allowed to fluctuate but within certain bounds, the key rate will remain close to a peak value. Hence, significant key rate can be achieved under this channel noise, by optimising the encoding parameters.

To understand the origin of these decaying oscillations, we analyse the effect of the channel at the logical level. As with the FBS channel, considering the effect at the logical level suffices to capture the effect of the noise on the key rate. Again, it can be well approximated by an amplitude damping and decoherence channel, but now with additional loss on the whole qubit, since the frequency-bin state will also be transformed by the channel (and partially ``leak'' outside the logical subspace).

\begin{figure*}[h]
    \centering
    \includegraphics[width=0.5\textwidth]{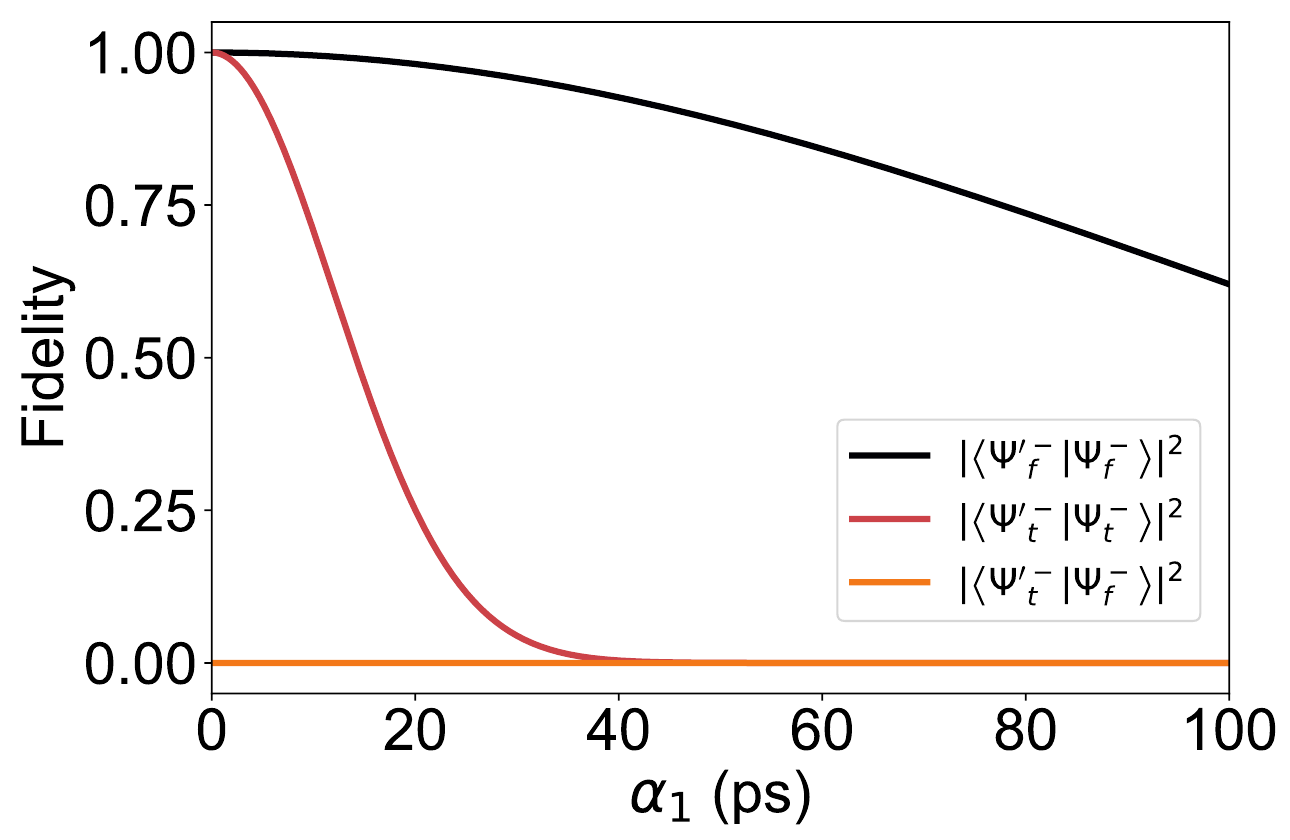}
    \caption{\textbf{Overlap fidelities at Bob given a state by Alice under the linear dispersion channel.} The state fidelities as a function of the linear dispersion parameter $\alpha_1$, for the states used for Fig.~\ref{fig:staterobustness} in the main text ($\tau_0 = 0$ ps, $\tau_1 = 220$ ps, $\sigma_t = 17$ ps, $\Omega_0 = 195.86$ THz, $\Omega_1 = 195.879$ THz, $\sigma_{\omega} = 1.1$ GHz). We have taken $\omega_0$ and $\Omega_0$ to be equal to the central frequency of the time-bin states.}
    \label{fig:linexample}
\end{figure*}

To see this, we calculate the fidelities at the logical level between Bob's measurement state and the state sent by Alice, i.e. $|\braket{\Psi'^-_x|\Psi^-_x}|^2$ for $x$ is $f$ or $t$, as
\begin{equation}
    \begin{split}
        |\braket{\Psi'^-_f|\Psi^-_f}|^2 = \exp{ \left(- \alpha_1^2 \sigma_{\omega}^2 \right)},
    \end{split}
\end{equation}
\begin{equation}
    \begin{split}
        |\braket{\Psi'^-_t|\Psi^-_t}|^2 = \exp{ \left(- \frac{\alpha_1^2}{\sigma_t^2} \right)},
    \end{split}
\end{equation}
and the bit error fidelity between the bit sent by Alice and Bob's measurement, i.e. $|\braket{\Psi'^-_t|\Psi^-_f}|^2$, as
\begin{equation} \label{eq:conjprob1}
    \begin{split}
        |\braket{\Psi'^-_t|\Psi^-_f}|^2= \frac{8\sigma_{\omega}^2\sigma_t^2}{(1+\sigma_{\omega}^2\sigma_t^2)^2}& \frac{1}{1-\exp{\left(-\frac{(\tau_0-\tau_1)^2}{2\sigma_t^2}\right)}} \frac{1}{1-\exp{\left(-\frac{(\Omega_0-\Omega_1)^2}{2\sigma_{\omega}^2}\right)}} \\
        &\times\exp{ \left(- \frac{\sigma_t^2(\Omega_0^2 + \Omega_1^2)}{(1+\sigma_{\omega}^2\sigma_t^2)} \right)} \exp{ \left(- \frac{\sigma_{\omega}^2[(\alpha_1 + \tau_0)^2+ (\alpha_1 + \tau_1)^2]}{(1+\sigma_{\omega}^2\sigma_t^2)} \right)} \\
        &\times \left[1 - \cos{\left(\frac{(\Omega_1-\Omega_0)(\tau_1-\tau_0)}{(1+\sigma_{\omega}^2\sigma_t^2)} \right)} \right].
    \end{split}
\end{equation}
In Fig.~\ref{fig:linexample}, we present these fidelities as a function of $\alpha_1$ for a set of states used in Fig.~\ref{fig:staterobustness} of the main text.

As for the FBS channel, we study the fidelities and accumulated phases in turn.
We observe that the bit error term $|\braket{\Psi'^-_t|\Psi^-_f}|^2$ is generally negligible across the range of $\alpha_1$, and thus the dominant contributions to key rate degradation come from subspace leakage and phase error. For example, the bit error rate of the states shown in Fig.~\ref{fig:staterobustness} in the main text (which have experimentally plausible parameters) is $\lesssim 10^{-4}$ for all $\alpha_1$. 

The fidelities $|\braket{\Psi'^-_t|\Psi^-_f}|^2$ and $|\braket{\Psi'^-_t|\Psi^-_f}|^2$ decay exponentially with $\sigma_{\omega}$ and $1/\sigma_t$, respectively, and with $\alpha_1$. We also note that $\sigma_t$ and $\sigma_{\omega}$ have an inverse effect to each other on the fidelity between their original and noise-affected states, as expected from the properties of the Fourier transform on the time-bin state.

Regarding the accumulated phase, the time-bin state accumulates no phase under the linear dispersion channel, as both bins are centred around the same frequency. In contrast, the frequency-bin qubit states are separated in frequency, and thus acquire different phases. This results in an overall phase on the antisymmetric frequency-bin state, which acquires a relative phase equivalent to $\alpha_1(\Omega_0 + \Omega_1)$, which contributes directly to logical decoherence between $\ket{0_L}$ and $\ket{1_L}$.

Collectively, these effects impart, to a good approximation, an amplitude damping and decoherence channel at the logical level (see SM, Sec.~\ref{sec:ampdampmodellogical}), just as for the FBS channel, but now with an additional global loss. This global loss arises since the $\ket{\Psi'^-_f}$ state is now also affected by the channel.
The effect of the channel can be written approximately as $\ket{0_L} \rightarrow \exp(-\alpha_1^2\sigma_{\omega}^2/2)\exp(i\alpha_1(\Omega_0 + \Omega_1) \ket{0_L}$ and $\ket{1_L} \rightarrow \exp(-\alpha_1^2/(2\sigma_t^2)) \ket{1_L}$.
For the states considered in Fig.~\ref{fig:linphaseKR}, $\exp(-\alpha_1^2/(2\sigma_t^2)) \leq \exp(-\alpha_1^2\sigma_{\omega}^2/2)$ for all $\alpha_1$. Thus, the overall loss on the protocol is $1 - \exp(-\alpha_1^2\sigma_{\omega}^2)$. Renormalising by this factor, and using the equivalence of quantum states under a global phase, the transformation can be rewritten as a map with the same form as that given by expression in Eq.~\eqref{eq:ampdampmap}. That is, we obtain 
\begin{equation}
    \ket{0_L} \rightarrow \ket{0_L}; \qquad \ket{1_L} \rightarrow \exp(-\alpha_1^2/(2\sigma_t^2))/\exp(-\alpha_1^2\sigma_{\omega}^2/2) \exp(i\alpha_1(\Omega_0 + \Omega_1) \ket{1_L},
\end{equation}
with inherent loss $1 - \exp(-\alpha_1^2\sigma_{\omega}^2)$.

This fully explains the decaying oscillations observed in Fig.~\ref{fig:linphaseKR}. The increasing level of asymmetric loss between logical states, along with the incurred global loss, gives rise to the decay in key rate with $\alpha_1$. Meanwhile, the relative phase $\alpha_1(\Omega_0 + \Omega_1)$ at the logical level leads to increasing phase error rate, which causes oscillations. The period of these oscillations with $\alpha_1$ is inversely proportional to $(\Omega_0 + \Omega_1)$, which explains how tuning $\Omega_1$ allows the key rate to be optimised under realistic fluctuations in $\alpha_1$.

\subsection{Quadratic (group velocity) dispersion (\texorpdfstring{$n=2$}{n=2})} \label{sec:n=2disp}
We now perform the same calculations for the $n=2$ case. The 1D overlap functions are now given by
\begin{equation}\label{eq:overlap2ff}
\begin{split}
    \braket{i_f|j_f'} = \int_{-\infty}^{\infty} d\omega &\exp (i\alpha_2 (\omega-\omega_0)^2) f_{\Omega_j, \sigma_{\omega}}(\omega) f_{\Omega_i, \sigma_{\omega}}(\omega)^* \\
    &= \sqrt{\frac{1}{1-i\alpha_2\sigma_{\omega}^2}} \exp{\left(i\alpha_2\omega_0^2\right)} \exp{\left( - \frac{(\Omega_j^2+\Omega_i^2)}{2\sigma_{\omega}^2}\right)} \exp{\left(\frac{(\Omega_j+\Omega_i-2i\alpha_2\omega_0\sigma_{\omega}^2)^2}{4\sigma_{\omega}^2(1-i\alpha_2\sigma_{\omega}^2)} \right)},
\end{split}
\end{equation}
\begin{equation} \label{eq:overlap2tt}
\begin{split}
    \braket{i_t|j_t'} = \int_{-\infty}^{\infty} d\omega &\exp (i\alpha_2 (\omega-\omega_0)^2) F_{\tau_j, \sigma_{t}}(\omega) F_{\tau_i, \sigma_{t}}(\omega)^* \\
    &= \sigma_{t}\sqrt{\frac{1}{\sigma_{t}^2-i\alpha_2}} \exp{\left(i\alpha_2\omega_0^2\right)} \exp{\left( - \frac{(\tau_j-\tau_i+2i\alpha_2\omega_0)^2}{4(\sigma_{t}^2-i\alpha_2)}\right)},
\end{split}
\end{equation}
and
\begin{equation}\label{eq:overlap2tf}
\begin{split}
    \braket{i_t|j_f'} &= \int_{-\infty}^{\infty} d\omega \exp (i\alpha_2 (\omega-\omega_0)^2) f_{\Omega_j, \sigma_{\omega}}(\omega) F_{\tau_i, \sigma_{t}}(\omega)^* \\
    &= \sqrt{\frac{2\sigma_{\omega}\sigma_t}{1+\sigma_{\omega}^2\sigma_t^2-2i\sigma_{\omega}^2\alpha_2}} \exp{\left(i\alpha_2\omega_0^2\right)} \exp{\left( - \frac{(\sigma_t^2-2i\alpha_2)\Omega^2}{2(1+\sigma_{\omega}^2\sigma_t^2-2i\sigma_{\omega}^2\alpha_2)}\right)} \\
    &\hspace{1cm} \times \exp{\left( - \frac{\sigma_{\omega}^2\tau_i^2}{2(1+\sigma_{\omega}^2\sigma_t^2-2i\sigma_{\omega}^2\alpha_2)}\right)} \exp{\left( \frac{i\Omega_j\tau_i}{(1+\sigma_{\omega}^2\sigma_t^2-2i\sigma_{\omega}^2\alpha_2)}\right)}\\
    &\hspace{1cm} \times \exp{\left( -\frac{2i\alpha_2\omega_0\Omega_j}{(1+\sigma_{\omega}^2\sigma_t^2-2i\sigma_{\omega}^2\alpha_2)}\right)} \exp{\left( \frac{2\sigma_{\omega}^2\alpha_2\omega_0\tau_i}{(1+\sigma_{\omega}^2\sigma_t^2-2i\sigma_{\omega}^2\alpha_2)}\right)}\exp{\left(- \frac{2\sigma_{\omega}^2\alpha_2^2\omega_0^2}{(1+\sigma_{\omega}^2\sigma_t^2-2i\sigma_{\omega}^2\alpha_2)}\right)}.
\end{split}
\end{equation}

As for $n=1$ dispersion, these directly correspond to the overlap parameters defined in Eq.~\eqref{eq:transformed1doverlaps} in SM, Sec.~\ref{sec:ququartchannel}, used to describe the action of the channel on an orthonormal ququart basis. 
Thus, they can be directly used to evaluate the key rate.

\begin{figure*}[h]
    \centering
    \includegraphics[width=\textwidth]{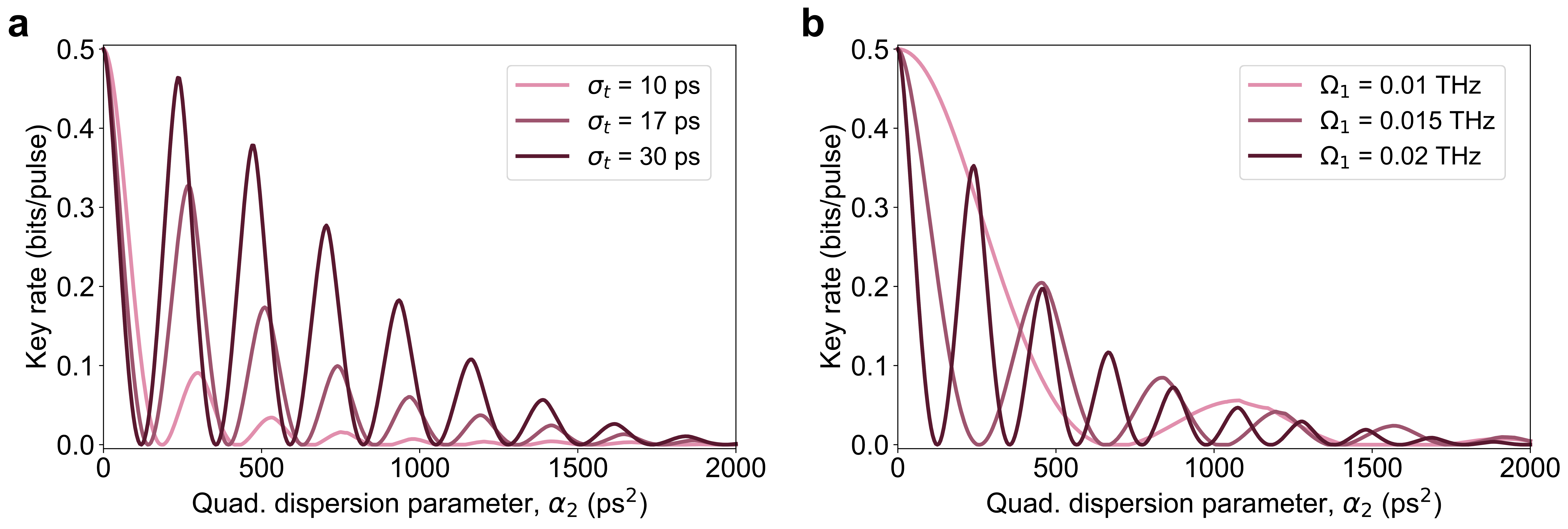}
    \caption{\textbf{Key rate as a function of the quadratic dispersion parameter.} The key rate as a function of the linear dispersion parameter $\alpha_2$ with varying encoding variables. (a) Key rate for varying Gaussian widths $\sigma_t = 10, 17, 30$ ps. The other encoding parameters are fixed, with $\Omega_0 = 19.86$ THz $\Omega_1 = 195.879$ THz, $\sigma_{\omega} = 1.1$ GHz, $\tau_0 = 0$ ps and $\tau_1 = 220$ ps, and $\omega_0$ and $\Omega_0$ are set equal to the central time-bin state frequency. (b) Key rate for varying $\Omega_1 = 0.01, 0.015, 0.02$ THz, with respect to the central time-bin frequency, keeping the other parameters from (a) the same, and setting $\sigma_t =$ 17 ps.}
    \label{fig:quadphaseKR}
\end{figure*}

We present the key rates as a function of $\alpha_2$ in Fig.~\ref{fig:quadphaseKR} for varying time-bin widths $\sigma_t$ and varying frequency-bin parameter $\Omega_1$. As for linear dispersion, we see in Fig.~\ref{fig:quadphaseKR}(a) that as the time-bin width decreases, the rate at which the key rate decays with $\alpha_2$ increases. We also find for $n=2$ that $\sigma_t$ has a small effect on the rate of oscillations. In Fig.~\ref{fig:quadphaseKR}(b), we show that by varying $\Omega_1$, as an illustrative example, we can again optimise the encoding such that $\alpha_2$ is allowed to fluctuate over a larger range whilst maintaining a non-zero key rate.

As for the $n=1$ case, we can gain intuition about the key rate dependencies on the encoding parameters by analysing the effect of the channel at the logical level.
We calculate the fidelities between Bob's measurement state and the state sent by Alice, which correspond to probabilities of Bob's measurement outcomes. To keep the expressions simpler, we take the case in which $\omega_0=0$ THz (i.e. $k(\omega)$ is expanded about the central frequency of the time-bin states). The fidelities are then given by
\begin{equation}\label{eq:2ff}
    \begin{split}
        |\braket{\Psi'^-_f|\Psi^-_f}|^2 =&\frac{1}{\left[1-\exp{\left(-\frac{(\Omega_0-\Omega_1)^2}{2\sigma_{\omega}^2}\right)} \right]^2} \frac{1}{1+\alpha_2^2\sigma_{\omega}^4} \exp{\left(-\frac{2\alpha_2^2\sigma_{\omega}^2(\Omega_0^2 + \Omega_1^2)}{1+\alpha_2^2\sigma_{\omega}^4}\right)} \\
        &\times\left[1+ \exp{\left(-\frac{(\Omega_0-\Omega_1)^2}{(1+\alpha_2^2\sigma_{\omega}^4)\sigma_{\omega}^2}\right)} - 2\exp{\left(-\frac{(\Omega_0-\Omega_1)^2}{2(1+\alpha_2^2\sigma_{\omega}^4)\sigma_{\omega}^2}\right)}\cos{\left(\frac{\alpha_2(\Omega_0-\Omega_1)^2}{2(1+\alpha_2^2\sigma_{\omega}^4)}\right)} \right],
    \end{split}
\end{equation}
\begin{equation}\label{eq:2tt}
    \begin{split}
        |\braket{\Psi'^-_t|\Psi^-_t}|^2 =&\frac{1}{\left[1-\exp{\left(-\frac{(\tau_0-\tau_1)^2}{2\sigma_{t}^2}\right)} \right]^2} \frac{\sigma_t^4}{\sigma_t^4+\alpha_2^2} \\
        &\times\left[1+ \exp{\left(-\frac{\sigma_t^2(\tau_0-\tau_1)^2}{(\sigma_{t}^4+\alpha_2^2)}\right)} - 2\exp{\left(-\frac{\sigma_t^2(\tau_0-\tau_1)^2}{2(\sigma_t^4+\alpha_2^2)}\right)}\cos{\left(\frac{\alpha_2(\tau_0-\tau_1)^2}{2(\sigma_t^4+\alpha_2^2)}\right)} \right],
    \end{split}
\end{equation}
\begin{equation}\label{eq:2tf}
    \begin{split}
        |\braket{\Psi'^-_t|\Psi^-_f}|^2 =&\frac{1}{\left[1-\exp{\left(-\frac{(\tau_0-\tau_1)^2}{2\sigma_{t}^2}\right)} \right]}\frac{1}{\left[1-\exp{\left(-\frac{(\Omega_0-\Omega_1)^2}{2\sigma_{\omega}^2}\right)} \right]} \frac{4\sigma_t^2\sigma_{\omega}^2}{(1+\sigma_t^2\sigma_{\omega}^2)^2+4\sigma_{\omega}^4\alpha_2^2} \\
        &\times \exp{\left(-\frac{[\sigma_t^2(1+\sigma_t^2\sigma_{\omega}^2)+4\sigma_{\omega}^2\alpha_2^2](\Omega_0^2+\Omega_1^2)}{(1+\sigma_t^2\sigma_{\omega}^2)^2+4\sigma_{\omega}^4\alpha_2^2}\right)} \exp{\left(-\frac{\sigma_{\omega}^2(1+\sigma_t^2\sigma_{\omega}^2)(\tau_0^2+\tau_1^2)}{(1+\sigma_t^2\sigma_{\omega}^2)^2+4\sigma_{\omega}^4\alpha_2^2}\right)}\\
        &\times\left[\exp{\left(-\frac{4\sigma_{\omega}^2\alpha_2(\Omega_0\tau_0+\Omega_1\tau_1)}{(1+\sigma_t^2\sigma_{\omega}^2)^2+4\sigma_{\omega}^4\alpha_2^2}\right)}+\exp{\left(-\frac{4\sigma_{\omega}^2\alpha_2(\Omega_0\tau_1+\Omega_1\tau_0)}{(1+\sigma_t^2\sigma_{\omega}^2)^2+4\sigma_{\omega}^4\alpha_2^2}\right)} \right.
        \\ &\hspace{2cm}\left.- 2\exp{\left(-\frac{2\sigma_{\omega}^2\alpha_2(\Omega_0+\Omega_1)(\tau_0+\tau_1)}{(1+\sigma_t^2\sigma_{\omega}^2)^2+4\sigma_{\omega}^4\alpha_2^2}\right)}\cos{\left(\frac{(1+\sigma_{\omega}^2\sigma_t^2)(\Omega_1-\Omega_0)(\tau_1-\tau_0)}{(1+\sigma_t^2\sigma_{\omega}^2)^2+4\sigma_{\omega}^4\alpha_2^2}\right)} \right].
    \end{split}
\end{equation}

\begin{figure*}[h]
    \centering
    \includegraphics[width=0.5\textwidth]{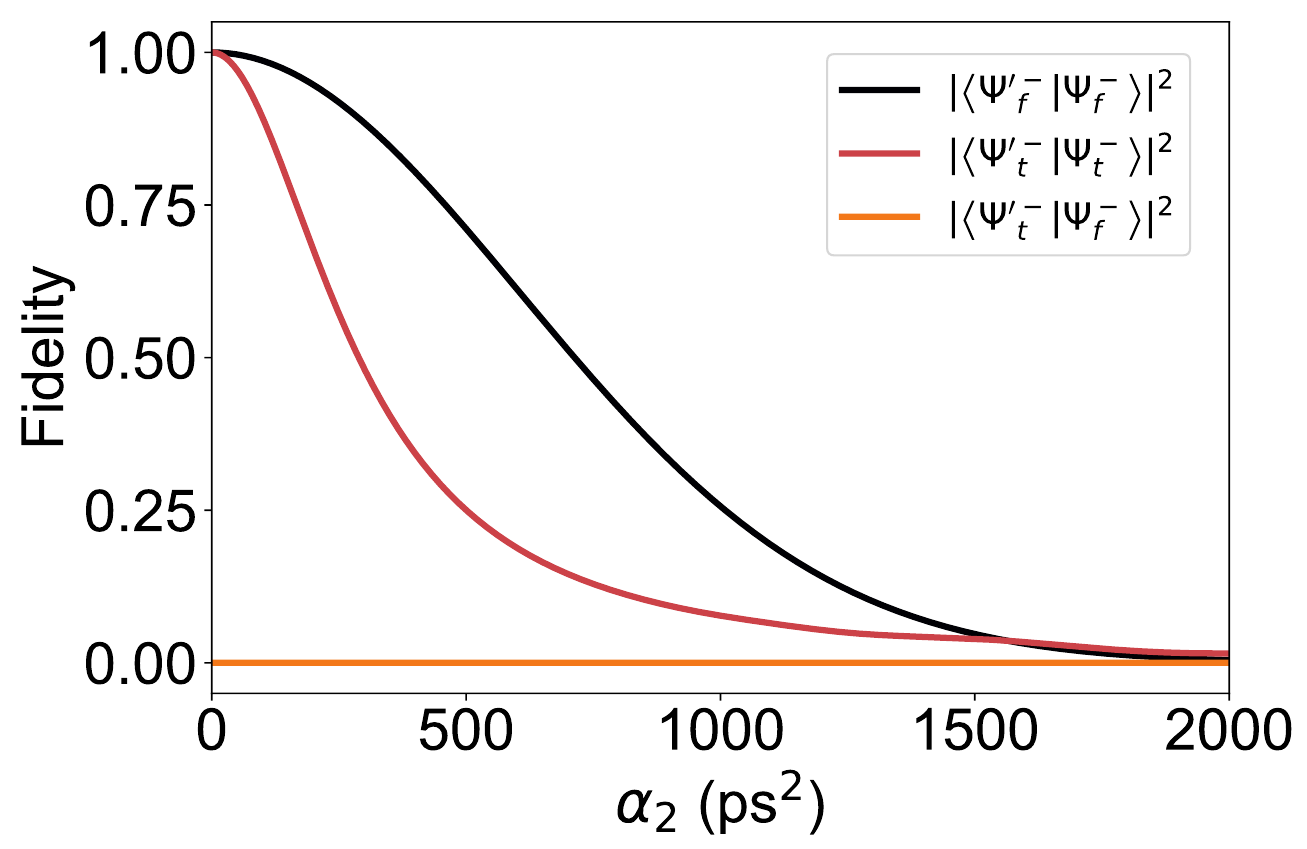}
    \caption{\textbf{Overlap fidelities at Bob given a state by Alice under the quadratic dispersion channel.} The state fidelities as a function of the quadratic dispersion parameter $\alpha_2$, for the Gaussian parameters of Fig.~\ref{fig:staterobustness} in the main text ($\tau_0 = 0$ ps, $\tau_1 = 220$ ps, $\sigma_t = 17$ ps, $\Omega_0 = 195.86$ THz, $\Omega_1 = 195.879$ THz, $\sigma_{\omega} = 1.1$ GHz). We have taken $\omega_0$ and $\Omega_0$ to be equal to the central frequency of the time-bin states.}
    \label{fig:quadexample}
\end{figure*}

We observe that $\sigma_t$ and $\sigma_{\omega}$ again play the same roles in the fidelities $|\braket{\Psi'^-_f|\Psi^-_f}|^2$ and $|\braket{\Psi'^-_t|\Psi^-_t}|^2$, with an inverse effect for $\sigma_t$ due to the Fourier transform. 
In Fig.~\ref{fig:quadexample}, we show these measurement probabilities as a function of $\alpha_2$ for states used in Fig.~\ref{fig:staterobustness} of the main text ($\tau_0=0$ ps, $\tau_1=220$ ps, $\sigma_t=17$ ps, $\Omega_0 = 195.86$ THz, $\Omega_1 = 195.879$ THz, and $\sigma_{\omega} = 1.1$ GHz, taking the central frequency of the time-bin states to equal $\Omega_0$). As for linear dispersion, we find that the bit error rate is negligible ($\lesssim 10^{-4}$) compared to the asymmetric loss. Further, we see that probabilities decay an order of magnitude more slowly for the absolute value of the $\alpha_2$ parameter than $\alpha_1$.

\begin{figure*}[h]
    \centering
    \includegraphics[width=\textwidth]{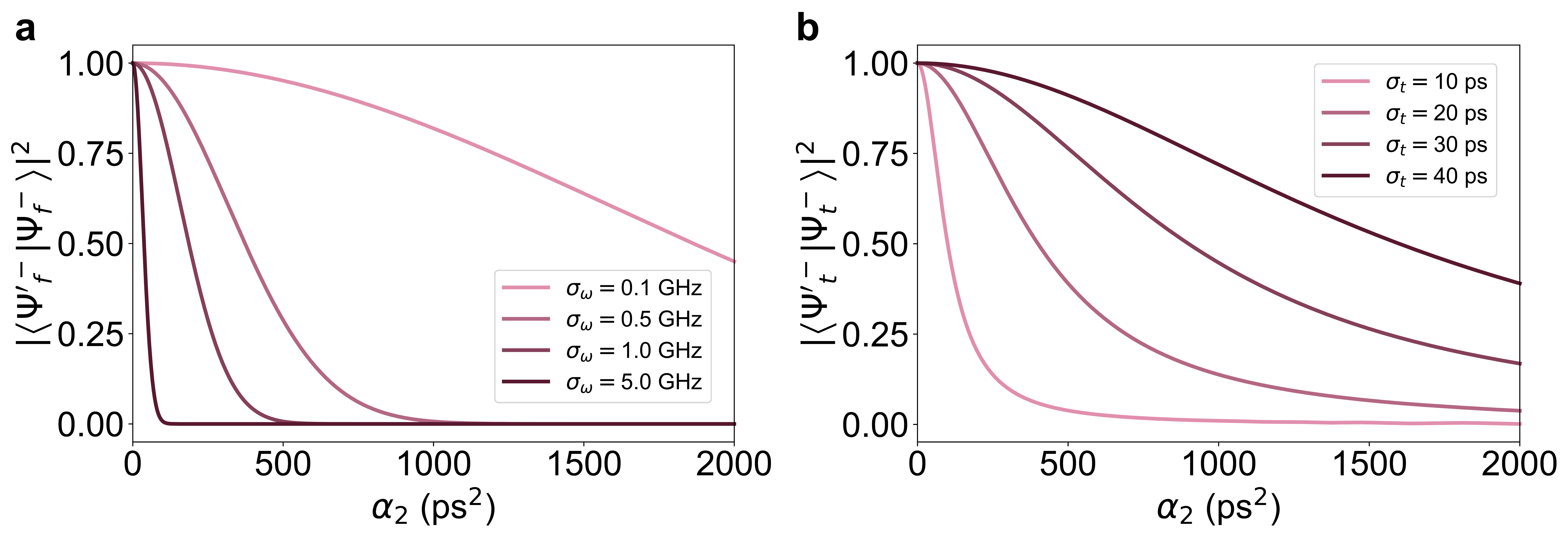}
    \caption{\textbf{Fidelity of the state arriving at Bob given a state sent down a quadratic dispersion channel by Alice for varying encoding bin widths.} (a) The frequency-bin state fidelity $|\braket{\Psi'^-_f|\Psi^-_f}|^2$, with a Gaussian state encoding, as a function of the quadratic dispersion parameter $\alpha_2$, for various frequency-bin widths $\sigma_{\omega}$. The fixed frequency-bin parameters are  $\Omega_0 = 195.86$ THz, $\Omega_1 = 195.94$ THz. (b) The time-bin state fidelity $|\braket{\Psi'^-_t|\Psi^-_t}|^2$ as a function of $\alpha_2$, for various time-bin widths $\sigma_t$. The time-bins are at times $\tau_0 = 0$ ps and $\tau_1 = 320$ ps. For both, the central time-bin state frequency is set equal to $\omega_0$ and $\Omega_0$.}
    \label{fig:GVDwidth}
\end{figure*}

\begin{figure*}[h]
    \centering
    \includegraphics[width=\textwidth]{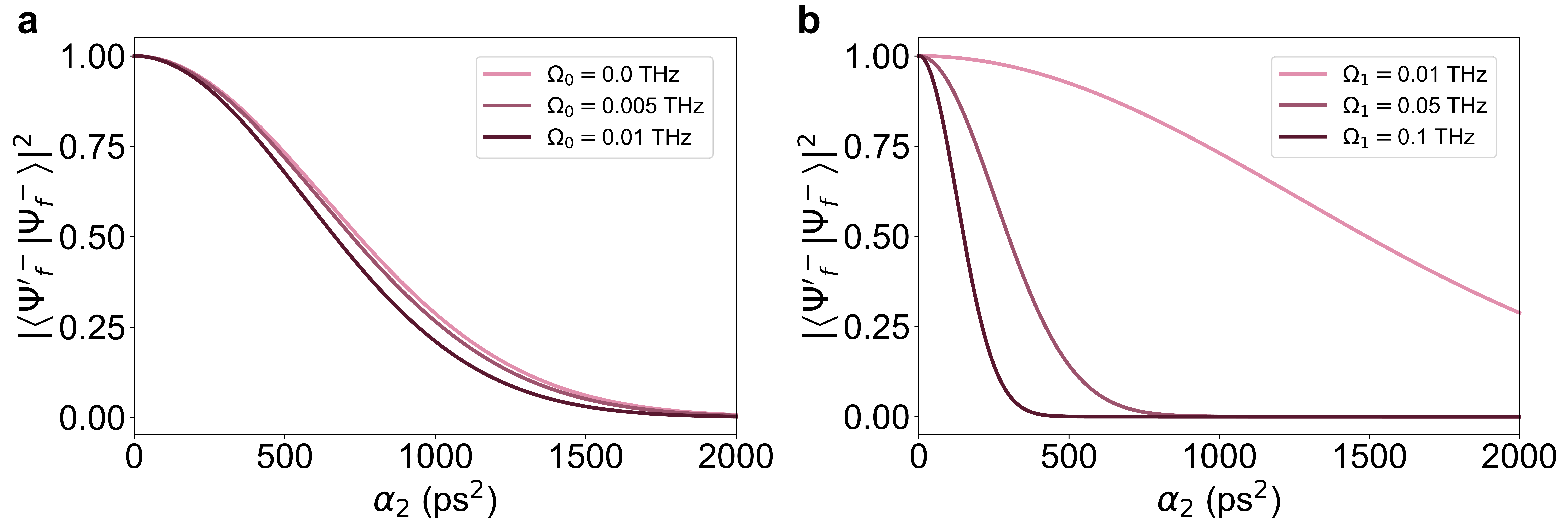}
    \caption{\textbf{Fidelity of the frequency-bin state arriving at Bob given a frequency-bin state sent down a quadratic dispersion channel by Alice for varying frequency-bin parameters $\Omega_0$ and $\Omega_1$.} (a) The frequency-bin state fidelity $|\braket{\Psi'^-_f|\Psi^-_f}|^2$, with a Gaussian encoding, as a function of the quadratic dispersion parameter $\alpha_2$, for $\Omega_0 = 0, 0.005, 0.1$ THz, relative to the central frequency of the time-bin states (which is set to be 195.86 THz). The fixed parameters are $\Omega_1 = 195.94$ THz and $\sigma_{\omega} = 1$ GHz. (b) The frequency-bin state fidelity as a function of $\alpha_2$, for $\Omega_1 = 0.01, 0.05, 0.1$ THz, relative to the central frequency of the time-bin states (again, set to be 195.86 THz). The fixed parameters are $\Omega_0 = 0$ THz (relative to the time-bin central frequency) and $\sigma_{\omega} = 1$ GHz.}
    \label{fig:GVDomega}
\end{figure*}

\begin{figure*}[h]
    \centering
    \includegraphics[width=\textwidth]{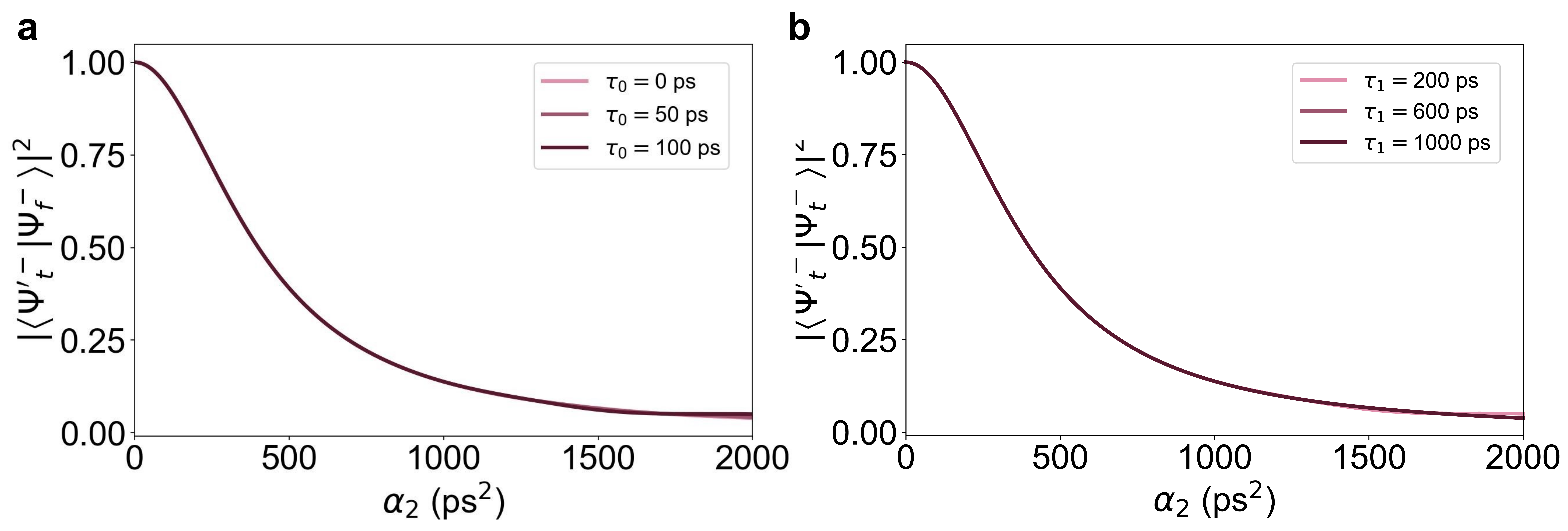}
    \caption{\textbf{Fidelity of the time-bin state arriving at Bob given a time-bin state sent down a quadratic dispersion channel by Alice for varying time-bin parameters $\tau_0$ and $\tau_1$.} (a) The time-bin state fidelity $|\braket{\Psi'^-_t|\Psi^-_t}|^2$, with Gaussian profiles, as a function of the quadratic dispersion parameter $\alpha_2$, for $\tau_0 = 0, 50, 100$ ps. The fixed parameters are $\tau_1 = 300$ ps and $\sigma_{t} = 20$ ps. (b) The time-bin state fidelity as a function of $\alpha_2$, for $\tau_1 = 200, 600, 1000$ ps. The fixed parameters are $\tau_0 = 0$ ps and $\sigma_{t} = 20$ ps.}
    \label{fig:GVDtau}
\end{figure*}

To understand the parameter dependence of Equations~\eqref{eq:2ff}~and~\eqref{eq:2tt} further, we visualise the dependence of the fidelities $|\braket{\Psi'^-_f|\Psi^-_f}|^2$ and $|\braket{\Psi'^-_t|\Psi^-_t}|^2$ on $\alpha_2$ for varying parameters in Figs.~\ref{fig:GVDwidth} - \ref{fig:GVDtau}. 

Fig.~\ref{fig:GVDwidth} shows the fidelities as a function of $\alpha_2$ for varying bin widths. As for linear dispersion, as the frequency-bin width $\sigma_{\omega}$ increases, the decay rate of frequency-bin state fidelity increases. Meanwhile, as the time-bin width $\sigma_t$ increases, the time-bin state fidelity decays less rapidly.

In Fig.~\ref{fig:GVDomega}, we present the fidelity of the frequency-bin state for varying $\Omega_0$ (a) and varying $\Omega_1$ (b) with $\sigma_{\omega}$ fixed. As $\Omega_0$ and $\Omega_1$ increase (relative to the central frequency of the time-bin state), the fidelity of the arriving state decreases. This can be understood from the exponential prefactor in Eq.~\eqref{eq:2ff}, which dominates the $\Omega_0$- and $\Omega_1$-dependence. The other exponential terms are approximately zero since $(\Omega_1-\Omega_0)\geq6\sigma_{\omega}$ is required for frequency-bins to form an approximate orthonormal basis. Thus, in the large $\alpha_2$ limit, the fidelity decays exponentially with increasing $\Omega_0$ and $\Omega_1$. For the small $\alpha_2$ limit, we can understand Eq.~\eqref{eq:2ff} using the heuristic explanation given in Section~\ref{sec:dispmain} of the main text.

In Fig.~\ref{fig:GVDtau}, we investigate the fidelity of the time-bin state for varying $\tau_0$ (a) and varying $\tau_1$ (b) with $\sigma_t$ fixed. The fidelity shows minimal dependence on $\tau_0$ and $\tau_1$. We can interpret this from the exponential terms in Eq.~\eqref{eq:2tt}. As for the frequency-bin states, these terms are approximately zero since $(\tau_1-\tau_0)\geq6\sigma_t$ is required to have time-bin distinguishability. Therefore, the fidelity falls approximately as $1/(1 + \alpha_2^2/\sigma_t^4)$, which is independent of $\tau_0$ and $\tau_1$. Small deviations can be seen in the large $\alpha_2$ limit, where the exponential terms become non-negligible. 

From the above analysis, we again see that the channel can be approximated by and amplitude damping and decoherence channel with a global logical qubit loss. Using an analogous method to the linear dispersion channel, the channel map can be approximately rewritten into the form of an amplitude damping and decoherence channel with an incurred global loss.

By considering the fidelities in Eqs.~\eqref{eq:2ff} and \eqref{eq:2tt}, we see that as $\alpha_2$ increases, the effective incurred loss and amplitude damping increases, giving rise to the key rate decay.
Similarly, by studying Equations~\eqref{eq:overlap2ff} and \eqref{eq:overlap2tt}, we can see that the logical decoherence phase now takes a more complex dependence on $\alpha_2$, with dependence on all encoding parameters $\vec{\boldsymbol{\Omega}}, \vec{\boldsymbol{\tau}}$. We can understand the decoherence phase with the same general intuition given for the $n=1$ case (but now for the time-bin states, there will also be an accumulated global phase relative to their untransformed states). This explains the oscillations in key rate, and the $\sigma_t$ dependence explains the varying oscillation rate in Fig.~\ref{fig:quadphaseKR}(a) found here, compared to the $n=1$ case.

Overall, our analysis shows high key rates can be achieved, even for large values of dispersion. For example, considering dispersion only and using the experimental value of $\alpha_2 \approx -1.58$ ps$^2$/km in fused silica at room temperature at 1530nm \cite{Kim1993}, the encoding in Fig.~\ref{fig:quadphaseKR}(a) can achieve key rates substantially larger than 0.1 bits/pulse over at least 200 km of fibre.

Additionally, our obtained key rates show that our protocol exhibits high tolerance to non-optimal encoding parameters (for example, resulting from uncertainties in the characterisation of $\alpha_2$). Dispersion parameter fluctuations are primarily dominated by temperature fluctuations. For instance, the encoding parameters used in Fig.~\ref{fig:quadphaseKR} operate with a central frequency of approximately 195.87 THz (1530 nm). Ref.~\cite{Kim1993} reports that (for their Fibre B) the second-order dispersion parameter $\alpha_2$ at 1530 nm varies from approximately 264 ps$^2$ at $-20^\circ$C to 314 ps$^2$ at $+60^\circ$C when converted to 200 km of fibre (and units converted from ps/(nm km)).
As seen from Fig.~\ref{fig:quadphaseKR}(a), this range lies well within the second oscillation peak of the key rate versus $\alpha_2$ curve for $\sigma_t = 30$ ps, for example, which spans from approximately 120 to 355 ps$^2$. This remains true even for large uncertainties, up to $\approx$ 185 ps$^2$, in the predicted average $\alpha_2$ (i.e., $\alpha_2 \approx 238 \pm 92$). We find that when $\alpha_2$ fluctuates randomly across the full extent of this peak, the overall key rate is approximately 0.15 bits/pulse (restricting fluctuations to just the expected range 264 to 314 ps$^2$ increases this to around 0.22 bits/pulse). Consequently, useful key rates ($\gtrsim$ 0.1 bits/pulse) can be maintained despite substantial uncertainty in the characterisation of $\alpha_2$. This highlights a strong tolerance to imperfect knowledge of the dispersion parameter.

\section{Multiple FBS pairs channel robustness} \label{sec:multiU}

We extend the FBS model from SM, Sec.~\ref{sec:CVnoise} further. We now include additional FBS operations acting on frequency bins outside of the $\{\ket{0_f}, \ket{1_f}\}$ basis states. With this, we further probe the robustness of our protocol to an increased level of noise.

For simplicity, let us highlight the new model considering just two pairs of bins, that are ordered in frequency $\Omega_0 < \Omega_1 < \Omega_2 <\Omega_3$ such that a single frequency pair FBS operation acts between $\Omega_0$ and $\Omega_1$ (with $\mu_1 = \Omega_1-\Omega_0$, width $\epsilon_1$ and $\theta_1, \phi_1$) and between $\Omega_2$ and $\Omega_3$ (with $\mu_2 = \Omega_3-\Omega_2$, width $\epsilon_2$ and $\theta_2, \phi_2$). In this case, we can write the action of the extended model as taking the general wavefunction $y(\omega)$ to the function $z(\omega)$ according to
\begin{equation}
    z(\omega) = \begin{cases}
    y(\omega) & \omega \leq \Omega_0 - \epsilon_1/2\\
    \cos \theta_1 y(\omega) - e^{-i\phi_1} \sin \theta_1 y(\omega+\mu_1) & \Omega_0 - \epsilon_1/2 < \omega \leq \Omega_0 + \epsilon_1/2\\
    y(\omega) & \Omega_0 + \epsilon_1/2 < \omega  \leq \Omega_1 - \epsilon_1/2\\
    \cos \theta_1 y(\omega) + e^{i\phi_1} \sin \theta_1 y(\omega-\mu_1) & \Omega_1 - \epsilon_1/2 < \omega \leq \Omega_1 +  \epsilon_1/2\\
    y(\omega) & \Omega_1 + \epsilon_1/2 < \omega  \leq \Omega_2 - \epsilon_2/2\\
    \cos \theta_2 y(\omega) - e^{-i\phi_2} \sin \theta_2 y(\omega+\mu_2) & \Omega_2 - \epsilon_2/2 < \omega \leq \Omega_2 + \epsilon_2/2\\
    y(\omega) & \Omega_2 + \epsilon_2/2 < \omega  \leq \Omega_3 - \epsilon_2/2\\
    \cos \theta_2 y(\omega) + e^{i\phi_2} \sin \theta_2 y(\omega-\mu_2) & \Omega_3 - \epsilon_2/2 < \omega \leq \Omega_3 +  \epsilon_2/2\\
    y(\omega) & \omega > \Omega_3 + \epsilon_2/2.
    \end{cases}
\end{equation} 
More generally, we can increase the number of pairs and remove the restriction that they are ordered.

\begin{figure*}[t]
    \centering
    \includegraphics[width=0.4\textwidth]{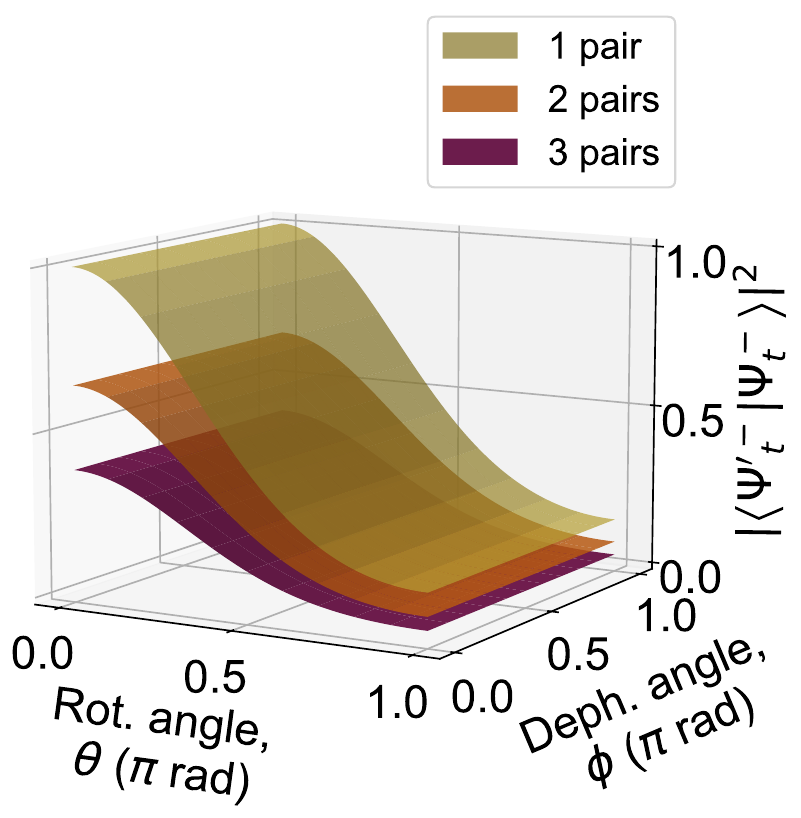}
    \caption{\textbf{Robustness of time-bin Bell-states under frequency beamsplitter operation, for varying numbers of operations.} (a) Probability of measuring the state $\ket{\Psi^-_t}$ as a function of unitary parameters $\theta$ and $\phi$, given the state $\ket{\Psi^-_t}$ with fixed parameters $\tau_0 = 0$ ps, $\tau_1 = 220$ ps, $\sigma_t = 17$ ps (equivalent to the state in Fig.~\ref{fig:staterobustness} in the main text) was sent down a channel applying a frequency beamsplitter operation $U(\theta,\phi)$ on the frequency bins $\Omega_0 = 195.860$ THz, $\Omega_1 = 195.879$ THz,  and varying the width of the frequency bins $\epsilon_1 = 3.0$ GHz (yellow curve) and an additional beamsplitter operation with $\theta=\frac{\pi}{2}, \phi=\frac{\pi}{2}$ on the frequency noise bins $\Omega_2 = 195.834$ THz, $\Omega_3 = 195.853$ THz, $\epsilon_2 = 3.0$ GHz (orange curve), and a further additional beamsplitter operation with $\theta=\frac{\pi}{2}, \phi=\frac{\pi}{2}$ on the frequency bin $\Omega_4 = 195.867$ THz, $\Omega_5 = 195.866$ THz, $\epsilon_3 = 3.0$ GHz (purple curve).}
    \label{fig:multiU}
\end{figure*}

In Fig.~\ref{fig:multiU}, we present the effect of this extended noise channel with FBS operations acting across one, two and three different pairs of frequency noise bins on the $\ket{\Psi^-_t}$ state (again, with Gaussian-shaped bins), with the parameters $\theta=\phi=\pi/2$ (equivalent to a bit-flip, Pauli $X$) for extra FBS pairs.
As one could predict, the fidelity $|\braket{\Psi'^-_t|\Psi^-_t}|^2$ is reduced by the action of these additional operations. 
However, we see that the fidelity is still greater than zero and non-negligible for a range of rotation and dephasing angles as the number of FBS operations increases. This allows a shared secret key to be established under less restrictive, noisier conditions, highlighting the utility of our scheme.

\section{Key rates of DV-encoded protocols} \label{sec:KR_DV_prots}

In this section, we discuss the key rates presented in Figs.~\ref{fig:noisycomp} and \ref{fig:losscomp} in the main text for the existing collective unitary noise-robust protocols in Refs.~\cite{Boileau, XBWang2005, efficient_collective_noise} and BB84~\cite{BB84_1}. Further, we provide additional figures of the noise resilience of these protocols over unitary channels including dephasing angle $\phi$ dependence, and present the security of Ref.~\cite{efficient_collective_noise} under a loss-only channel. Further, we compare the key rate to the rates predicted within their work.

\subsection{Key rates in collective unitary channels of BB84 and Wang} \label{sec:KR_U_prots}

\begin{figure*}[h]
    \centering
    \includegraphics[width=0.8\textwidth]{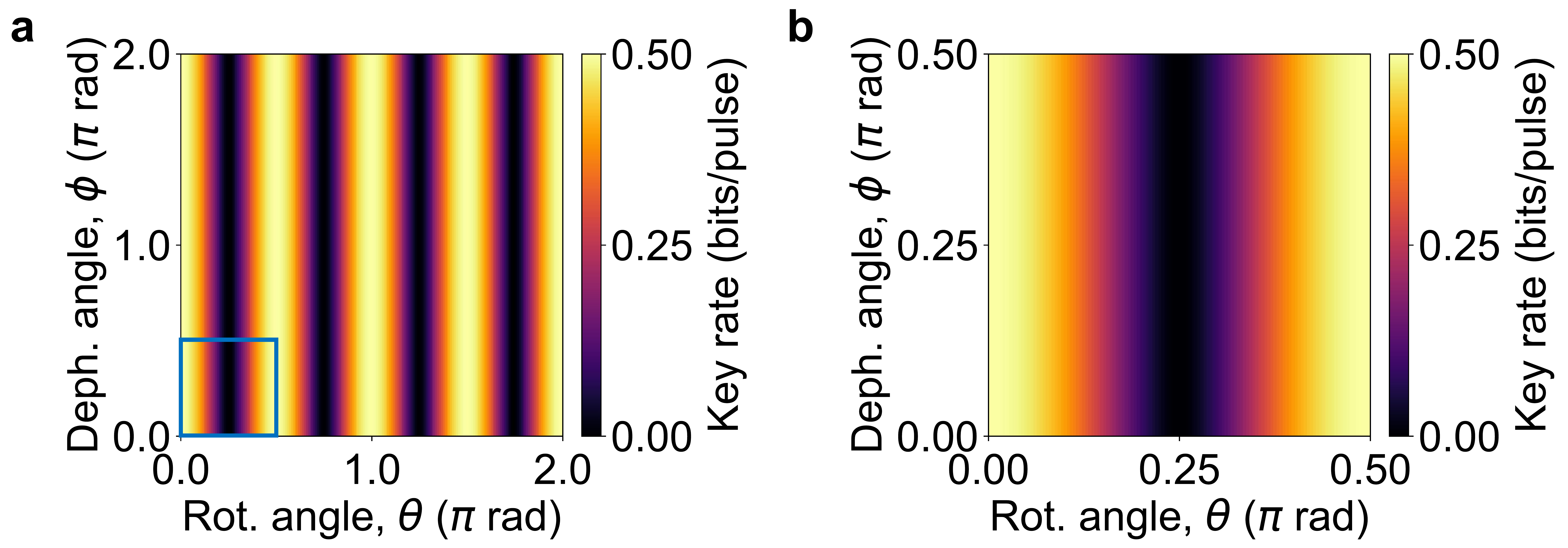}
    \caption{\textbf{Key rate of Wang (2005)~\cite{XBWang2005} under a collective unitary channel with rotation angle $\theta$ and dephasing angle $\phi$.} Key rate of Ref.~\cite{XBWang2005} for unitary parameters $\theta$ and $\phi$ over range (a) $[0,2\pi)$ and (b) $[0, \pi/2)$. This is symmetric (mirrored) about the $\pi/2$ line and about $\pi$ in both $\theta$ and $\phi$. Further, the key rate is independent of the dephasing angle $\phi$. The blue square in (a) indicates the range of the simulation for (b).}
    \label{fig:Wang_U}
\end{figure*}

Ref~\cite{XBWang2005} demonstrated that their encoding was perfectly robust against dephasing-only ($\phi$) noise and outlined its increased robustness to rotation noise ($\theta$) in terms of bit error rates. They showed that the bit error rate goes as $\sin^2 \theta/(\cos^4 \theta + \sin^4 \theta)$, compared to $\sin^2 \theta$ for BB84. 
From  Fig.~\ref{fig:Wang_U}, we see that Ref~\cite{XBWang2005} is indeed perfectly resilient to dephasing-only noise, demonstrated by the independence of the key rate on $\phi$. As shown in Fig.~\ref{fig:noisycomp} in the main text, the key rate of Ref~\cite{XBWang2005} shows a trend matching that predicted by the simple bit error rate analysis. The key rate for BB84 is lower than the key rate for Ref~\cite{XBWang2005} for all $\theta$ parameters, thus demonstrating the predicted improvement.

\begin{figure*}[h]
    \centering
    \includegraphics[width=0.8\textwidth]{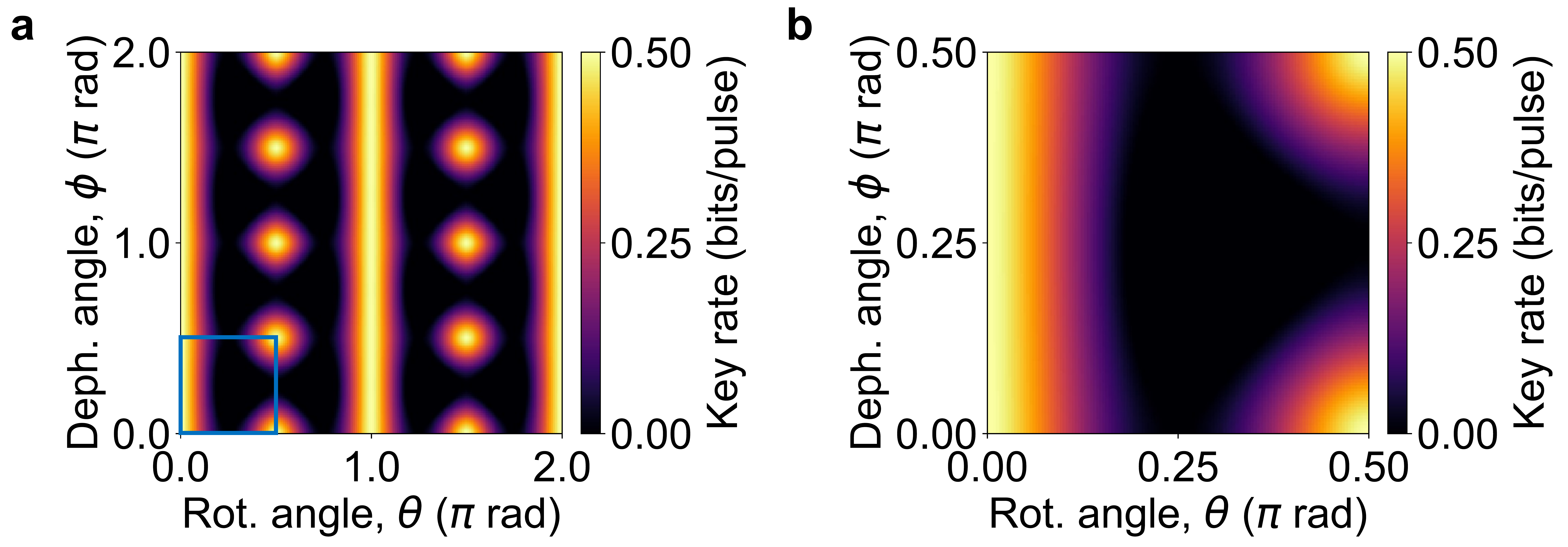}
    \caption{\textbf{Key rate of BB84 under a collective unitary channel with rotation angle $\theta$ and dephasing angle $\phi$.} Key rate of BB84 for unitary parameters $\theta$ and $\phi$ over range (a) $[0,2\pi)$ and (b) $[0, \pi/2)$. The key rate is symmetric (mirrored) about the $\pi/2$ line and about $\pi$ in both $\theta$ and $\phi$. The blue square in (a) highlights the range of the simulation covered in (b).}
    \label{fig:BB84_U}
\end{figure*}

For $0<\theta<\pi/4$ and $3\pi/4<\theta<\pi$, for both BB84 and Ref.~\cite{XBWang2005}, Alice and Bob's outcomes are correlated. For $\pi/4<\theta<3\pi/4$, before post-processing, Alice and Bob's outcomes will be anticorrelated. In post-processing, the bit value is flipped so that Alice and Bob have matching bits. However, it is this effect that gives rise to the zero key rate when $\theta$ is not constant but allowed to vary over its full range (Sec.~\ref{sec:dispmain} in the main text), as would be expected experimentally without stabilisation techniques.

\subsection{Key rate comparison of the 4-photon protocols Boileau \textit{et al}. (2004) and Li \textit{et al}. (2008)} \label{sec:4photoncomp}

Ref.~\cite{Boileau} predicted that their two protocols would show B92-like security \cite{B92}. 
Instead, our analysis shows BB84-like scaling with loss. This realisation was aided by the improved understanding of security analysis and standard techniques now available. 

\begin{figure*}[h]
    \centering
    \includegraphics[width=0.7\textwidth]{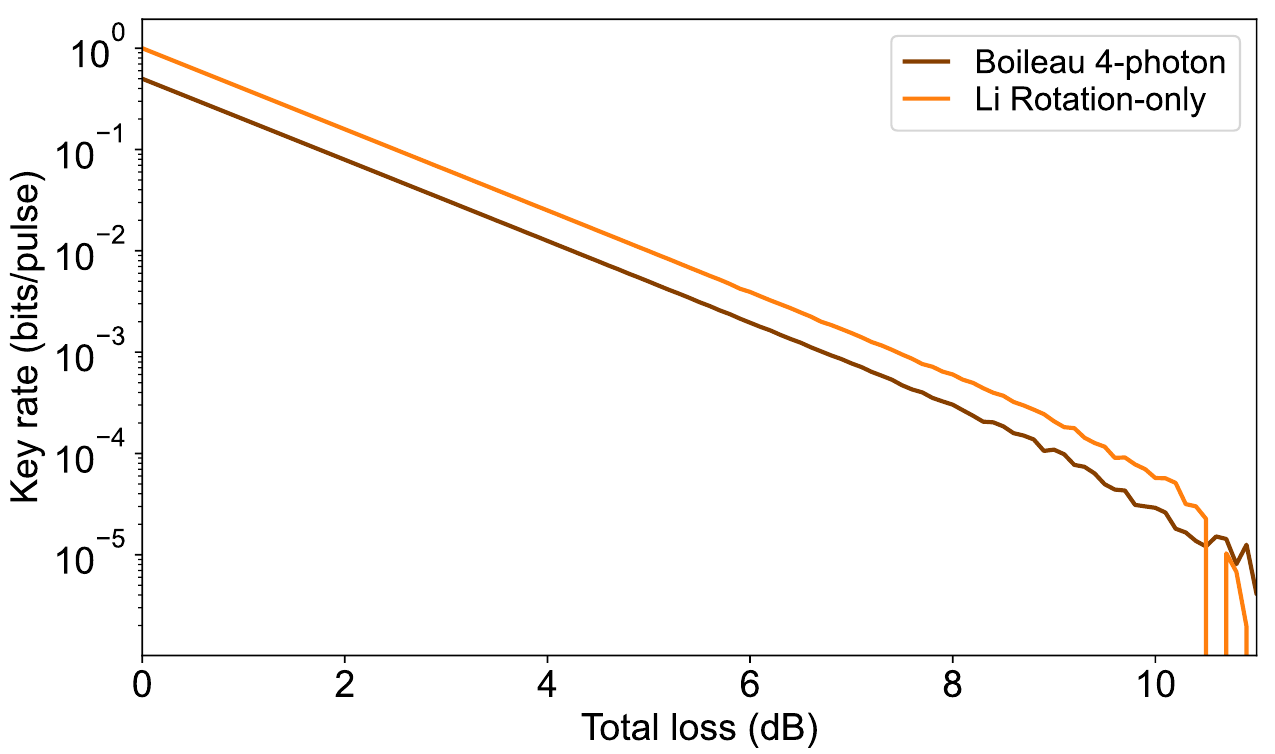}
    \caption{\textbf{Key rate for the 4-photon protocols in Li \textit{et al}. (2008) and Boileau \textit{et al}. (2004) as a function of loss.} The key rate as a function of loss for two 4-photon protocols that are robust to collective unitary noise. The protocol by Li \textit{et al}. (2008) achieves double the key rate of Boileau \textit{et al}. (2004) due to its efficient measurement scheme, at the cost of general unitary noise robustness.}
    \label{fig:li_losscomp}
\end{figure*}

In Fig.~\ref{fig:li_losscomp}, we present the key rate for the protocol in Ref.~\cite{efficient_collective_noise} compared to the 4-photon protocol of Ref.~\cite{Boileau}. Ref.~\cite{efficient_collective_noise} outlined that their efficient measurement technique (that does not discard outcomes) would double their key rate. We observe this in Fig.~\ref{fig:li_losscomp} for the rotation-only protocol,verifying the claim. It should be noted, that for the dephasing-only approach, we do not observe this improvement. This is due to extracting key from both bases in our analysis. For the dephasing-only protocol, the bit values at Bob can only be determined for $Z$ measurements and not for the $X$ measurements, whereas key can be extracted from either measurement basis for the rotation-only scheme.

\begin{figure*}[h]
    \centering
    \includegraphics[width=0.7\textwidth]{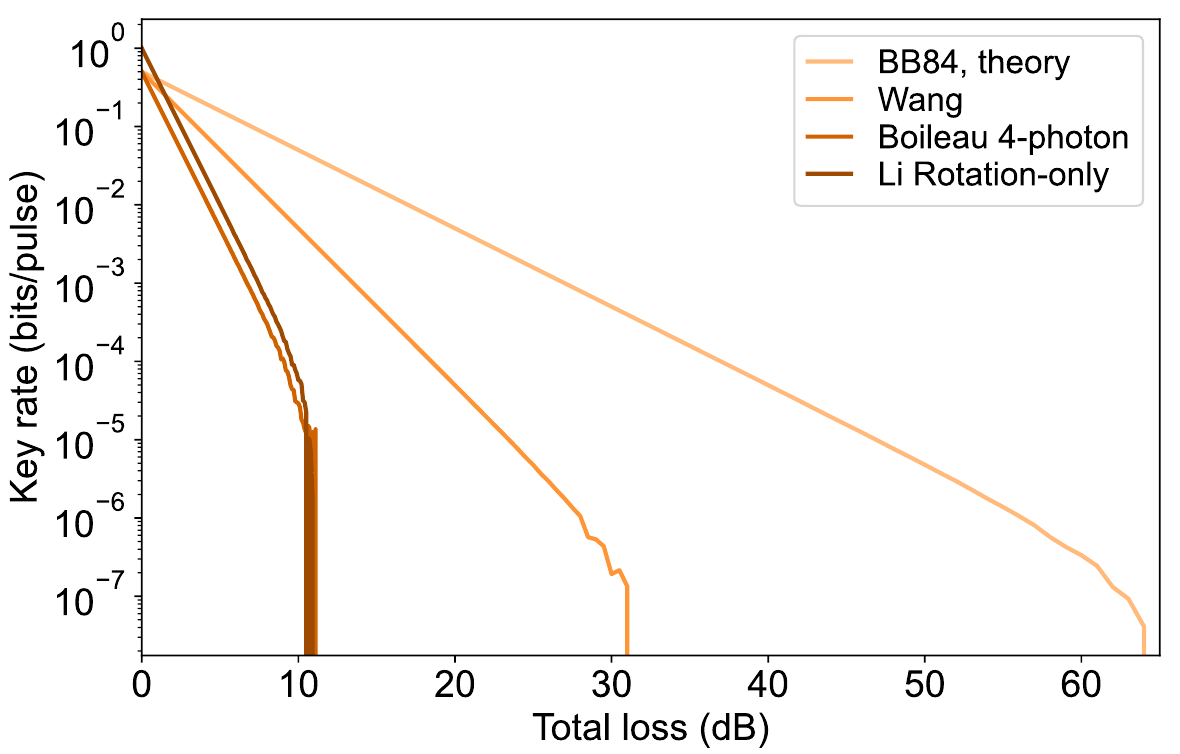}
    \caption{\textbf{Key rate for collective unitary noise-robust protocols as a function of loss.} At high loss values, when the key rate is small, instabilities appear in the numerical solver. Additionally, the smaller the dimension of the logical state, the smaller the key rate can be before the instabilities appear.}
    \label{fig:kr_instabilities}
\end{figure*}

For the 4-photon protocols in Fig.~\ref{fig:li_losscomp}, instability and a reduction in key rate beyond the expected $\eta^4$ scaling appears at higher losses. This is attributed to the numerical instabilities of the key rate simulation, which uses a coordinate descent solver, at small key rate values. We justify this claim in Fig.~\ref{fig:kr_instabilities} by simulating BB84 and Ref.~\cite{XBWang2005} at high losses. As for Refs.~\cite{Boileau} and \cite{efficient_collective_noise}, we also observe instabilities and a drop below the expected scaling once the key rate becomes small (on the order of $10^{-6}$). Additionally, the high dimension of the 4-photon states used, adds extra uncertainty into the simulation, giving rise to these instabilities at larger values of key rate (on the order of $10^{-4}$).

\section{Security analysis method} \label{sec:securityframework}
In this section, we outline the technique used to perform a rigorous security analysis of the protocols considered in the main text (except TF-QKD, for which we used the analytically derived key rates from existing literature \cite{Lutkenhaus2018_TFQKDsecurity}).
Specifically, the numerical analysis used Version 1 of the Open QKD Security package \cite{burniston_2024_14262569, Winick2018reliablenumerical}.
First, we rephrase the method from Ref~\cite{Winick2018reliablenumerical} for completeness. 
We then outline the corresponding mathematical elements of each protocol required to perform this analysis.

\subsection{Security framework} 
In a typical prepare-and-measure protocol, Alice prepares a state from a set ${\ket{\varphi_i}}$ with probability $p_i$, effectively sending a mixed state $\sum\limits_i p_i \ket{\varphi_i}\bra{\varphi_i}$ through a quantum channel to Bob, who then measures the incoming state.
It is often more convenient to describe this in the \textit{entanglement-based picture}, where Alice prepares instead a purification $\ket{\psi}_{AA'} = \sum\limits_i \sqrt{p_i}\ket{i}_{A}\ket{\varphi_i}_{A'}$. 
She then sends the system $A'$ to Bob and measures the system $A$ in the computational basis, thus preparing the state $\ket{\varphi_i}$ when the outcome $i$ is observed.
During the transmission over the channel, the system $A'$ is transformed into system $B$ received by Bob, and their shared state is generally mixed, described by $\rho_{AB}$.

The description of this state contains information about potential tampering on the channel, and the statistics observed by Alice and Bob limit the possible action an eavesdropper could have had. 
Since the system $A$ is kept securely by Alice and inaccessible to the eavesdropper, Alice's partial state has the additional constraint $\rho_{A}=\trace[B]{\rho_{AB}} = \trace[A']{\proj{\psi}_{AA'}}$. 
Typically, security analyses based on numerical methods consist in an optimisation over all possible bipartite quantum states $\rho_{AB}$ that are compatible with the observed statistics, and maximise the amount of information leaked to an eavesdropper. 

Here, we consider the security in the \textit{device-dependent} case, where Alice's and Bob's systems are completely characterised, and the operations they perform are POVMs denoted by $P_A^i, P_B^j$ for Alice and Bob respectively. 
After the measurement, Alice and Bob assign to each of their outcomes $i, j$, an announcement value $\tilde{a}, \tilde{b}$ and a secret value $\bar{a}, \bar{b}$, which are stored respectively in registers $\tilde{A}, \tilde{B}, \bar{A}, \bar{B}$. 
To simplify the notation, we directly label the measurement operators using the announcement and secret values instead of the outcome index, and we denote them $P_A^{\tilde{a}\bar{a}}, P_B^{\tilde{b}\bar{b}}$. 

The action of measuring the system and recording the announcement and secret value in new registers can be seen as a channel $\mathcal{A}(\rho_{AB})$ described by the Kraus operators $K_{\tilde{a}} \otimes K_{\tilde{b}}$, where:
\begin{subequations}
    \begin{align}
    K_{\tilde{a}} = \sum_{\bar{a}} \sqrt{P_A^{\tilde{a}\bar{a}}} \ket{\tilde{a}}_{\tilde{A}}\ket{\bar{a}}_{\bar{A}}\\
    K_{\tilde{b}} = \sum_{\bar{b}} \sqrt{P_B^{\tilde{b}\bar{b}}} \ket{\tilde{b}}_{\tilde{B}}\ket{\bar{b}}_{\bar{B}}. 
\end{align}
\end{subequations}

After each transmission and measurement, Alice and Bob publicly communicate their announcement values. 
Depending on the announced values, the round may be discarded, for instance if Alice's and Bob's basis choice do not match, or if the outcome measured by Bob is inconclusive. 
This can be seen as a post-selection map: 
\begin{equation}
    \Pi = \sum_{(\tilde{a}, \tilde{b}) \in \mathrm{A}} \proj{\tilde{a}}_{\tilde{A}}\otimes \proj{\tilde{b}}_{\tilde{B}} \otimes \mathbb{I}_{AB\bar{A}\bar{B}}, 
\end{equation}
where $\mathrm{A}$ is the ensemble of accepted announcement values. 

The secret key is defined by either Alice or Bob using the publicly available announcement values, and their secret value using a key map. 
The key map is given by either a function $g(\tilde{a}, \tilde{b},\bar{a})$ (direct reconciliation) or $g(\tilde{a}, \tilde{b},\bar{b})$ (reverse reconciliation). 
The value of the secret bit is written in a register $R$ using the isometry $V$, where for instance in the case of direct reconciliation: 
\begin{equation}
    V = \sum_{\tilde{a}\tilde{b}\bar{a}} \ket{g(\tilde{a}, \tilde{b}, \bar{a})}_R \otimes \proj{\tilde{a}}_{\tilde{A}} \otimes \proj{\tilde{b}}_{\tilde{B}} \otimes \proj{\bar{a}}_{\bar{A}}. 
\end{equation}

Then, following again the notation from Ref.~\cite{Winick2018reliablenumerical}, we define two quantum channels:
\begin{subequations}
    \begin{align}
    &\mathcal{G}(\rho_{AB}) = V \Pi \mathcal{A}(\rho_{AB}) \Pi V^{\dagger}, \\
    &\mathcal{Z}(\sigma) = \sum\limits_k \big(\proj{k}_R \otimes \mathbb{I}_{A\tilde{A}\bar{A}B\tilde{B}\bar{B}}\big)\sigma \big(\proj{k}_R \otimes \mathbb{I}_{A\tilde{A}\bar{A}B\tilde{B}\bar{B}}\big), 
    \end{align}
\end{subequations}
where $\mathcal{Z}(\sigma)$ is a pinching channel and $\ket{k}_R$ is the canonical basis for system $R$. 

Finally, a lower bound on the asymptotic secure key rate is given by
\begin{equation}
\label{eq:keyrate}
    \mathcal{D}\big(\mathcal{G}(\rho_{AB}) || (\mathcal{Z}\circ \mathcal{G}) (\rho_{AB})\big) - p_{\text{concl}} \text{leak}_{EC}, 
\end{equation}
where $p_{\text{concl}}=\trace{\mathcal{A}(\rho_{AB}) \Pi}$ is the conclusive rate of the protocol, $\text{leak}_{EC}$ is the amount of information required for error correction in conclusive rounds, and $\mathcal{D(\rho || \tau)}$ is the quantum divergence (also called the Umegaki relative entropy) defined as \cite{Tomamichel2016}
\begin{equation}
    \mathrm{D(\rho || \tau)} = \trace{\rho \log \rho} -\trace{\rho\log\tau}. 
\end{equation}

While the second half of Eq.~\eqref{eq:keyrate} can be computed directly from Alice's and Bob's observed statistics (in this paper, we consider the error correction at the Shannon limit), the first half involves the actual state $\rho_{AB}$ shared between them, which results from the interaction between Alice's prepared state and the communication channel. 
Since the structure of this state is not known directly, it rather has to be optimised over all possible quantum states compatible with the observed statistics so as to minimise the achievable key rate. 
The benefit of the formulation of Eq.~\eqref{eq:keyrate} is that it corresponds to a convex optimisation problem which can be solved efficiently using well known techniques \cite{Boyd_Vandenberghe_2004}. 
Since any solver with a finite precision will eventually return a result which is slightly higher than the true minimum, Ref.~\cite{Winick2018reliablenumerical} also introduced an extra step to convert such a suboptimal solution into a valid lower bound on the key rate using duality arguments. 

Using this framework, the security analysis of QKD protocols is reduced to a numerical computation of a lower bound on the achievable secure key rate. 
For each protocol we are considering in this paper, we simply need to specify the states Alice is preparing ($\ket{\varphi_i}$), the measurements Bob is performing ($P_B^{\tilde{b}\bar{b}}$), the post-selection map ($\Pi$), the key map ($g(\tilde{a}, \tilde{b},\bar{a})$ or $g(\tilde{a}, \tilde{b},\bar{b})$) and a reasonable channel model to simulate statistics. 
In practice, the post-selection map can be included implicitly in the definition of the Kraus operators, or sometimes also in the definition of the key map where values that are not mapped are implicitly discarded. 
The other quantities are summarised in the next section for each protocol. 

\subsection{Protocol description}


\subsubsection{States used in existing schemes} \label{sec:syndromestates}
Here, we define several states that are useful to describe concisely the existing protocols under study later on. 
First, single qubit states are defined in the well known $\cZ$ and $\cX$ bases as $\ket{0}, \ket{1}$ and $\ket{\pm}=\frac{1}{\sqrt{2}}\big(\ket{0}\pm\ket{1}\big)$ respectively. 

Next, two-qubit states are constructed similar to a repetition code \cite{PhysRevA.52.R2493} where the first system corresponds to the reference qubit, and the second system is repeated. 
The single qubit system can be seen as a physical qubit (i.e. the discrete encoding of a single photon), while any state of more qubits can be seen as the encoding of a single logical qubit. 
We indicate if the repetition is correct or flipped using a lower-script $s_0$, which takes the value of $0$ or $1$ respectively. 
Here, we also indicate the number of physical qubits in the logical state (i.e., the number of photons) using a super-script. 
Note that it is omitted whenever there is only one photon for simplicity. 
In the $\cZ$ basis, this defines the states:
\begin{equation}
\ket{x}_{s_0}^{(2)} = \ket{x}\ket{x + s_0}; x \in \set{0,1}, s_0 \in \set{0,1}, 
\end{equation}
where the addition between binary digits $x$ and $s_0$ is taken modulo 2. 

Similarly, the $\cX$ basis states are defined after performing a Hadamard on the $\cZ$ basis states:
\begin{align}
\ket{+}_{s_0}^{(2)} = (H \otimes H)\ket{0}_{s_0}^{(2)} \\
\ket{-}_{s_0}^{(2)} = (H \otimes H)\ket{1}_{s_0}^{(2)}
\end{align}

Then, we define logical states made of four physical qubits as a repetition of two-qubit states. 
Hence, the way the states are defined can be seen as a concatenation of repetition codes. 
We keep track of the \textit{syndrome} of the code, i.e. whether the repeated bit is correct or flipped, at each stage. 
The first bit of the syndrome corresponds to a flip in the inner two-qubit logical state encoding, while the second bit corresponds to a flip in the outer logical encoding. 

In the $\cZ$ basis, this defines the following states:
\begin{equation}
    \ket{x}_{s_0s_1}^{(4)} = \ket{x}_{s_0}^{(2)}\ket{x + s_1}_{s_0}^{(2)}; x \in \set{0,1}, s=s_0s_1 \in \set{00,01,10,11}
\end{equation}
and similarly for states in the $\cX$ basis. 
Here the basis change is at the level of the physical qubit and not at the level of the logical qubit, e.g. $\ket{+}_0^{(2)} = \ket{++} \neq \frac{1}{\sqrt{2}}\big(\ket{0}_0^{(2)}+\ket{1}_0^{(2)}\big)$. 
In the rest of this section, we denote the syndrome by $s=s_0s_1$, taking values in the ensemble $\set{00,01,10,11}$, or equivalently if written in decimal $\set{0,1,2,3}$. 

In some of the protocols under study, the information is encoded in the syndrome rather than in the value of the reference bit. 
Therefore, it is more convenient to consider the following states: 
\begin{align}
    \ket{s}_{\cZ}^{(4)} =& \frac{1}{\sqrt{2}}\big(\ket{0}_{s}^{(4)}+\ket{1}_{s}^{(4)}\big), \\
    \ket{s}_{\cX}^{(4)} =& \frac{1}{\sqrt{2}}\big(\ket{+}_{s}^{(4)}+\ket{-}_{s}^{(4)}\big). 
\end{align}

For each protocol under study in this paper, we summarise Alice's prepared states and associated announcement and secret values in Table~\ref{table:alice_states}. 
The 3-photon version of the protocol in Ref.~\cite{Boileau} is obtained by tracing out the fourth photon of the 4-photon scheme, which corresponds to Alice preparing mixed states. 
In the implementation, we consider instead a purification of these mixed states by duplicating each signal state with an extra qubit register, and assigning each purified state to the same announcement and secret value. 

\begin{table*}
    \centering
    \begin{tabular}{|c|c|c|c|c|c|c|c|c|}
    \hline
            \hspace{0.08cm}$\tilde{\boldsymbol{a}}$\hspace{0.08cm} & \hspace{0.08cm}$\bar{\boldsymbol{a}}$\hspace{0.08cm} & \hspace{0.08cm}\textbf{BB84}\hspace{0.08cm} & \hspace{0.08cm}\textbf{Our prot.}\hspace{0.08cm} & \hspace{0.08cm}\textbf{Wang}\hspace{0.08cm} & \hspace{0.08cm}\textbf{Boileau (4 photons)}\hspace{0.08cm} & \hspace{0.08cm}\textbf{Li (dephasing)}\hspace{0.08cm} & \hspace{0.08cm}\textbf{Li (rotation)}\hspace{0.08cm} \\ 
            \hline
            
            0&0
                & $\ket{0}$ 
                & $\ket{0_L}$ 
                & $\ket{10}$ 
                & $\frac{1}{\sqrt{2}}\big(\ket{2}_{\cZ}^{(4)}-\ket{3}_{\cZ}^{(4)}\big)$ 
                & \hspace{0.08cm}$\frac{1}{\sqrt{2}}\big(\ket{0}_{\cX}^{(4)}-\ket{1}_{\cX}^{(4)}\big)$\hspace{0.08cm} 
                & \hspace{0.08cm}$\frac{1}{\sqrt{2}}\big(\ket{0}_{\cZ}^{(4)}+\ket{1}_{\cZ}^{(4)}\big)$\hspace{0.08cm} \\
            \hline
            
            0&1
                & $\ket{1}$ 
                & $\ket{1_L}$ 
                & $\ket{01}$ 
                & $\frac{1}{\sqrt{2}}\big(\ket{1}_{\cZ}^{(4)}-\ket{3}_{\cZ}^{(4)}\big)$ 
                & $\frac{1}{\sqrt{2}}\big(\ket{2}_{\cX}^{(4)}-\ket{3}_{\cX}^{(4)}\big)$ 
                & $\frac{1}{\sqrt{2}}\big(\ket{2}_{\cZ}^{(4)}-\ket{3}_{\cZ}^{(4)}\big)$\\
            \hline

            1&0
                & $\ket{+}$ 
                & $\ket{+_L}$ 
                & $\ket{\psi^+}$ 
                & $\frac{1}{\sqrt{2}}\big(\ket{1}_{\cZ}^{(4)}-\ket{3}_{\cZ}^{(4)}\big)$ 
                & $\frac{1}{\sqrt{2}}\big(\ket{0}_{\cX}^{(4)}-\ket{2}_{\cX}^{(4)}\big)$ 
                & $\frac{1}{\sqrt{2}}\big(\ket{0}_{\cZ}^{(4)}+\ket{2}_{\cZ}^{(4)}\big)$\\
            \hline

            1&1
                & $\ket{-}$ 
                & $\ket{-_L}$ 
                & $\ket{\psi^-}$ 
                & $\frac{1}{\sqrt{2}}\big(\ket{2}_{\cZ}^{(4)}-\ket{1}_{\cZ}^{(4)}\big)$ 
                & $\frac{1}{\sqrt{2}}\big(\ket{1}_{\cX}^{(4)}-\ket{3}_{\cX}^{(4)}\big)$ 
                & $\frac{1}{\sqrt{2}}\big(\ket{1}_{\cZ}^{(4)}-\ket{3}_{\cZ}^{(4)}\big)$\\
            \hline
            
    \end{tabular}
    \caption{\textbf{Prepared states and Alice's information.} All states are prepared with uniform probability, and for each state, Alice records the associated announcement value $\tilde{a}$ and secret value $\bar{a}$. } 
    \label{table:alice_states}
\end{table*}

\subsubsection{States used in our scheme} \label{sec:ququartencoding}
As discussed in the preceding subsection, existing noise-resilient protocols typically define logical qubit states using multiple identical physical qubits. 
In contrast, our approach employs two physically distinct qubit types (frequency-bin and time-bin).
In the main text, we considered the security at the level of the logical qubit. 
However, to carry out a fully rigorous security analysis, it is necessary to account for the full two-photon structure of the encoded states. 
This motivates the formulation of the states in terms of a ququart basis, as we now present.

As described in Sec.~\ref{sec:protocol}, our encoding is built from four continuous DoF states: $\ket{0_f}, \ket{1_f}, \ket{0_t}, \ket{1_t}$, which are not mutually orthogonal. Nevertheless, these four states span a four-dimensional Hilbert space. Thus, they can be reformulated in terms of an orthonormal basis $\{ \ket{0}, \ket{1}, \ket{2}, \ket{3} \}$ (e.g., via Gram-Schmidt orthogonalisation).
To define the orthonormal ququart basis, we express each basis state as a linear combination of the original frequency- and time-bin states. 
To do this, we first define overlap parameters of the non-orthogonal states by 
\begin{equation}
    \alpha_{ij} = \braket{i_f|j_t} 
\end{equation}
(with $i,j \in \{0,1\}$) for mixed qubit types and 
\begin{equation}
    \beta_x = \braket{0_x|1_x}
\end{equation}
(where $x$ is $f$ or $t$) for qubits of the same type. 
These can typically be found analytically. For example, the parameters $\alpha_{ij}$ for the Gaussian states defined in SM, Sec.~\ref{sec:gaussstates} are given by Eq.~\eqref{eq:1doverlapft}.

With these parameters defined, we can find basis states given by
\begin{subequations} \label{eq:onbasis}
\begin{align}
    \ket{0} &= \frac{\ket{0_f} + \ket{1_f}}{\sqrt{2(1+\beta_f)}},\\
    \ket{1} &= \frac{\ket{0_f} - \ket{1_f}}{\sqrt{2(1-\beta_f)}},\\
    \ket{2} &= N_2 \left[\ket{0_t} - A \ket{0_f} - B \ket{1_f}\right],\\
    \begin{split}
    \ket{3} &= N_3 \left[ \ket{1_t} - |N_2|^2(\beta_t - A^* \alpha_{01} - B^*\alpha_{11})\ket{0_t} \right. \\
    & \left. \hspace{2cm} + \left[|N_2|^2(\beta_t - A^* \alpha_{01} - B^*\alpha_{11})A - C\right] \ket{0_f} + \left[|N_2|^2(\beta_t - A^* \alpha_{01} - B^*\alpha_{11})B - D\right]  \ket{1_f} \right],
    \end{split}
\end{align}
\end{subequations}
where $A, B, C, D$ are given by 
\begin{subequations} \label{eq:ququartfacs}
\begin{align}
    A &= \frac{\alpha_{00} - \alpha_{10}\beta_f}{1-\beta_f^2},\\
    B &= \frac{\alpha_{10} - \alpha_{00}\beta_f}{1-\beta_f^2},\\
    C &= \frac{\alpha_{01} - \alpha_{11}\beta_f}{1-\beta_f^2},\\
    D &= \frac{\alpha_{11} - \alpha_{01}\beta_f}{1-\beta_f^2},
\end{align}
\end{subequations}
and $N_2$ and $N_3$ are normalisation factors given by
\begin{equation}
    N_2 = \left( 1 - \frac{|\alpha_{00}|^2 + |\alpha_{10}|^2 - (\alpha_{00}\alpha_{10}^* + \alpha_{10}\alpha_{00}^*)\beta_f}{1-\beta_f^2} \right)^{-1/2}
\end{equation}
and
\begin{equation}
\begin{split}
    N_3 &= \left(1 - 2\beta_t |N_2|^2(\beta_t - \text{Re}(A^* \alpha_{01} + B^*\alpha_{11})*)  \right. \\
    & \hspace{1cm} + 2\text{Re}(\left[|N_2|^2(\beta_t - A^* \alpha_{01} - B^*\alpha_{11})A - C\right] \alpha_{01}^*) + 2\text{Re}(\left[|N_2|^2(\beta_t - A^* \alpha_{01} - B^*\alpha_{11})B - D\right] \alpha_{11}^*) \\
    & \hspace{2cm} - 2\text{Re}(|N_2|^2(\beta_t - A \alpha_{01}^* - B\alpha_{11}^*)\left[|N_2|^2(\beta_t - A^* \alpha_{01} - B^*\alpha_{11})A - C\right] \alpha_{00}^*)\\
    & \hspace{2cm} - 2\text{Re}(|N_2|^2(\beta_t - A \alpha_{01}^* - B\alpha_{11}^*)\left[|N_2|^2(\beta_t - A^* \alpha_{01} - B^*\alpha_{11})B - D\right] \alpha_{10}^*) \\
    &\hspace{2cm} +2\beta_f\text{Re}(\left[|N_2|^2(\beta_t - A^* \alpha_{01} - B^*\alpha_{11})A - C\right]\left[|N_2|^2(\beta_t - A^* \alpha_{01} - B^*\alpha_{11})B - D\right]) \\
    &\hspace{2cm} + |N_2|^4|\beta_t - A^* \alpha_{01} - B^*\alpha_{11}|^2\\
    &\hspace{2cm} + \left. |\left[|N_2|^2(\beta_t - A^* \alpha_{01} - B^*\alpha_{11})A - C\right]|^2 + |\left[|N_2|^2(\beta_t - A^* \alpha_{01} - B^*\alpha_{11})B - D\right]|^2 \right)^{-1/2},
\end{split}
\end{equation}
respectively.
Here, we have constructed $\ket{0}$ and $\ket{1}$ directly from the frequency-bin states, and $\ket{2}$ and $\ket{3}$ are orthogonalised combinations involving the time-bin states.

These expressions can be inverted to recover the states $\ket{i_x}$ in terms of the ququart basis states. This allows the logical qubit states defined in Equations~\eqref{eq:bellstate}-\eqref{eq:psiperp} in Sec.~\ref{sec:protocol} to be rewritten into a discrete representation that fully captures the two-photon nature of the encoding, despite the continuous properties of the physical qubit states.

\subsubsection{Measurement}
Typically, Bob's measurement is designed to discriminate which state Alice prepared, possibly with some inconclusive outcome. 
Here, it is essentially composed of projectors onto the states that Alice prepared, or states related to them. 

We define a few additional four-photon projectors: 
\begin{align}
    \Pi_{s, \cZ} =& \proj{0}_s^{(4)} + \proj{1}_s^{(4)} \\
    \Pi_{s, \cX} =& \proj{+}_s^{(4)} + \proj{-}_s^{(4)}
\end{align}
They are used to discriminate the states $\ket{s}_{\cZ}$ and $\ket{s}_{\cX}$ (defined in the SM, Sec.~\ref{sec:syndromestates}), respectively. 
Those are not projectors onto these states though, but rather a pinched version since the coherence terms are missing. 
They are easier to implement as they are composed only of $\cZ$ or $\cX$ basis measurements on the physical qubits. 

We assume that a photon loss can always be detected at the receiver, which will output a special ``no-click'' character, $\varnothing$ . 
This is a common approach (see for instance Refs.~\cite{Winick2018reliablenumerical, wang2021numericalsecurityproofdecoystate}, which are based on an argument from Ref.~\cite{PhysRevA.89.012325}), and it is modelled by considering that the channel is moving the signal state onto a dimension $\ket{\perp}$, orthogonal to the signal space. 
In practice, it means that all POVM matrices are defined with an extra dimension to account for the no-click outcome. 
For instance in a qubit protocol (where $\oplus$ represents here the direct sum of vectors) : $\proj{+} \oplus 0 = \frac{1}{2}\begin{pmatrix}1&1&0 \\ 1&1&0\\ 0&0&0 \end{pmatrix}$, and the projector $\proj{\perp} = \begin{pmatrix}0&0&0\\0&0&0\\0&0&1\end{pmatrix}$ reads the orthogonal dimension to determine if a no-click event has occurred. 
We use the same notation $\proj{\perp}$ for the loss outcome in all the protocols under study to keep the notation simple, even though the dimension of the actual operator is not always the same, and depends on the dimension of the signal space. 
In our protocol, the space $\ket{\perp}$ also includes all states that lie outside of the ququart subspace defined in the previous section, in addition to typical photon loss.

For each protocol under study, we summarise Bob's measurement operators and associated announcement value and secret value in Table~\ref{table:bob_povm}. 
The 3-photon version of the protocol in Ref.~\cite{Boileau} is obtained by removing the measurement on the fourth system of the 4-photon scheme, all the POVMs are otherwise identical. 

\begin{table*}
    \centering
    \begin{tabular}{|c|c|c|c|c|c|}
    \hline
                \hspace{0.08cm}$\tilde{\boldsymbol{b}}$\hspace{0.08cm} & \hspace{0.08cm}$\bar{\boldsymbol{b}}$\hspace{0.08cm} & \textbf{BB84} & \textbf{Our protocol} & \textbf{Wang} & \hspace{0.08cm}\textbf{Boileau, Li (dephasing, rotation)}\hspace{0.08cm} \\ 
            \hline
            
            0&0 
                & $\frac{1}{2}\proj{0}$ 
                & $\frac{1}{2}\proj{0_L}$ 
                & $\frac{1}{2}\proj{10}$ 
                & $\frac{1}{2}\Pi_{0, \cZ}$ \\
            \hline
            
            0&1
                & $\frac{1}{2}\proj{1}$ 
                & $\frac{1}{2}\proj{1_L}$ 
                & $\frac{1}{2}\proj{01}$ 
                & $\frac{1}{2}\Pi_{1, \cZ}$ \\
            \hline

            0&2
                & N/A
                & N/A
                & N/A
                & $\frac{1}{2}\Pi_{2, \cZ}$\\
            \hline

            0&3
                & N/A
                & N/A
                & N/A
                & $\frac{1}{2}\Pi_{3, \cZ}$\\
            \hline

            1&0
                & \hspace{0.08cm}$\frac{1}{2}\proj{+}$\hspace{0.08cm} 
                & $\frac{1}{2}\proj{+_L}$ 
                & $\frac{1}{2}\proj{\psi^+}$ 
                & $\frac{1}{2}\Pi_{0, \cX}$\\
            \hline

            1&1
                & $\frac{1}{2}\proj{-}$ 
                & $\frac{1}{2}\proj{-_L}$ 
                & $\frac{1}{2}\proj{\psi^-}$ 
                & $\frac{1}{2}\Pi_{1, \cX}$\\
            \hline

            1&2
                & N/A
                & N/A
                & N/A
                & $\frac{1}{2}\Pi_{2, \cX}$ \\
            \hline

            1&3
                & N/A
                & N/A
                & N/A
                & $\frac{1}{2}\Pi_{3, \cX}$\\
            \hline

            2&0 
                & N/A
                & \hspace{0.08cm}rest in ququarts\hspace{0.08cm}
                & \hspace{0.08cm}$\proj{00} + \proj{11}$\hspace{0.08cm}
                & rest\\
            \hline

            3&0 
                & $\proj{\perp}$
                & $\proj{\perp}$
                & $\proj{\perp}$
                & $\proj{\perp}$\\
            \hline
            
    \end{tabular}
    \caption{\textbf{Bob's measurement and information.} For each measurement, Bob records the announcement value $\tilde{b}$ and secret value $\bar{b}$ associated with the outcome. The announcement $\tilde{b}=2$ always corresponds to inconclusive rounds, while other values may correspond to conclusive or inconclusive rounds depending on the protocol. The key map defined in Table~\ref{table:keymap} clarifies which sets of values $(\tilde{a}, \tilde{b}, \bar{b})$ correspond to conclusive rounds. The notation ``rest'' indicates any other operator such that overall the sum of all operators is identity as to define a valid measurement. }
    \label{table:bob_povm}
\end{table*}

\subsubsection{Key map}
Usually, writing the key map is straightforward for most protocols. 
This is the case for instance for the BB84 protocol where the secret bit is usually defined to be Alice's secret value whenever the announcement values match. 
This corresponds to Bob trying to guess Alice's secret bit with his measurement whenever their basis choices match. 
However here, in the four-photon protocols under study, it is actually easier to define the key map from Bob's perspective. 
This is because the information encoded in Alice's states seems to be in the syndrome of the logical encoding, rather than only in either her secret bit or basis choice. 
As a result, it is only after Bob records his outcome and relates it to Alice's announcement value that the round can be classified as conclusive or inconclusive. 

In Table~\ref{table:keymap}, we summarise all the possible values of $(\tilde{a}, \tilde{b}, \bar{b})$ resulting in a value $r=0$ or $r=1$. 
Any other combination is implicitly assumed to be inconclusive. 
The classical reconciliation (error correction and privacy amplification) is then performed between the classical registers $R$ for Bob and $\bar{A}$ for Alice. 

\begin{table*}
    \centering
    \begin{tabular}{|c|c|c|}
    \hline
        $\boldsymbol{r}$ & 0 & 1 \\ 
    \hline
        \textbf{BB84} 
            & \hspace{0.12cm}$(0,0,0), (1,1,0)$\hspace{0.12cm} 
            & \hspace{0.12cm}$(0,0,1), (1,1,1)$\hspace{0.12cm} \\
    \hline
        \textbf{Our protocol} 
            & $(0,0,0), (1,1,0)$
            & $(0,0,1), (1,1,1)$ \\
    \hline
        \textbf{Wang} 
            & $(0,0,0), (1,1,0)$
            & $(0,0,1), (1,1,1)$ \\
    \hline
        \textbf{Boileau} 
            & \makecell{$(0,0,2), (0,1,2)$, \\ $(1,0,3),(1,1,3)$}
            & \makecell{$(0,0,1), (0,1,1)$, \\ $(1,0,2),(1,1,2)$} \\
    \hline
        \hspace{0.08cm}\textbf{Li (dephasing)}\hspace{0.08cm}
            & \makecell{$(0,1,0), (0,1,1)$ \\ $(1,1,0), (1,1,2)$}
            & \makecell{$(0,1,2), (0,1,3)$,\\$(1,1,1),(1,1,3)$} \\
    \hline
        \textbf{Li (rotation)} 
            & \makecell{$(0,0,0),(0,0,1)$,\\ $(0,1,0),(0,1,1)$,\\ $(1,0,0),(1,0,2)$, \\ $(1,1,0),(1,1,2)$}
            & \makecell{$(0,0,2),(0,0,3)$,\\$(0,1,2),(0,1,3)$,\\ $(1,0,1),(1,0,3)$, \\$(1,1,1),(1,1,3)$} \\
    \hline
    \end{tabular}
    \caption{\textbf{Key map $r=g(\tilde{a}, \tilde{b}, \bar{b})$}. This defines the secret bit from the announcement values and Bob's secret value. }
    \label{table:keymap}
\end{table*}

\subsection{Channel models}

\subsubsection{Modelling noise on the ququart basis states} \label{sec:ququartchannel}

The FBS and dispersion channels considered here act unitarily on the physical qubit states $\ket{i_x}$ (i.e. they preserve the norm and can in principle be reversed through appropriate inverse operations). However, these operations do not necessarily preserve the physical qubits within the ququart subspace. That is, while the continuous operation is unitary on the continuous DoFs, states that initially lie within the ququart encoding may be mapped partially outside of this subspace. 

The action of a unitary in the continuous DoFs on the ququart basis (defined in SM, Sec.~\ref{sec:ququartencoding}) can then be written as 
\begin{equation}
    \ket{i} \mapsto \ket{i'} =  \sum_{j=0}^{3} a_{ij} \ket{j} + a_{i\perp} \ket{\perp_i}
\end{equation}
where $a_{ij} = \braket{j|i'}$ are the overlap amplitudes within the ququart subspace and $\ket{\perp_i}$ denotes a normalised state orthogonal to the ququart subspace. This component is generally different for each basis state $\ket{i}$, with corresponding amplitude $a_{i\perp}$.

In SM, Sec.~\ref{sec:ququartencoding}, we expressed the ququart basis states in terms of the underlying frequency-bin and time-bin states. Thus, to evaluate the action of the channel explicitly (i.e. the coefficients $a_{ij}$), we can compute the overlaps between transformed and untransformed frequency- and time-bin states. We define these overlaps by the coefficients:
\begin{equation}\label{eq:transformed1doverlaps}
    \gamma_{ij} = \braket{i_f|j_t'}, \qquad \delta_{ij} = \braket{i_t|j_f'}, \qquad \kappa_{ij} = \braket{i_f|j_f'}, \qquad \lambda_{ij} = \braket{i_t|j_t'}.
\end{equation}

The coefficients $a_{ij}$ fully describe the channel's effect on the orthonormal basis within the ququart space. They can then be determined analytically in terms of the above overlap parameters using the definition of $a_{ij}$ given above and the expressions for the ququart states $\ket{i}$ in terms of the physical frequency- and time-bin qubit states, given in Eqs.~\eqref{eq:onbasis}. For example, the coefficients for $i=0$ are given by:
\begin{subequations}
\begin{align}
    a_{00} &= \frac{\kappa_{00} + \kappa_{01} + \kappa_{10} + \kappa_{11}}{2(1+\beta_f)},\\
    a_{01} &= \frac{\kappa_{00} + \kappa_{01} - \kappa_{10} - \kappa_{11}}{2\sqrt{1-\beta_f^2}},\\
    a_{02} &= \frac{N_2^*}{\sqrt{2(1+ \beta_f)}}\left[ \delta_{00} + \delta_{01} - A^*(\kappa_{00}+\kappa_{01}) - B^*(\kappa_{10}+\kappa_{11}) \right],\\
    \begin{split}
    a_{03} &= \frac{N_3^*}{\sqrt{2(1+ \beta_f)}}\left[ \delta_{10} + \delta_{11} - |N_2|^2(\beta_t -A\alpha_{01}^* -B\alpha_{11}^*)(\delta_{00} + \delta_{11}) \right.\\ 
    &\hspace{2cm} + \left[|N_2|^2(\beta_t -A\alpha_{01}^* -B\alpha_{11}^*)A^* - C^* \right](\kappa_{00} +\kappa_{01}) \\
    &\hspace{2cm} + \left. \left[|N_2|^2(\beta_t -A\alpha_{01}^* -B\alpha_{11}^*)B^* - D^* \right](\kappa_{10} +\kappa_{11}) \right].
    \end{split}
\end{align}
\end{subequations}

The relevant parameters in Eq.~\eqref{eq:transformed1doverlaps} can be derived analytically for specific encodings, such as the Gaussian states used here. 
For the FBS channel, these expressions are given in SM, Sec.~\ref{sec:CVnoise} (e.g., Eqs.~\eqref{eq:1doverlap} and~\eqref{eq:1dconjoverlap}). For the dispersion channels, the overlaps are given in SM, Sec.~\ref{sec:dispanalysis} by Eqs.~\eqref{eq:overlapff1}-\eqref{eq:overlapft1} for linear dispersion, and by Eqs.~\eqref{eq:overlap2ff}-\eqref{eq:overlap2tf} for quadratic dispersion.

The final step to perform the analysis is to express how this action modifies the transmitted density matrices. 
Since the ququart encoding spans a four-dimensional Hilbert space, but the channel may introduce leakage into orthogonal components, the full action is described by an isometry $V: \mathbb{C}^4 \mapsto \mathbb{C}^8$, mapping the original ququart space to an extended space that includes the leakage subspace spanned by the $\ket{\perp_i}$ states. 
In our protocol, however, we do not resolve individual $\ket{\perp_i}$ states but instead group all outcomes lying outside the ququart subspace (including photon loss) into a single inconclusive outcome, represented by a projector onto $\ket{\perp}$.

The effective transformation of a two-ququart input state $\rho$ (assuming collective noise) under a continuous DoF unitary operation is then given by
\begin{equation}
    \rho \rightarrow 
    \left( \begin{array}{@{}c|c@{}}
   \begin{matrix}
      \nu = (F\otimes F)\rho(F\otimes F)^{\dagger}
   \end{matrix} 
      & \begin{matrix}
      0 \\
      ... \\
      0
   \end{matrix} \\
   \cmidrule[0.4pt]{1-2}
   \begin{matrix}
      0 & ... & 0 
   \end{matrix} & \begin{matrix}
      1-tr(\nu)
   \end{matrix} \\
\end{array} \right)
\end{equation}
where $\nu$ is the density matrix within the ququart subspace and $F$ is a $4\times4$ matrix with elements $F_{ij} = a_{ij}$ ($i,j \in \{0,1,2,3\}$), which represents the action within the ququart subspace. The diagonal term $1 - tr(\nu)$ gives the total probability of leakage into the inconclusive subspace.

We have thus arrived at a full description of the channel's action on our encoding, which can be used directly in the numerical framework of Ref.~\cite{Winick2018reliablenumerical}.

\subsubsection{Modelling noise at the logical level} \label{sec:ampdampmodellogical}

As outlined in Section~\ref{sec:FBSeffect} and SM, Sec.~\ref{sec:dispanalysis}, the effect of the FBS and dispersion channels on our encoding are well approximated by amplitude damping and decoherence channels at the logical level. The fidelities of the channel-transformed states with the states measured by Bob determine the amplitude damping, while the accumulated phases correspond to the decoherence phase for each set of noise parameters $(\vec{\boldsymbol{\epsilon}}, \theta, \phi)$ for the FBS and $\alpha_n$ for dispersion. 

In Section~\ref{sec:noisycomp}, we state that the key rates obtained using the full two-photon ququart treatment of our protocol (as discussed above) closely match those predicted by this simplified consideration of the noise at the logical level only. We confirmed this by simulating the key rate for a BB84 protocol subject to the amplitude damping and decoherence channel, which acts on the logical basis states as:
\begin{equation}\label{eq:ampdamp}
    \ket{0_L} \rightarrow \ket{0_L} \qquad \ket{1_L}\rightarrow \sqrt{\eta_0}e^{i\Delta}\ket{1_L}
\end{equation}
where, $\eta_0$ and $\Delta$ are the amplitude damping and decoherence parameters, respectively. These parameters are extracted from the fidelities and accumulated phases of our encoding for each set of noise parameters.

Expressed as a quantum channel on density operators, the transformation is:
\begin{equation}\label{eq:ampdampmap}
    \rho \rightarrow E_0 \rho E_0^{\dagger} + E_1 \rho E_1^{\dagger}
\end{equation}
where $E_0$ and $E_1$ are the channel Kraus operators given by
\begin{equation}
    E_0 = \begin{pmatrix}
            0 & 0 & 0 \\
            0 & 0 & 0 \\
            0 & \sqrt{1-\eta_0} & 0
        \end{pmatrix}
        \qquad
    E_1 = \begin{pmatrix}
            1 & 0 & 0 \\
            0 & \sqrt{\eta_0} e^{i\Delta} & 0 \\
            0 & 0 & 1
        \end{pmatrix}.
\end{equation}
The Kraus operators are $3\times3$ matrices, with the dimensions corresponding to the $\ket{0_L}, \ket{1_L}$ states, and the inconclusive state $\ket{\perp}$, respectively. Here, the state $\rho_{\perp}$ corresponds to both loss and two-photon states lying outside the logical subspace.


\subsubsection{Modelling noise on the existing schemes}

Modelling the noise on the exisiting schemes simply requires describing the action on an $N$-qubit state. To analyse the collective unitary noise-robust protocols \cite{XBWang2005, Boileau, efficient_collective_noise} (given in Sec.~\ref{sec:litrev} of the SM) when encoded in frequency-bins, we use the noise models outlined in Appendix~\ref{sec:tradeoff}. Combining both the collective unitary noise channel and the loss channel from Appendix~\ref{sec:tradeoff}, the general model used is given by
\begin{equation}
    \rho_N \rightarrow p^N U^{\otimes N} \rho_N U^{\dagger \otimes N} + (1-p^N)\rho_{\perp},
\end{equation}
where $\rho_{\perp}$ is the state that gives an inconclusive outcome due to loss, $p$ is the single photon loss probability, and $N$ is the number of photons in the logical state $\rho_N$. Here $U$ is some unknown unitary, parameterised by the FBS noise parameters $\theta$ and $\phi$ as
\begin{equation} \label{eq:U}
    U = \begin{pmatrix}
        \cos \theta & -e^{-i\phi} \sin \theta \\
        e^{i\phi} \sin \theta & \cos \theta
    \end{pmatrix}.
\end{equation}
Here we have defined $U$ only over the space that does not result in a conclusive outcome (i.e., does not include the dimension for $\ket{\perp}$). An extra row and column is added to the operation $U^{\otimes N}$ to account for this additional dimension.

\section{Key rate dependence of our protocol on \texorpdfstring{$\phi$}{phi}}\label{sec:KRphi}

In Fig.~\ref{fig:Ourprot_U}, we present the key rate of our protocol under the FBS channel for the full range of rotation angles $\theta$ and dephasing angles $\phi$. We observe that there is negligible dependence on $\phi$ for the parameters shown. Hence, the key rate can be most easily analysed for the parameters chosen by considering a fixed $\phi$ and varying $\theta$ only, as in the main text.

\begin{figure*}[h]
    \centering
    \includegraphics[width=\textwidth]{FBS_2D_scan_subfigs_cropped.pdf}
    \caption{\textbf{Key rate of our protocol under FBS channel with rotation angle $\theta$ and dephasing angle $\phi$.} Key rate of our encoding for unitary parameters $\theta$ and $\phi$ for varying FBS noise bin width, $\epsilon$, for (a) $\epsilon= 0.01 \times \frac{1}{\sigma_t} = 0.6$ GHz, (b) $\epsilon= 0.05 \times \frac{1}{\sigma_t} = 3.0$ GHz, and (c) $\epsilon= 0.11 \times \frac{1}{\sigma_t} = 6.6$ GHz . The other parameters match those of Fig.~\ref{fig:staterobustness} in the main text, i.e. the encoding parameters are $\tau_0=0$ ps, $\tau_1=220$ ps, and $\sigma_t = 17$ ps, with a central frequency of 195.860 THz, and FBS noise parameters $\Omega = 195.860$ THz, $\Omega +\mu = 195.879$ THz.}
    \label{fig:Ourprot_U}
\end{figure*}

As also shown in Fig.~\ref{fig:noisycomp} of the main text, the smaller the value of $\epsilon$, the larger the lower bound on key rate across the range $\theta$ (found for $\theta = \pi$).

We note that we observed solver instabilities at three individual, random points in both Figs.~\ref{fig:Ourprot_U}(a) and (c), which resulted in zero key rates. To enable a smooth visualisation of the key rate as a function of both $\theta$ and $\phi$, we simulated the key rate at four neighbouring points offset by $\pi\times10^{-4}$ in $\theta$ and $\phi$ on either side of the unstable $(\theta, \phi)$ pairs. The average of these surrounding values was then used to estimate the key rate at the affected points shown in the plots.

\section{Multi-photon security considerations} \label{sec:multiphotonsec}

In Sec.~\ref{sec:security} of the main text, we discussed the security of our protocol by considering Alice transmitting only single logical qubits and the effect of the channel at the logical level. However, our logical encoding specifically employs two photons per logical qubit, each carrying some signature of Alice's encoding choice in the continuous domain. Additionally, potential practical implementations of our scheme, such as the proposal in Appendix~\ref{app:gen&meas}, may lead to the preparation of multiple logical qubits per generation event at Alice, in a similar manner to common deployments of BB84 using weak coherent pulses \cite{Lo2005}.
Consequently, alternative attack strategies, such as the photon-number splitting attack (PNS) \cite{XBWang2005_PNS}, should also be taken into consideration.

In SM, Sec.~\ref{sec:securityframework}, we present a comprehensive analysis of our protocol taking into full account the two-photon nature of our encoding. Here, we will provide a intuitive explanation for the security of our protocol when the two-photon nature of our encoding is considered.
Additionally, while a comprehensive security analysis of our protocol under multiple logical qubit generation events lies beyond the scope of this work and warrants a dedicated study, we will outline how the standard decoy-state methods used in BB84 can be directly applied to our protocol, allowing security to maintained under such PNS attacks.

In the following, we divide the considerations into two sections. The first considers the effect of splitting the multi-photon state. The second considers splitting off pairs of photons produced in multi-pair emission events (analogous to PNS attacks for single-photon BB84 schemes).

\subsection{Splitting the multi-photon states}

First, we consider the case that Alice prepares a single copy of the ideal two-photon logical state she wishes to send and Eve splits off one of the two photons. Eve then has two options: she can either inject an extra photon back into the channel, or not. The security is manifest in a measurement implementation-specific manner, and we highlight two representative approaches that illustrate how the protocol resists this eavesdropping strategy.

In the first case, consider the measurement scheme described in Appendix~\ref{app:gen&meas}, where Bob interferes the incoming state with a locally prepared logical state. Destructive interference (yielding an inconclusive result) occurs if the states match; constructive interference (yielding a conclusive detection of the opposite basis state) occurs if they differ. If Eve removes a photon, the incoming state is altered and fails to destructively interfere, even when it should -- leading to conclusive outcomes in cases where none are expected. Similarly, since the two-photon state Bob generates is entangled, if Eve injects an extra photon back into the channel, the arriving state will not be entangled, and thus will also fail to destructively interfere. This results in a high error rate, making Eve’s presence readily detectable.

Alternatively, assume Bob’s detectors have basic photon-number-resolving (PNR) capabilities, allowing them to distinguish between single-photon and multi-photon arrivals. Since the logical states are defined over \textit{both} photons sent by Alice, if Eve removes one of the photons Bob will receive only a single photon, which he labels as an inconclusive outcome. This ensures Eve’s interference cannot contribute to the key. If instead Eve injects an extra photon back into the channel, the security of our scheme can be understood with two key observations.
First, while the overall logical states are orthogonal, the individual single-photon states that comprise them are not (e.g., $\braket{0_t|0_f} \neq 0$). This prevents Eve from perfectly distinguishing the state even with full basis information. Second, as described in SM, Section \ref{sec:securityframework}, prepare-and-measure protocols can be reformulated into an entanglement-based interpretation. In this picture, any attempt by Eve to intercept and re-inject a photon reduces the entanglement between Alice and Bob, thereby introducing detectable errors.

\subsection{Splitting multi-pair generation events}

We now study the case of multiple logical qubit generations. In this scenario, Eve can intercept pulses containing more than two photon pairs, extract one logical qubit (photon pair), and forward the rest to Bob; pulses with only one logical qubit are discarded. This attack can be counteracted using a standard decoy-state BB84-type scenario \cite{Lo2005}, which is directly translatable to our protocol. In this method, the intensities of the pulses sent by Alice are varied such that fewer or greater than one logical qubit state are prepared per pumping event on average (labelled as signal and decoy states, respectively). By measuring the yield of signal versus decoy states at Bob, Alice and Bob can detect the eavesdropper's presence, since an attack will increase the multi-pair yield, and thus decoy state yield, beyond an expected bound set by the yield of the signal state \cite{Hwang2003}.

Such a method is applicable to the experimental proposal in Appendix~\ref{app:gen&meas}, for example, where increasing the pump power increases the number of parametrically down-converted photon pairs. Our protocol then requires just one adjustment from standard decoy-state BB84. In conventional decoy-state BB84, Bob's detectors must distinguish between one and more than one photon arriving at him, since logical states contain a single photon. In our scheme, where logical states consist of two photons, Bob must instead distinguish between two and more than two photons arriving at him. This capability is readily available in commercial photon-number-resolving detectors, so the modification introduces no experimental limitations. With this minor adaptation, standard decoy-state techniques apply directly to our scheme.

\section{Comparison of our encoding to alternative continuous DoF encodings}

\subsection{Bell-states across different frequency-bins}\label{sec:diffbinencoding}
In Appendix~\ref{app:compencoding}, we presented an alternative encoding in the frequency domain, using Bell-states on two pairs of frequency bins, and justified why our proposed encoding allows for higher robustness under reduced noise assumptions. We discussed qualitatively the effect of the FBS operation when it acts on frequency bins that are not the frequency bins chosen for the encoding (in the worst case, when the noise acts on only one of the encoding bins with another bin outside the space). Here, we provide a mathematical analysis of this discussion. 

First, we demonstrate analytically the effect of this change of the FBS operation on our encoding. As described in the Appendix~\ref{app:compencoding}, we show there is only an additional incurred loss with negligible incurred bit error rate. Subsequently, we will apply the same analysis to the alternative encoding to show that a bit error rate will be incurred.

To show this, we consider that an FBS noise operation acts across the frequency-bin $\ket{0_f}$ and a third bin, which we label as a qudit state $\ket{2_f}$, and another FBS acts across bin $\ket{1_f}$ and a fourth bin, $\ket{3_f}$. The FBS acts unitaries across these frequency bins, which we label $U_{02}(\theta_1, \phi_1)$ and $U_{13}(\theta_2, \phi_2)$, respectively. The unitary operation $U_{ab,M}(\theta, \phi)$ on states $\ket{a_f}_M$ and $\ket{b_f}_M$ (in mode $M$) takes the form
\begin{align}
    &U_{ab,M}\ket{a_f}_M = \cos \theta \ket{a_f}_M + e^{i\phi}\sin\theta \ket{b_f}_M,\\
    &U_{ab,M}\ket{b_f}_M = -e^{-i\phi} \sin\theta \ket{a_f}_M + \cos \theta \ket{b_f}_M.
\end{align}

Acting these FBS unitary operations on the frequency-bin Bell-state $\ket{0_L}=\ket{\Psi^-_f}_{01,MN}$ results in the state
\begin{equation} \label{eq:U02U13}
\begin{split}
    U_{02,M} \otimes U_{13,N}\ket{\Psi^-_f}_{01,MN} &= \frac{1}{\sqrt{2}}[(\cos \theta_1 \ket{0_f}_M + e^{i\phi_1}\sin \theta_1 \ket{2_f}_M)\otimes(\cos \theta_2 \ket{1_f}_N + e^{i\phi_2}\sin \theta_2 \ket{3_f}_N) \\
    &\hspace{2cm} - (\cos \theta_2 \ket{1_f}_M + e^{i\phi_2}\sin \theta_2 \ket{3_f}_M)\otimes(\cos \theta_1 \ket{0_f}_N + e^{i\phi_1}\sin \theta_1 \ket{2_f}_N)] \\
    &= \cos \theta_1 \cos \theta_2 \ket{\Psi^-_f}_{01,MN} + e^{i \phi_1}\sin \theta_1 \cos \theta_2 \ket{\Psi^-_f}_{21,MN} \\
    &\hspace{2cm} + e^{i\phi_2}\sin \theta_2 \cos \theta_1 \ket{\Psi^-_f}_{03,MN} + e^{i(\phi_1 + \phi_2)} \sin \theta_1 \sin \theta_2 \ket{\Psi^-_f}_{23,MN}
    \end{split}
\end{equation}
at Bob, where  $\ket{\Psi^-_f}_{ab,MN} = \frac{\ket{a_f}_M \ket{b_f}_N - \ket{b_f}_M \ket{a_f}_N}{\sqrt{2}}$, as described in Appendix~\ref{app:compencoding}. 

We now consider the effect of this on our encoding. Under the previous noise model (Section~\ref{sec:FBS} in the main text), we saw that the frequency-bin state was unaffected and the time-bin state had a reduced fidelity. Under the new model, the time-bin state will show a reduction in fidelity as before, but reduced further by the double FBS operation (see Fig~\ref{fig:multiU}), and the overlap of the time-bin state with the frequency-bin state will remain negligible. However, now the frequency-bin Bell-state will also ``leak'' outside of the logical subspace. The overlap of the transformed frequency-bin Bell-state with its untransformed version $\ket{\Psi^-_f}_{01,MN}$ is reduced from 1 to $\cos \theta_1 \cos \theta_2$ [Eq~\eqref{eq:U02U13}].
To summarise the effect of this revised model on our encoding, in the $Z_L$ basis, there will be no bit error rate, but now an incurred loss for the $\ket{0_L}$ state of $l_0 = (1-\cos^2\theta_1 \cos^2\theta_2)$ will arise. The incurred loss for $\ket{1_L}$ is found using the methods in Sections~\ref{sec:CVnoise} and~\ref{sec:multiU}.

We now compare this to the effect of the revised double FBS operation on the alternative encoding (outlined in Appendix~\ref{app:compencoding}). We will demonstrate the claim that this encoding can give rise to both an incurred loss and a bit error rate, in contrast to our encoding.

First, let us assume that each FBS acts between each one of the encoding frequency-bins with another frequency-bin that lies outside of the chosen encoding space (i.e. the FBS applies unitaries $U_{04}, U_{15}, U_{26}, U_{37}$). In this case, there is no bit error rate, but each of the two logical qubit basis states has an incurred loss of $l = (1-\cos^2\theta_1 \cos^2\theta_2)$, as seen for our encoding, where $\theta_1$ and $\theta_2$ are the unitary rotation angles that will be different for each frequency-bin pair.

In contrast, if we assume now that the double FBS acts on the worst pairing of bins, such that it applies unitaries $U_{02}(\theta_1, \phi_1)$ and $U_{13}(\theta_2, \phi_2)$ on the frequency-bin pairs $\{\ket{0_f},\ket{2_f}\}$ and $\{\ket{1_f},\ket{3_f}\}$, as before, then a bit error rate will also be induced. The effect of $U_{02} \otimes U_{13}\ket{\Psi^-_f}_{01}$ is given by Eq.~\eqref{eq:U02U13}. The effect of $U_{02} \otimes U_{13}\ket{\Psi^-_f}_{23}$ is given by 
\begin{equation}
\begin{split}
    U_{02,M} \otimes U_{13,N}\ket{\Psi^-_f}_{23,MN} &= \cos \theta_1 \cos \theta_2 \ket{\Psi^-_f}_{23,MN} - e^{i \phi_2}\sin \theta_2 \cos \theta_1 \ket{\Psi^-_f}_{21,MN} \\
    &\hspace{2cm} - e^{i\phi_1}\sin \theta_1 \cos \theta_2 \ket{\Psi^-_f}_{03,MN} + e^{-i(\phi_1 + \phi_2)} \sin \theta_1 \sin \theta_2 \ket{\Psi^-_f}_{01,MN}
\end{split}
\end{equation}
From this, we see that there is an incurred loss of $l_0 = l_1 = 1- \cos^2\theta_1\cos^2\theta_2 - \sin^2\theta_1 \sin^2\theta_2$ for both logical basis states. However, this additionally demonstrates the accumulation of a non-negligible bit error rate of $e_{bit} = \sin^2\theta_1 \sin^2\theta_2$ (not normalised for the loss rate), and phase error rate of $e_{ph} = \sin^2(\phi_1 + \phi_2)\sin^2\theta_1\sin^2\theta_2$.

Thus, we have shown that the we can achieve better robustness under fewer assumptions on the noise, using our encoding, compared to an alternative encoding. In contrast to our encoding, the alternative encoding will acquire a bit error rate, that can lead to the abortion of the protocol under sufficiently large noise parameters $(\theta_1, \theta_2)$.

\subsection{Bell-states of different symmetries}
We present another alternative encoding. However, we explain that this encoding does not give rise to a secure QKD protocol, and thus cannot be employed.

Using the ideas in Ref.~\cite{XBWang2005}, with the conjugate bases encoding of our scheme, one could consider an alternative continuous DoF encoding in which the key states are given by $\ket{\Psi^-_f}$ and $\ket{\Psi^+_f}$, and the test states are given by $\ket{\Psi^-_t}$ and $\ket{\Psi^+_t}$ (where $\ket{\Psi^+_x} = (\ket{0_x}_M\ket{1_x}_N - \ket{1_x}_M\ket{0_x}_N)/\sqrt{2}$ and $x$ is $f$ or $t$). Bob could then perform a simpler measurements by performing a Bell-state measurement in either the frequency or the time basis. However, the states of opposite symmetry are always orthogonal (i.e. the symmetric frequency-bin Bell-states are orthogonal to the antisymmetric time-bin Bell-states, and vice versa, as is evident from the equations in Section~\ref{sec:overlap}). This renders the encoding completely insecure, since it forbids Alice and Bob from gaining phase information about the channel, to verify the security of the channel. 

From a geometrical picture, these states comprise a 4-dimensional, rather than 2-dimensional space. Eve can make measurements in the higher dimensional space that leave the states in the lower dimensional subspace on which Bob measures undisturbed.

\section{Collective noise with differences in noise parameters}
Typical experimental implementations of protocols using logical states with multiple photons will expect small deviations of the noise parameters between the photons in the logical state. 
These deviations are anticipated when the photons are temporally separated or spatially routed into two different fibre cores, for example, so that they can be distinguished.
In this section, we analyse the impact of these small experimental deviations in the FBS noise parameters and the dispersion noise between the two photons of our encoding, with Gaussian amplitudes, on the fidelity and key rate. 

\subsection{FBS noise deviations}
Representing the FBS parameters as $\vec{\boldsymbol{\epsilon}} = (\Omega, \mu, \epsilon, \theta, \phi)$ and difference in these parameters between the two photons as $\vec{\boldsymbol{\delta}} = (\delta_{\Omega}, \delta_{\mu}, \delta_{\epsilon}, \delta_{\theta}, \delta_{\phi})$, we model the FBS channel now as applying
\begin{equation}
    \rho \rightarrow (\hat{F}_{\vec{\boldsymbol{\epsilon}}}\otimes\hat{F}_{\vec{\boldsymbol{\epsilon}}+\vec{\boldsymbol{\delta}}})\rho (\hat{F}_{\vec{\boldsymbol{\epsilon}}}^{\dagger}\otimes\hat{F}_{\vec{\boldsymbol{\epsilon}}+\vec{\boldsymbol{\delta}}}^{\dagger}), 
\end{equation}
where $\hat{F}_{\vec{\boldsymbol{\epsilon}}}$ is the FBS channel with parameters $\vec{\boldsymbol{\epsilon}}$.

\begin{figure*}[h]
    \centering
    \includegraphics[width=\textwidth]{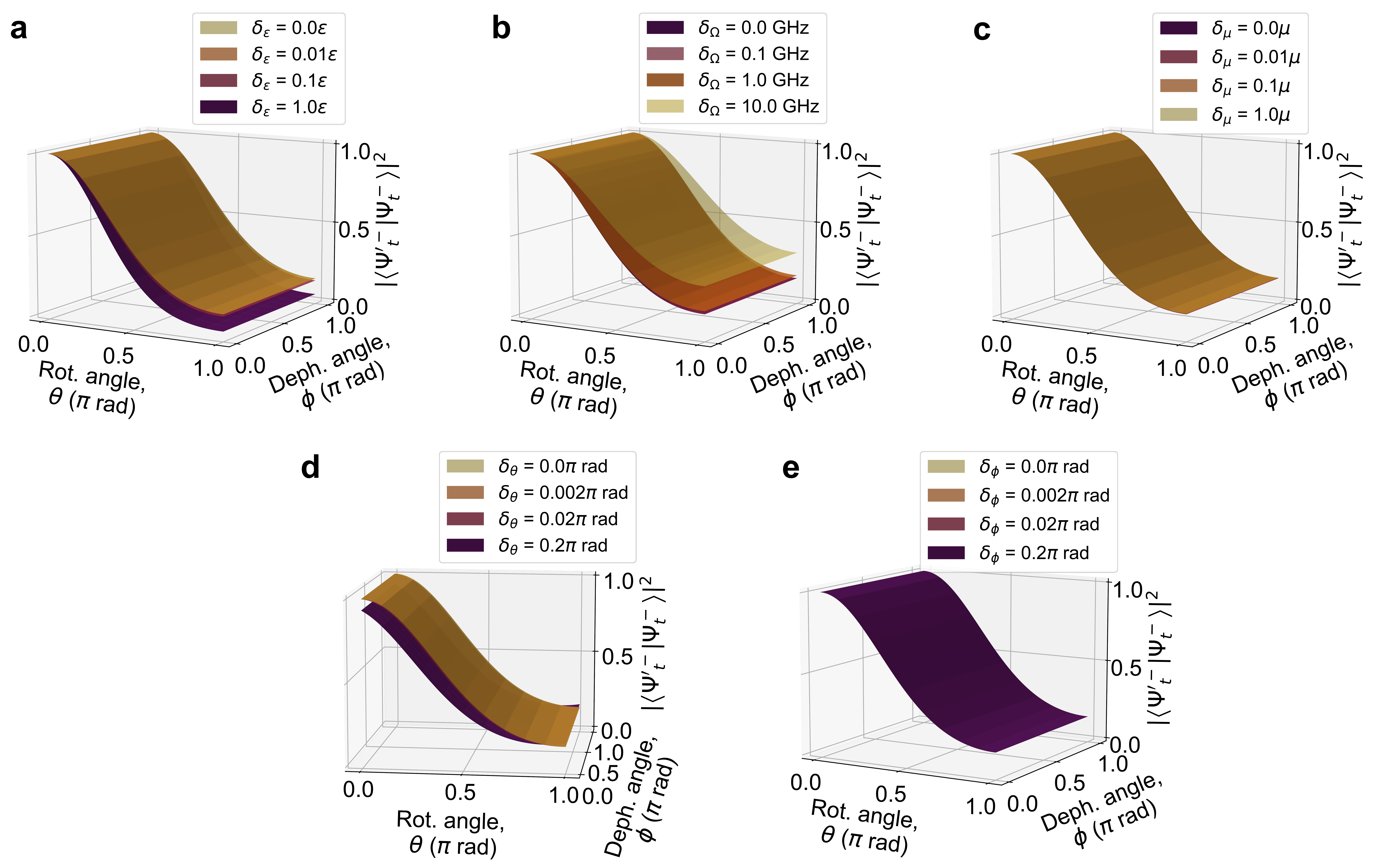}
    \caption{\textbf{Robustness of the time-bin Bell-states under FBS noise with deviations between the noise parameters on the two photons.} Fidelity between the arriving state $\ket{\Psi'^-_t}$ sent by Alice through the FBS channel with deviating noise parameters on the two photons and the state $\ket{\Psi^-_t}$ measured by Bob. We use values close to the experimentally reported values in Refs.~\cite{finco2024timebinentangledbellstate, Clementi2023}, with $\tau_0=0$ ps, $\tau_1=220$ ps, and $\sigma_t = 17$ ps and a central frequency of 195.860 THz, with FBS noise parameters on the first photon given by $\Omega = 195.860$ THz, $\mu = 0.019$ THz, $\epsilon = 3.0$ GHz. The fidelity is plotted as a function of $\theta$ and $\phi$ for different values of $\delta_x$ (with all other $\delta$'s being zero) where $x$ is (a) the noise bin width $\epsilon$, (b) the first noise bin central frequency $\Omega$, (c) the noise bin separation $\mu$, (d) the unitary rotation angle $\theta$, and (e) the unitary dephasing angle $\phi$. $\delta_{\epsilon}$ and $\delta_{\mu}$ are given as fractions of the value of the noise parameters on the first photon, whilst $\delta_{\Omega}$, $\delta_{\theta}$ and $\delta_{\phi}$ are given as absolute values.}
    \label{fig:unequalFBS}
\end{figure*}

In Fig.~\ref{fig:unequalFBS}, we present the fidelity of the transformed time-bin states compared to the time-bin states sent by Alice and measured by Bob, for varying deviations in each of the noise parameters (including $\vec{\boldsymbol{\delta}}=0$ plots for comparison to perfect collective noise). 
The results in Fig.~\ref{fig:unequalFBS} show that the deviations only have a marginal impact on the fidelity, even under relatively large deviations on the noise parameters between the two photons. 
Experimentally, the photons are expected to be transmitted in close temporal and spatial proximity, meaning the $\vec{\boldsymbol{\delta}}$ values are expected to be small.
Additionally, under these deviations, the bit error rate remains negligible.
Similarly, the global phase accumulated by the time-bin state remains small, such that it only contributes to a small reduction in key rate.
It is only for large values of $\delta_{\phi}$ that large global phases, resulting in significant reductions in key rate, are seen.
However, as described above, this is unlikely to be expected in typical experimental scenarios.

In addition to the effect of these parameter deviations on the time-bin states, we must consider the effect on the frequency-bin states.
We assume $\epsilon$ is always sufficiently large to contain the encoded bins, and that the deviations of $\Omega$ and $\mu$ are considered still contain either all or none of the encoded bins, such that a unitary is applied to the frequency bins.
From Section~\ref{sec:diffbinencoding} we know that deviations in $\Omega$ or $\mu$ to states bins outside of the encoded space simply result in a higher rate of inconclusive outcomes.
Considering then the effect of deviations in the $\theta$ parameter, with $\phi=0$, on the frequency-bin state, we assume the action of the FBS on the frequency-bin Bell-states is a unitary with rotation angle $\theta$ on one frequency-bin qubits ($U_f(\theta, \phi=0)$), and a unitary with rotation angle $\theta +\delta_{\theta}$ on the other ($U_f(\theta+\delta_{\theta}, \phi=0)$).
Using the small angle approximation, up to $\mathcal{O}(\delta^2)$, this transforms the two-photon frequency-bin state to
\begin{equation}
    U_f(\theta, \phi=0)\otimes U_f(\theta+\delta_{\theta}, \phi=0) \ket{\Psi^-_f} = (1-\delta_{\theta}^2/2)\ket{\Psi^-_f} -\delta_{\theta}\ket{\Phi^+_f}.
\end{equation}
From this, we see there is a small increase in inconclusive outcomes due to rotating the state into $\ket{\Phi^+_f}$, with a rate given by $\delta_{\theta}^2$. 
$\ket{\Phi^+_f}$ will always be orthogonal to an antisymmetric state, due to the orthogonality of opposite symmetry states. 
In this case, the time-bin state arriving is not perfectly antisymmetric due to the difference in noise parameters between the two qubits. 
However, since the deviations are typically small, and the time-bins will be broad in frequency, whilst the $\ket{\Phi^+_f}$ is made from narrow frequency bins, the overlap between the two states will remain small. 
Hence, there is negligible increase in the bit error rate of our encoding.

Bringing these effects together, we see that experimental deviations in the noise parameters between the two photons of the logical state has only a small affect on the achievable key rates of our protocol, due to a marginal increase in inconclusive rates.

\subsection{Dispersion noise deviations}
For the dispersion noise, we assume that one photon sees the dispersion parameter $\alpha_n$ and the other photon sees $\alpha_n +\delta_{\alpha}$ for a fixed $n$, where $\delta_{\alpha}$ is the experimental deviation.

\begin{figure*}[h]
    \centering
    \includegraphics[width=\textwidth]{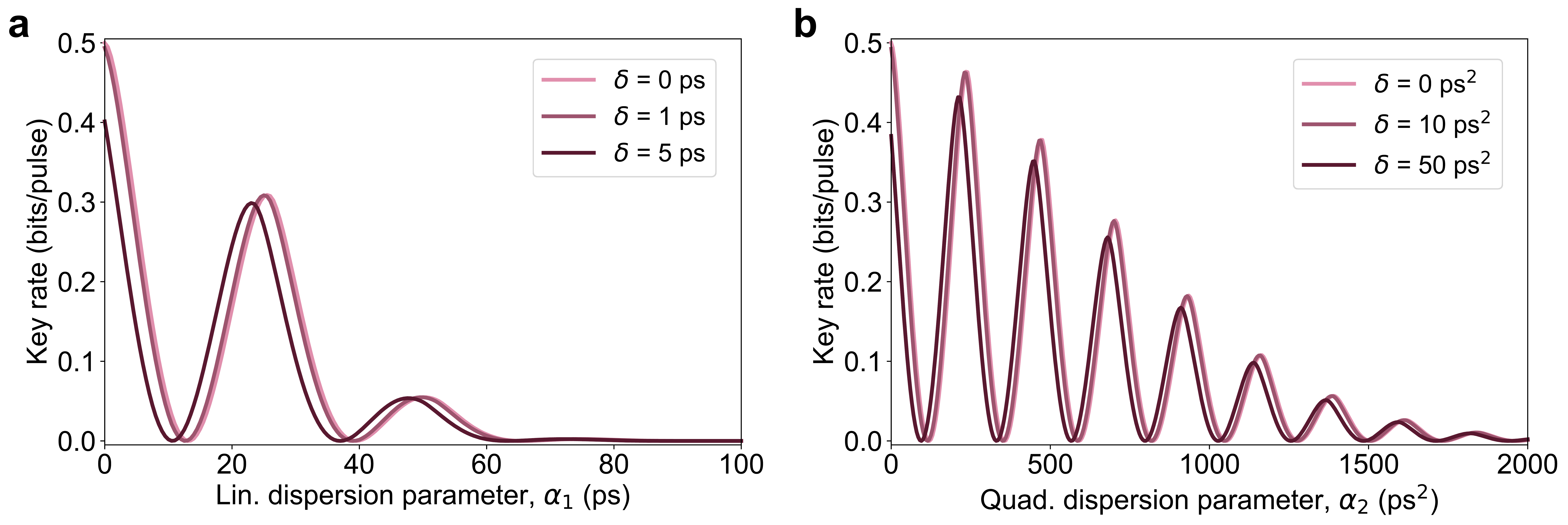}
    \caption{\textbf{Robustness of the time-bin Bell-states under dispersion noise with deviations between the noise parameters on the two photons.} Fidelity between the arriving state $\ket{\Psi'^-_t}$ sent by Alice through the dispersion channel with deviating noise parameters on the two photons and the state $\ket{\Psi^-_t}$ measured by Bob. We use values close to the experimentally reported values in Refs.~\cite{finco2024timebinentangledbellstate, Clementi2023}, with $\tau_0=0$ ps, $\tau_1=220$ ps, and $\sigma_t = 30$ ps and a central frequency of 195.860 THz. The key rate is plotted as a function of (a) linear dispersion, $\alpha_1$, and (b) quadratic dispersion $\alpha_2$, with the dispersion parameters offset by an absolute fixed value $\delta$.}
    \label{fig:unequalGVD}
\end{figure*}

We present the effect of this in Fig.~\ref{fig:unequalGVD}. We see that this offset in the dispersion parameter causes a translation of the key rate with $\alpha_n$. As discussed for the FBS model, typically experimental deviations between the two photons are small. Therefore, this also has a minimal effect on the key rate of our encoding.
Additionally, this can be accounted for in the optimisation of the encoding parameters.
Thus, our protocol performs well under and is high applicability to experimentally expected scenarios.

\section{Frequency-dependent channel loss}
In many practical set-ups, for example, in optical fibres, the absorption of the light is dependent on the frequency of the light \cite{Paschottapropagation_losses}. Since we encode photons in the frequency domain across different frequency modes, we should briefly consider the effect of this loss on our encoding.

The logical states are encoded in antisymmetric Bell-states, hence both photons in each state are equally superposed across the different frequencies. Therefore, the effect of loss should be the same on both photons in the logical state, and it is quick to see that the effect of the channel for the frequency-bin Bell-states is simply loss with a probability given by the average of the loss for the two frequencies of the encoding. For the time-bin Bell-states, the average loss will be given by the loss probability of the central frequency of the encoding. This will give rise to an amplitude damping effect on the key rate, which can be small if the encoded frequencies are appropriately chosen.


\end{document}